# Understanding liquid-jet atomization cascades via vortex dynamics

A. Zandian[1], W. A. Sirignano[1]† and F. Hussain[2]

[1]Department of Mechanical and Aerospace Engineering, University of California, Irvine, CA 92697, USA

[2]Department of Mechanical Engineering, Texas Tech University, Lubbock, TX 79409, USA

Temporal instabilities of a planar liquid jet are studied using direct numerical simulation (DNS) of the incompressible Navier-Stokes equations with level-set (LS) and volume-of-fluid (VoF) surface tracking methods. $\lambda_2$ contours are used to relate the vortex dynamics to the surface dynamics at different stages of the jet breakup, namely, lobe formation, lobe perforation, ligament formation, stretching, and tearing. Three distinct breakup mechanisms are identified in the primary breakup, which are well categorized on the parameter space of gas Weber number ($We_g$) versus liquid Reynolds number ($Re_l$). These mechanisms are analyzed here from a vortex dynamics perspective. Vortex dynamics explains the hairpin formation, and the interaction between the hairpins and the Kelvin-Helmholtz (KH) roller explains the perforation of the lobes, which is attributed to the streamwise overlapping of two oppositely-oriented hairpin vortices on top and bottom of the lobe. The formation of corrugations on the lobe front edge at high $Re_l$ is also related to the location and structure of the hairpins with respect to the KH vortex. The lobe perforation and corrugation formation are inhibited at low $Re_l$ and low $We_g$ due to the high surface tension and viscous forces, which damp the small scale corrugations and resist hole formation. Streamwise vorticity generation - resulting in three-dimensional instabilities - is mainly caused by vortex stretching and baroclinic torque at high and low density ratios, respectively. Generation of streamwise vortices and their interaction with spanwise vortices produce the liquid structures seen at various flow conditions. Understanding the liquid sheet breakup and the related vortex dynamics are crucial for controlling the droplet size distribution in primary atomization.

**Key words:**

---

## 1. Introduction

Earlier computational works on the breakup of liquid streams at higher Weber number ($We$) and Reynolds number ($Re$) (i.e. in the atomization range) focused on the surface dynamics using either volume-of-fluid or level-set methods (Shinjo & Umemura 2010; Desjardins & Pitsch 2010; Herrmann 2011). More recently, Jarrahbashi & Sirignano (2014) and Jarrahbashi *et al.* (2016) numerically simulated the temporal behavior of round jets and Zandian *et al.* (2017) computed the temporal behavior of planar jets with additional data analysis that related the vorticity dynamics to the surface dynamics. Zandian *et al.* (2017) presented several significant accomplishments: (i) three breakup mechanisms were identified and their zones of occurrence were specified on the gas-phase Weber number ($We_g$) versus liquid-phase Reynolds number ($Re_l$) map; (ii) the most important actions in each of the three breakup domains were explained; (iii) the effects of density ratio, viscosity ratio, and sheet thickness on the breakup domains were described; (iv) characteristic times for each of these breakup domains were correlated with key parameters; and (v) the same breakup domains were shown to apply for round jets and planar jets with a very similar $We_g$ versus $Re_l$ map.

† Email address for correspondence: sirignan@uci.edu

In recent years, a number of analyses for spatially developing instability and breakup of liquid streams have appeared. They do add interesting and useful information; however, all of those analyses are at relatively low values of Weber number ($We_g < 100$). That is, although some of those works are described as "atomization" studies, they all fit better under the classical characterization of "wind-induced capillary instabilities" given by Ohnesorge (1936) and Reitz & Bracco (1986). Ling *et al.* (2017) use three-dimensional direct numerical simulation (DNS) and resolve the smaller scales; they treat air-assisted injection of a planar sheet and give detailed discussion about the challenge of numerical accuracy. While they mention briefly vortex dynamics and the use of the $\lambda_2$ method, little detail is given. Zuzio *et al.* (2016) include "preliminary" results for sheet breakup in their 3D DNS analysis. The other papers give analyses that are linear (Otto *et al.* 2013), two-dimensional inviscid (Matas *et al.* 2011), two-dimensional (Fuster *et al.* 2013; Agbaglah *et al.* 2017), or three-dimensional large-eddy simulations (Agbaglah *et al.* 2017). Of course, these methods cannot resolve the smaller structures that form during the cascade process of the breakup. An analysis with spatial development offers some advantage with practical realism over temporal analysis. At the same time, the additional constraints given by the boundary conditions remove generality in the delineation of the important relevant physics. For these reasons, we follow the path with temporal-instability analysis in the classical atomization (high $We_g$ range) provided by Jarrahbashi & Sirignano (2014), Jarrahbashi *et al.* (2016), and Zandian *et al.* (2017). The goal is to reveal and interpret the physics in the cascade process known as atomization. Note that some spatial development is provided when the temporal analysis covers a domain that is several wavelengths in size. Using linear theory, relations between spatially developing results and temporal results have been demonstrated for single-phase flows (Gaster 1962) and two-phase flows (Fuster *et al.* 2013).

Jarrahbashi *et al.* (2016) showed that important spray characteristics, e.g. droplet size and spray angle, differed in different ranges of $We$, $Re$, and density ratio. Therefore, further studies of the breakup mechanisms are needed to better understand the causes of these differences. Consequently, there are unresolved questions to be addressed in this paper: What are the details of the liquid dynamics in each breakup domain? What causes the difference in the breakup cascade? What roles do surface tension, liquid viscosity, and gas density (i.e. pressure) play? How do the roles of streamwise vorticity (i.e. hairpin vortices) differ in the three breakup domains? How does the behavior of a jet flow into an alike fluid (e.g. water into water or air into air) compare with liquid-jet flow into gas? The answers to these questions would be crucial in understanding and controlling the ligament and droplet size distribution in the primary atomization of liquid jets.

Vortex dynamics concepts can shed further light on surface deformation of a liquid jet in the primary atomization process - a cascade involving the formation of smaller and smaller liquid structures. The Kelvin-Helmholtz (KH) instability at the liquid-gas interface promotes the growth of spanwise vorticity waves forming coherent vortices. These vortices evolve into hairpins with counter-rotating streamwise legs (Bernal & Roshko 1986). The streamwise and spanwise vortical waves combine to produce different surface structures, e.g. lobes, bridges, and ligaments, which eventually break up into droplets.

The link between the vortex dynamics and surface dynamics in primary atomization is important, but rarely explored and poorly understood; hence this study is an attempt to fill that gap.

There have been several studies of the jet instabilities from the vortex dynamics perspective. Most of them, however, do not address density and viscosity discontinuities. These studies have mainly focused on understanding and relating vortex stretching (Pope 1978), vortex tilting (Lasheras & Choi 1988), and baroclinic effects (Schowalter *et al.* 1994) to the three-dimensional liquid jet instabilities. Earlier experimental studies in this field (Widnall & Sullivan 1973; Widnall *et al.* 1974; Breidenthal 1981; Jimenez 1983; Bernal & Roshko 1986; Lasheras & Choi 1988; Liepmann & Gharib 1992; Schowalter *et al.* 1994) have been followed and reproduced in more detail by numerical simulations (Ashurst & Meiburg 1988; Martin & Meiburg 1991; Comte *et al.* 1992; Collis *et al.* 1994; Schoppa *et al.* 1995; Brancher *et al.* 1994; Danaila *et al.* 1997; Shinjo & Umemura 2010; Jarrahbashi & Sirignano 2014; Jarrahbashi *et al.* 2016). The vorticity dynamics of planar mixing layers and jets flowing into like-density fluids have been reviewed by Jarrahbashi & Sirignano (2014) and Jarrahbashi *et al.* (2016).

In the first studies of the role of streamwise vorticity in round liquid jets flowing into a gas, Jarrahbashi & Sirignano (2014) and Jarrahbashi *et al.* (2016) showed how lobe and ligament formation mechanisms relate to augmentation of streamwise vorticity. They also showed that a natural mode number of lobes exists for a given configuration and cannot be changed by weak forcing (Jarrahbashi & Sirignano 2014); however, strong forcing can produce lobes of different mode numbers.

Empirical evidence (Lefebvre 1989) has long been available that spray character differs significantly for differing values of $Re$ and $We$. Jarrahbashi *et al.* (2016) showed that different breakup mechanisms result in differing spray angles and droplet-size distributions. Thus, we see that, for control of spray character, it is very valuable to understand the details of the cascade processes for each of the identified atomization domains. Control and optimization, although not addressed in this study, motivate the detailed exploration and the behavioral characterizations reported here.

### 1.1. *Liquid jet breakup domains*

Jarrahbashi *et al.* (2016) found three distinct physical domains in round liquid jets, including hole formation and lobe stretching, and explained surface wave dynamics, vortex dynamics and their interactions. The perforations were correlated with the fluid motion induced by the hairpin and helical vortices. They also found that the hole formation process is dominated by inertia rather than capillary forces, and the hole merging was related to the slower development of hairpin vortices and lobe shape.

Zandian *et al.* (2017) identified three mechanisms for liquid sheet surface deformation and breakup, which were well categorized on a gas Weber number ($We_g$) versus liquid Reynolds number ($Re_l$) map, shown in figure 1. The red symbols on this diagram indicate the results for a lower density ratio of 0.05, not present in the original diagram of Zandian *et al.* (2017), but recently added following our new data. The liquid structures seen in either one of these breakup domains are sketched in figure 2, where the evolution of a liquid lobe is shown from a top view. At high $Re_l$, the liquid sheet breakup characteristics change based on a modified Ohnesorge number, $Oh_m \equiv \sqrt{We_g}/Re_l$, as follows: (i) at high $Oh_m$ and high $We_g$, the lobes become thin and puncture, creating holes and bridges. Bridges break as perforations expand and create ligaments. Ligaments then stretch and break into droplets by capillary action. This domain is indicated as Atomization Domain II in figure 1, and its mechanism was called *LoHBrLiD* based on the cascade of structures seen in this domain ($Lo \equiv$ Lobe, $H \equiv$ Hole, $Br \equiv$ Bridge, $Li \equiv$ Ligament, and $D \equiv$ Droplet); and (ii) at low $Oh_m$ and high $Re_l$, holes are not seen at early times; instead, many corrugations form on the lobe front edge and stretch into ligaments. This mechanism is called $LoCLiD$ ($C \equiv$ Corrugation) and occurs in Atomization Domain III (see figure 1), and results in ligaments and droplets without the hole and bridge formations. The third mechanism follows a $LoLiD$ process and occurs at low $Re_l$ and low $We_g$ (Atomization Domain I), but with some difference in details from the $LoCLiD$ process. The main difference between the two ligament formation mechanisms at high and low $Re_l$'s is that at higher $Re_l$ the lobes become corrugated before stretching into ligaments. Hence, each lobe may produce multiple ligaments, which are typically thinner and shorter than those at lower $Re_l$. At low $Re_l$, on the other hand, because of the higher viscosity, the entire lobe stretches into one thick, usually long ligament.

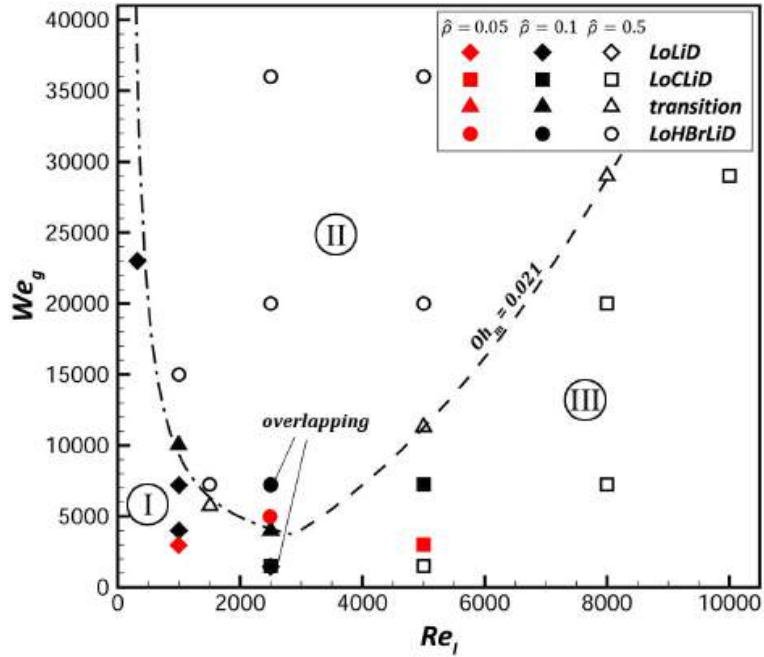

FIGURE 1. The breakup characteristics based on $We_g$ and $Re_l$, showing the $LoLiD$ mechanism (Atomization Domain I) denoted by diamonds, the $LoHBrLiD$ mechanism (Atomization Domain II) denoted by circles, the $LoCLiD$ mechanism (Atomization Domain III) denoted by squares, and the transitional region denoted by triangles. The cases with density ratio of 0.1 ($\hat{\rho} = 0.1$) are shaded. The $\hat{\rho} = 0.1$ and $\hat{\rho} = 0.5$ cases that overlap at the same point on this diagram are noted. $-\cdot-\cdot-$, transitional boundary at low $Re_l$; and $---$, transitional boundary at high $Re_l$ (Zandian *et al.* 2017). The red symbols denote the cases for lower $\hat{\rho} = 0.05$, added to the original diagram.

There is also a transitional region in the $We_g$-$Re_l$ map, in which both lobe/ligament stretching and hole-formation mechanisms occur simultaneously. The transitional region at low $Re_l$ follows a hyperbolic relation, i.e. $We_g = A/Re_l$, shown by the dash-dotted line in figure 1; while in the high $Re_l$ limit, it follows a parabolic relation, i.e. $We_g = B^2 Re_l^2$, shown by the dashed line in figure 1. The constant $B$ is a critical $Oh_m$ at high $Re_l$, $B \approx 0.021$ (Zandian *et al.* 2017).

### 1.2. *Objectives*

Our objectives for the planar jet are to (i) explain the mechanisms of surface deformation and breakup in the three domains introduced by Zandian *et al.* (2017) using more sophisticated data analysis for the vortex dynamics (i.e. the $\lambda_2$ method); (ii) determine the importance of streamwise vorticity (i.e. hairpin vortices) in the breakup mechanisms; (iii) identify the generation mechanisms for the streamwise vorticity; and (iv) learn the differences in generation and role of the spanwise and streamwise vorticity at low and high density ratios.

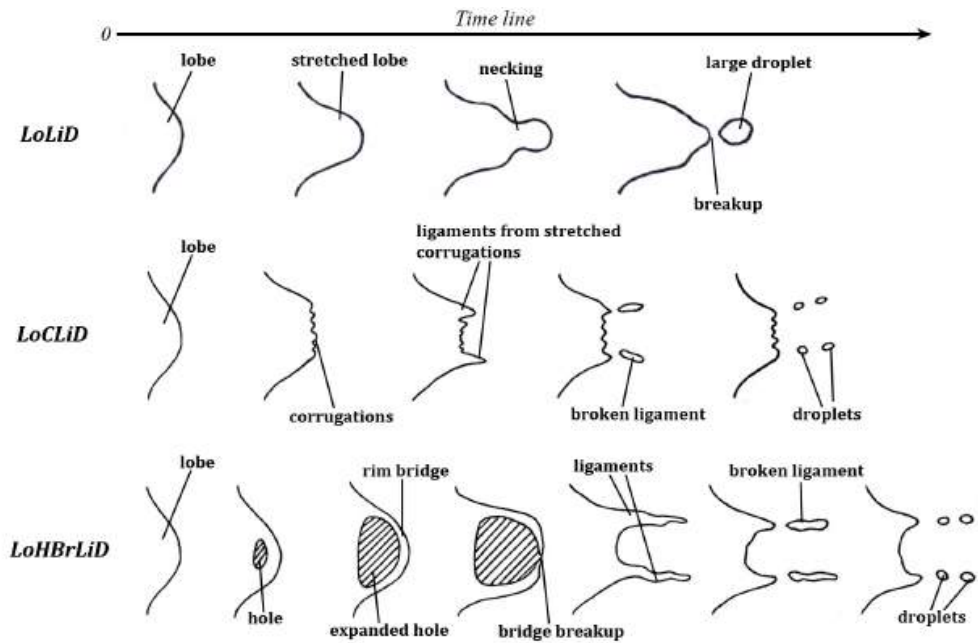

FIGURE 2. Cascade of structures for the *LoLiD* (top), *LoCLiD* (center), and *LoHBrLiD* (bottom) processes; sketch showing top view of a liquid lobe undergoing these processes. The gas flows on top of these structures from left to right, and time increases to the right.

In §2, the numerical methods and the most important flow parameters are presented along with the calculations for vortex identification. Section 3 is devoted to our numerical results and their analyses. The vortex structures are tracked in time to explain the liquid structures in the three breakup mechanisms. The hole formation, corrugation formation, and lobe/ligament stretching are analyzed in §3.2, §3.3, and §3.4, respectively. The streamwise vorticity generation is studied in detail in §3.5. The similarities between the vortex dynamics of planar and circular jets are discussed in §3.6. Finally, our findings and conclusions are summarized in §4.

## 2. Numerical Modeling

### 2.1. *Governing Equations*

The governing equations are the continuity, momentum and level-set/volume-of-fluid equations. Since both the gas and the liquid are incompressible, the continuity equation is

$$\boldsymbol{\nabla} \cdot \boldsymbol{u} = 0, \tag{2.1}$$

where $\boldsymbol{u}$ is the velocity vector. The momentum equation, including the viscous diffusion and surface tension forces and neglecting the gravitational force, is

$$\frac{\partial(\rho \boldsymbol{u})}{\partial t} + \boldsymbol{\nabla} \cdot (\rho \boldsymbol{u}\boldsymbol{u}) = -\boldsymbol{\nabla} p + \boldsymbol{\nabla} \cdot (2\mu \boldsymbol{D}) - \sigma \kappa \delta(d)\boldsymbol{n}, \tag{2.2}$$

where $p$, $\rho$ and $\mu$ are the pressure, density and dynamic viscosity of the fluid, respectively. $\boldsymbol{D}$ is the rate of deformation tensor,

$$\boldsymbol{D} = \frac{1}{2}\left[(\boldsymbol{\nabla}\boldsymbol{u}) + (\boldsymbol{\nabla}\boldsymbol{u})^T\right]. \tag{2.3}$$

The last term in equation (2.2) is the surface tension force per unit volume

$\boldsymbol{F} = -\sigma\kappa\delta(d)\boldsymbol{n}$; where $\sigma$ is the surface tension coefficient, $\kappa$ is the surface curvature, $\delta(d)$ is the Dirac delta function, $d$ is the distance from the interface, and $\boldsymbol{n}$ is the unit vector normal to the liquid/gas interface.

The level-set function $\phi$ is defined as a smooth distance function, which enables evaluation of the density, viscosity and surface tension at any distance from the interface in either gas or liquid zones. This method was developed by Osher and his coworkers (Zhao *et al.* 1996; Sussman *et al.* 1998; Osher & Fedkiw 2001). In this algorithm, the interface $\Gamma$ is the zero level set of $\phi$,

$$\Gamma = \{\boldsymbol{x} \mid \phi(\boldsymbol{x},t) = 0\}.$$

$\phi < 0$ in the liquid region and $\phi > 0$ in the gas region are taken. $\boldsymbol{u}$ is continuous across the interface. Since the interface moves with the fluid particles, the evolution of $\phi$ is then given by

$$\frac{\partial \phi}{\partial t} + \boldsymbol{u} \cdot \boldsymbol{\nabla}\phi = 0, \tag{2.4}$$

which is called the level-set equation. If the initial distribution of the level set is a signed distance function, after a finite time of being convected by a nonuniform velocity field, it will not remain a distance function. Therefore, the level-set function should be re-initialized in such a way that it will be a distance function, without changing the zero level set (position of the interface). This is achieved by solving the following differential equation (Sussman *et al.* 1998);

$$\frac{\partial d}{\partial \tau} = \text{sign}(\phi)(1 - |\boldsymbol{\nabla}d|), \tag{2.5}$$

with the initial condition

$$d(\boldsymbol{x}, 0) = \phi_0(\boldsymbol{x}),$$

where $\tau$ is a pseudo time. The steady solutions of equation (2.5) are distance functions. Furthermore, since $\text{sign}(0) = 0$, then $d(\boldsymbol{x}, t)$ has the same zero level set as $\phi(\boldsymbol{x})$.

Since the density and viscosity are constant in each fluid, they take on two different values depending on the sign of $\phi$; we can write

$$\rho(\phi) = \rho_l + (\rho_g - \rho_l)\text{H}(\phi), \tag{2.6}$$

and

$$\mu(\phi) = \mu_l + (\mu_g - \mu_l)\text{H}(\phi), \tag{2.7}$$

where $\text{H}(\phi)$ is a smoothed Heaviside function, and the subscripts $g$ and $l$ refer to gas and liquid, respectively. Using these expressions, the governing equation for the fluid velocity $\boldsymbol{u}$ (equation 2.2) can be written as a single equation containing both liquid and gas properties.

At low density ratios, we use a transport equation similar to the level-set equation (2.4) for the volume fraction $f$, also called the volume-of-fluid (VoF) variable, in order to describe the temporal and spatial evolution of the two phase flow (Hirt & Nichols 1981);

$$\frac{\partial f}{\partial t} + \boldsymbol{u} \cdot \boldsymbol{\nabla}f = 0, \tag{2.8}$$

where the VoF-variable $f$ represents the volume of (liquid phase) fluid fraction as follows:

$$f(t) = \begin{cases} 0, & \text{outside of liquid pahse} \\ 0 < f < 1, & \text{at the interface} \\ 1, & \text{inside the liquid phase}. \end{cases} \tag{2.9}$$

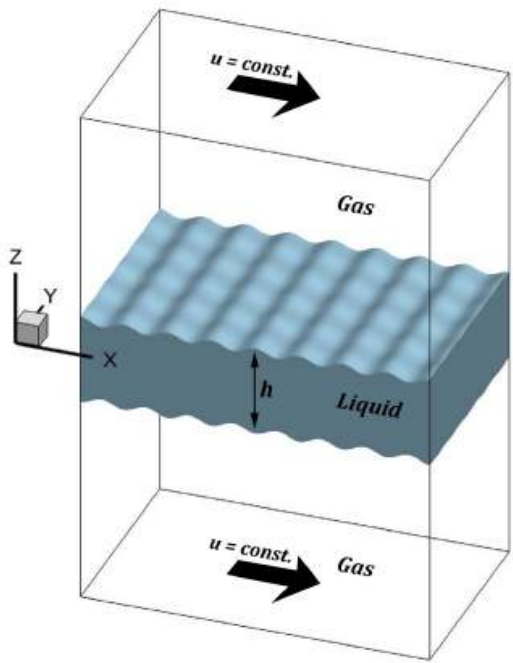

FIGURE 3. The computational domain with the initial liquid and gas zones.

The normal direction of the fluid interface is found where the value of $f$ changes most rapidly. With this method, the free surface is not defined sharply; instead, it is distributed over the height of the cell. Since fluid properties are required every time step in order to solve the Navier-Stokes equations, the density and viscosity change continuously based on equations (2.6) and (2.7). However, in these equations, the argument of the heaviside function is replaced by $f$, in the VoF method.

The formulation for the fully conservative momentum convection and volume fraction transport, the momentum diffusion, and the surface tension are treated explicitly. To ensure a sharp interface of all flow discontinuities and to suppress numerical dissipation of the liquid phase, the interface is reconstructed at each time step by the PLIC (piecewise linear interface calculation) method proposed by Rider & Kothe (1998). The liquid phase is transported on the basis of its reconstructed distribution. The capillary effects in the momentum equations are represented by a capillary tensor as introduced by Scardovelli & Zaleski (1999).

### 2.2. *Flow Configuration*

The computational domain, shown in figure 3, consists of a cube, which is discretized into uniform-sized cells. The liquid segment, which is a sheet of thickness $h$ ($h = 50$ μm for the thin sheet and 200 μm for the thick sheet in this study), is centered at the middle of the box. The domain size in terms of the sheet thickness $h$ is $16h \times 10h \times 10h$, in the $x$, $y$ and $z$ directions, respectively, for the thin sheet, and $4h \times 4h \times 8h$ for the thick sheet. The liquid segment is surrounded by gas zones on top and bottom. The gas moves in the positive $x$- (streamwise) direction, with a constant velocity ($U = 100$ m/s) at the top and bottom boundaries, and its velocity diminishes to the interface velocity with a boundary layer thickness obtained from 2D full-jet simulations. In the liquid, the velocity decays to zero at the center of the sheet with a hyperbolic tangent profile. For more detailed description of the initial conditions, see figure 12 of Zandian *et al.* (2016).

Our study involves a temporal computational analysis with a relative velocity between the two phases. Due to friction, the relative velocity decreases with time. Furthermore, the domain is several wavelengths long in the streamwise direction so that some spatial development occurs. In order to reduce the dependence of the results on details of the boundary conditions, specific configurations (e.g., air-assist or air-blast atomization) are avoided. However, calculations are made with the critical non-dimensional parameters in the ranges of practical interest.

The liquid/gas interface is initially perturbed symmetrically on both sides with a sinusoidal profile and predefined wavelength and amplitude obtained from the 2D full-jet simulations (see Zandian *et al.* 2016). Two analyses without forced or initial surface perturbations - the full-jet 2D simulations (figure 11 of Zandian *et al.* 2016) and the initially non-perturbed 3D simulations (figure 13 of Zandian *et al.* 2016, and figure 51 of the current article) - show KH wavelengths in the moderate range of 80–125 μm over a wide range of $Re_l$, $We_g$ and $\hat{\rho}$ studied here. In order to expedite the appearance and growth of the KH waves, initial perturbations with wavelength of 100 μm are imposed on the interface, with a small amplitude of 4 μm. The amplitude of the perturbations is small enough so that any other subharmonic wavelengths would have a chance to form and grow. As shown by Zandian *et al.* (2017) and Jarrahbashi & Sirignano (2014), at higher $Re_l$ and lower $\hat{\rho}$ a variety of smaller and larger wavelengths also form; at very low $We_g$, the waves also merge to create longer wavelengths. Both streamwise ($x$-direction) and spanwise ($y$-direction) perturbations are considered in this study. The amplitude of the spanwise perturbations is reduced to 2.5 μm to let the other subharmonic waves appear naturally. At higher $Re_l$, smaller spanwise waves (formed first as corrugations) appear superimposed on the initial perturbations (Zandian *et al.* 2017). Similarly, the waves merge to create larger waves (lobes) at lower $We_g$. Periodic boundary conditions for all components of velocity as well as the level-set/VoF variable are imposed on the four sides of the computational domain; i.e. the $x$- and $y$-planes.

The most important dimensionless groups in this study are the Reynolds number ($Re$), the Weber number ($We$), and the gas-to-liquid density ratio ($\hat{\rho}$) and viscosity ratio ($\hat{\mu}$), as defined below. The initial KH wavelength-to-sheet-thickness ratio ($\Lambda$) is also an important parameter that defines the relative length of the initial perturbations;

$$Re = \frac{\rho_l U h}{\mu_l}, \quad We = \frac{\rho_l U^2 h}{\sigma}, \quad \hat{\rho} = \frac{\rho_g}{\rho_l}, \quad \hat{\mu} = \frac{\mu_g}{\mu_l}, \quad \Lambda = \frac{\lambda}{h}. \quad (2.10)$$

The sheet thickness $h$ is considered as the characteristic length, and the relative gas–liquid velocity $U$ as the characteristic velocity. Theoretically, if the flow field is infinite in the streamwise direction (as in our study), a Galiliean transformation shows that only the relative velocity between the two streams is consequential. Spatially developing flow fields, however, are at most semi-infinite so that both velocities at the flow-domain entry and their ratio (or their momentum ratio) are important. A wide range of $Re$ and $We$ at high and low density ratios is covered in this research. The main six cases studied in this article are presented in table 1. The cases are called in a "Dnx" format, where "n" presents the Domain number (1, 2 or 3), and "x" presents the range of the case ("a" for high density ratio, and "b" for low density ratios). Two cases – one at high density ratio and fairly high $We_g$, and one at low $\hat{\rho}$ and low $We_g$ – are studied in each domain to clearly show the effects of density ratio on the vortex dynamics of each atomization process. The "b" cases are more practical and in the ranges usually seen in most atomization applications. The range of parameters for other cases presented in this article is indicated in their corresponding figure captions.

### 2.3. *Numerical Methods*

Direct numerical simulation (DNS) is done by using an unsteady three-dimensional finite-volume solver to solve the Navier-Stokes equations for the planar incompressible liquid sheet segment (initially stagnant), which is subject to instabilities due to a gas stream that flows past it on both sides. The level-set (LS) and volume-of-fluid (VoF) methods are used for liquid/gas interface tracking. The temporal deformation of the liquid-gas interface is predicted, resulting in three-dimensional instabilities, that can lead to ligament formation and sheet breakup. The original code containing the LS subroutines was developed and applied by Dabiri *et al.* (2007), and later by Jarrahbashi & Sirignano (2014), Jarrahbashi *et al.* (2016), and Zandian *et al.* (2016). Because of the weakness of the level-set method in mass conservation at low density ratios, the VoF method is used at low gas densities.

| Case | $Re_l$ | $We_g$ | $\hat{\rho}$ | $\hat{\mu}$ |
|------|--------|--------|------|------|
| D1a | 320 | 23000 | 0.1 | 0.002 |
| D1b | 1000 | 3000 | 0.05 | 0.01 |
| D2a | 5000 | 20000 | 0.5 | 0.006 |
| D2b | 2500 | 5000 | 0.05 | 0.01 |
| D3a | 5000 | 7250 | 0.5 | 0.006 |
| D3b | 5000 | 3000 | 0.05 | 0.01 |

TABLE 1. The main cases studied with their dimensionless parameter values

A uniform staggered grid is used with the mesh size of 2.5 μm and a time step of 5 ns – finer grid resolution of 1.25 μm is used for the case with higher $We_g$ ($We_g = 115\,000$ and $Re_l = 320$) or higher $Re_l$ ($Re_l = 5000$ and $We_g = 3000, 7250, 20\,000$, and $36\,000$). Third-order accurate QUICK scheme is used for spatial discretization and the Crank-Nicolson scheme for time marching. The velocity-pressure coupling is established using the SIMPLE algorithm. The initial velocity profile and some other initial estimations, e.g. initial surface perturbation wavelength, are obtained from a 2D planar full-jet simulation with the same schemes and methodology as those of Zandian *et al.* (2016).

The grid independency tests were performed previously by Jarrahbashi & Sirignano (2014), Jarrahbashi *et al.* (2016), and Zandian *et al.* (2016). They showed that the errors in the size of the ligaments, penetration length of the liquid jet and the magnitude of the velocity computed using different mesh resolutions were within an acceptable range. The effects of the mesh size, the thickness of the fuzzy zone between the two phases, where properties have large gradients to approximate the discontinuities, and mass conservation of the level-set method have been previously addressed by Jarrahbashi & Sirignano (2014). The effects of mesh resolution on the most important flow parameters, e.g. surface structures, velocity and vorticity profiles, are studied in detail in §2.4. The domain-size independency was also checked in both streamwise and spanwise directions to make sure that the resolved wavelengths were not affected by the domain length or width. The normal dimension of the domain was chosen such that the top and bottom boundaries remain far from the interface at all times, so that the surface deformation is not directly affected by the boundary conditions. In addition, accuracy tests and validation with experiments and other numerical approaches were performed previously (Dabiri *et al.* 2007; Jarrahbashi & Sirignano 2014), and will not be repeated here.

### 2.4. *Effects of mesh resolution*

The computational grid should be able to resolve well the boundary layer of the gas and liquid near the interface in order to capture the frequency and growth of the interfacial instabilities accurately (Fuster *et al.* 2013). More importantly, the requirement for sufficient numerical resolution to compute the formation of holes and corrugations on the lobes indeed is a stricter restriction on the mesh size in the current study. To assess the effects of grid resolution on the numerical results of the DNS simulation of the atomization, three grid resolutions are considered in this section: M1, M2, and M3, with $\Delta = 5$ μm, 2.5 μm, and 1.25 μm, respectively.

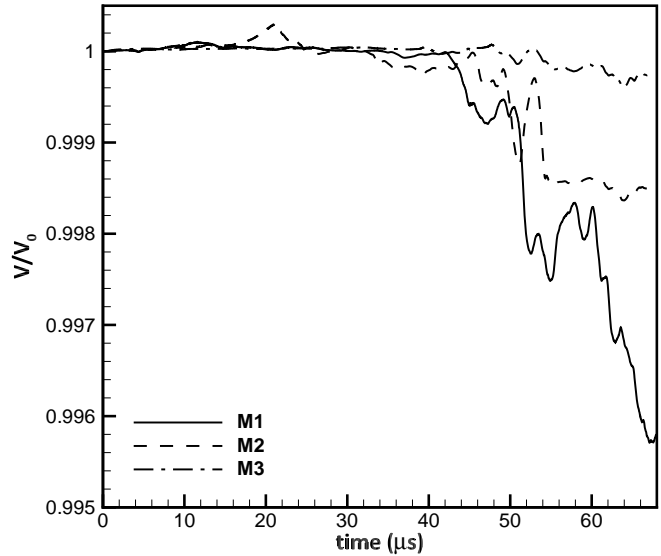

FIGURE 4. Mass conservation check for different mesh resolutions.

Figure 4 shows the mass conservation error in different mesh resolutions. The vertical axis measures the instantaneous volume of the liquid phase $V$ normalized by the initial volume of the liquid in the domain $V_0$. The coarse grid M1 clearly has the largest mass loss with an error of around 0.5% near the end of the computation. The M2 and M3 grids, however, both have fair mass conservation with low errors of 0.15% and $< 0.1$%, respectively. The results shown in figure 4 are for a case with $\hat{\rho} = 0.05$. At higher density ratios, the mass conservation is significantly better.

A close-up view of the different liquid surface structures found in different atomization domains are shown in figure 5 for different mesh resolutions of the same cases and at the same time steps. Figures 5($a$–$c$) show the effects of mesh resolution on depicting the hole formation in Domain II. Physically, holes are formed on the liquid lobes only when the lobe thickness is very small ($O(10)$ nm) and the disjoining pressure becomes active (Ling *et al.* 2017). Here in the simulations, holes appear when the thickness of the liquid lobe (sheet) decreases to about the cell size $\Delta$. Since mechanisms of sheet rupture, such as disjoining pressure, are absent in the present study, and because of this numerical cutoff criterion, the initial perforation of the liquid sheets is highly dependent on the mesh resolution. As can be seen in figures 5($a$–$c$), the M1 grid does not have enough resolution to capture the hole formation accurately. The holes form much sooner and expand much faster in the M1 grid compared to the other two finer grids. As seen in figure 5($c$), the large perforation seen on the lobe is a result of merging of several smaller holes on the liquid sheet, which can be captured only by a high resolution grid such as M3. The entire size of the hole is fairly similar for the M2 and M3 grids. The lobe rim is much thicker for finer resolutions. Consequently, a further increase of mesh resolution will only delay the pinch-off point but will not affect the ligaments formed from the expansion of the holes. Hole formation and rim dynamics are well captured by the M2 and M3 grids.

At high $Re_l$, in Domain III of figure 1, corrugations form on the lobe rim. There are typically three or four such small scale corrugations (per wavelength) on the lobe rims, each having a size (thickness) of around 15–25 µm. Clearly, the M1 grid does not have enough resolution to capture these corrugations, and only the larger ligaments are resolved (figure 5$d$). The M2 and M3 grids both resolve the corrugations correctly with very similar sizes and structures (figures 5$e,f$). Figures 5($g,h$) compare the droplet formation and the cascade of structures for the M2 and M3 grids. The general aspects of the process, e.g. size and location of the lobes, ligaments, and holes and thickness of the rims, are fairly similar for both grids. However, as described before, because of the numerical cut-off criterion, the holes form and expand slightly sooner ($< 1$ µs) for the M2 grid. The size of the resolved droplets and ligaments are also slightly smaller in the M2 grid. This is similar to what was seen by Ling *et al.* (2017) and Jarrahbashi & Sirignano (2014) in their studies of the grid resolution. It is well known that the size of corrugations and the resulting ligaments and droplets decrease as $Re_l$ and $We_g$ increase. In the current study, the M3 grid is used only for those higher $Re_l$ and $We_g$, and for the rest of the cases the M2 grid is adequate to capture the physics of the cascade process. As shown in figure 5($i$) for a high $Re_l = 5000$ and high $We_g = 36\,000$, the M3 grid resolution is fine enough to capture the smallest holes, ligaments, and droplets. The radii of curvature of the corrugations and holes and ligaments can go down to around 6 µm, which can be resolved well only by the M3 grid with $\Delta = 1.25$ µm.

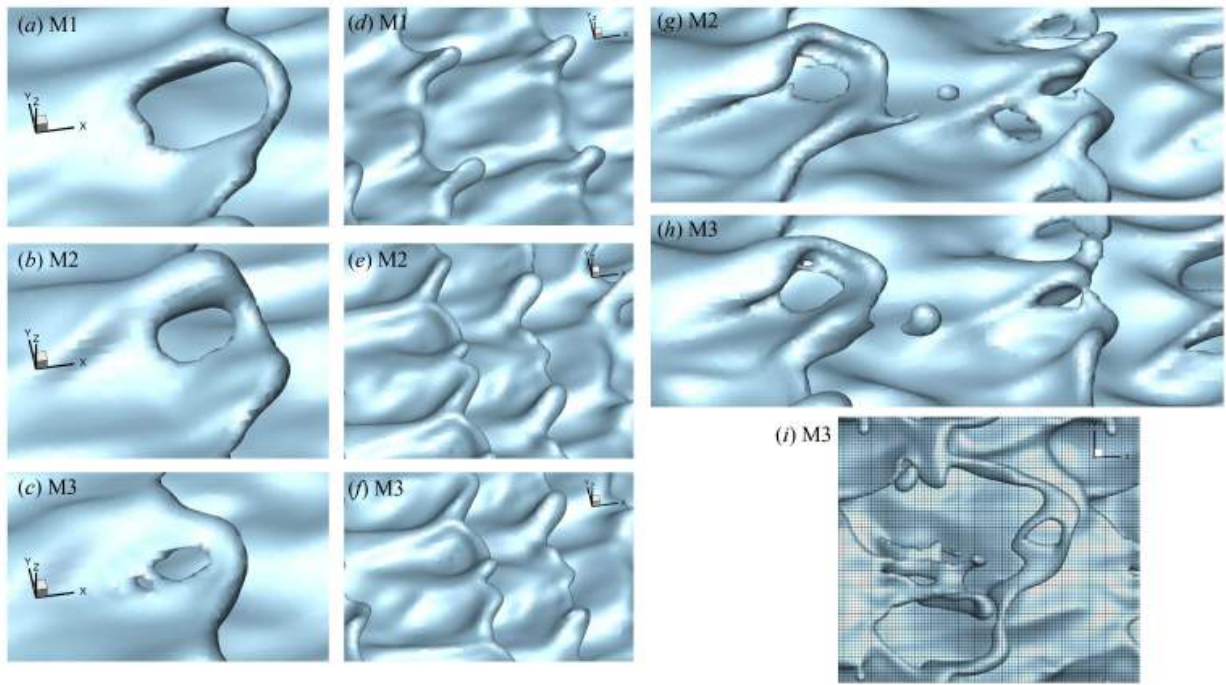

FIGURE 5. Close-up view of the liquid surface structures formed at the wave crest for different mesh resolutions. Hole formation in Domain II, for $Re_l = 2500$ and $We_g = 7250$ at 90 µs ($a$–$c$); corrugation formation in Domain III, for $Re_l = 5000$ and $We_g = 7250$ (Case D3a) at 90 µs ($d$–$f$); droplet formation in Domain II, for $Re_l = 2500$ and $We_g = 7250$ at 95 µs ($g$–$h$). Comparison of the relative sizes of holes, ligaments and droplets with the grid size for $Re_l = 5000$ and $We_g = 36\,000$ at 13 µs ($i$). $\hat{\rho} = 0.1$, $\hat{\mu} = 0.0066$ and $\Lambda = 0.5$ for all images.

Since the major part of this study is concerned with the vortex dynamics of the cascade process, it is important that we resolve the velocity and vorticity in the gas-liquid mixing layer correctly. A comparison of the streamwise velocity and spanwise vorticity profiles near the wave crest at an early time ($t = 10$ µs) and a later time ($t = 70$ µs) are shown in figure 6. These profiles are obtained at the front-most tip of one of the lobe crests at two distinct times. All three grids resolve the gas and liquid boundary layers correctly at early times, with the M1 profile slightly shifted towards the sheet center near the interface (figure 6a). The gas vorticity layer $\delta_g$ imposed in this simulation is initially $\approx 200$ μm, and the velocity gradient, hence the vorticity magnitude near the interface increases with time (see figure 6c). At later times (figure 6c,d), the M1 grid fails to capture the velocity profile correctly and is also unable to resolve the small scale velocity fluctuations and the small vortices near the interface. The M2 and M3 grids, however, resolve the velocity fluctuations similarly with very little difference, and vividly portray the location and amplitude of those fluctuations. The liquid sheet bulk velocity increases in the meantime.

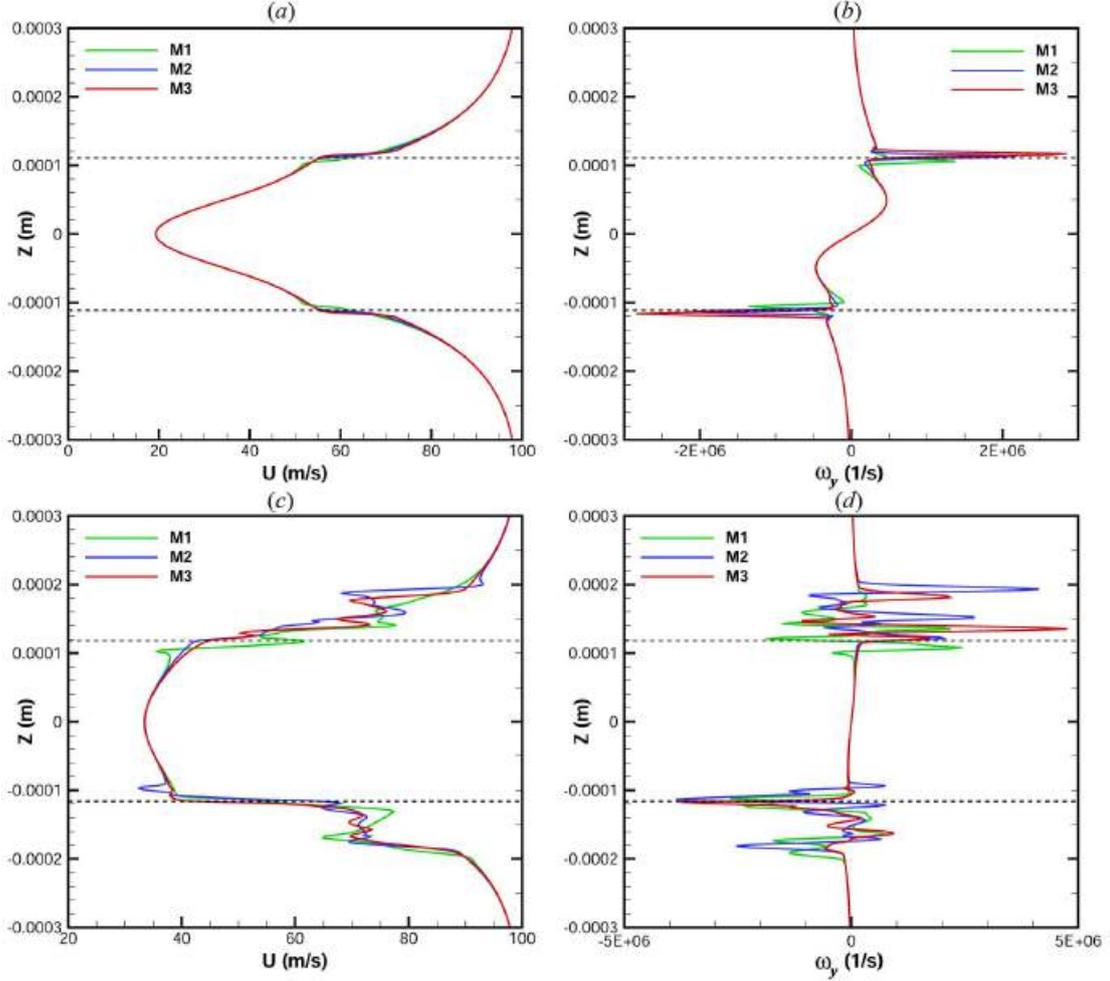

FIGURE 6. The streamwise velocity profile and the spanwise vorticity profile near the wave crest for Case D3a with different mesh resolutions at an early time $t = 10$ μs $(a),(b)$, and at a later time $t = 70$ μs $(c),(d)$. The gas-liquid interface locations are denoted by the dashed lines.

In terms of calculating the vorticity, figure 6(b) shows that the magnitude of the spanwise vorticity obtained from M1 grid is significantly lower than that of M2 and M3 even at early times. The location of the vorticity peak is also closer to the sheet center and inside the liquid phase because the velocity gradient and the vorticity magnitude cannot be predicted correctly with the M1 grid. At later times (figure 6d), this difference between the vorticity magnitudes of M1 and M2/M3 becomes even more noticeable, where the M1 grid cannot resolve the small vortices. The M2 and M3 grids have much better consistency in resolving the number of peaks, their locations and their magnitudes; however, the M3 grid is required for capturing all the vortices correctly at such high $Re_l$'s. The spanwise vorticity is not equally distributed between the two phases, and it mainly sits on the gas side since the vorticity thickness (and the boundary layer thickness) is larger in the gas phase compared to the liquid phase. The inclination of the vortices towards the gas phase is more pronounced at lower density ratios (Hoepffner *et al.* 2011).

Experiments and simulations show that as KH waves amplify, the convective velocity of the waves becomes very close to the Dimotakis speed (Dimotakis 1986) defined as $U_D = (U_l + \sqrt{\hat{\rho}} U_g)/(1 + \sqrt{\hat{\rho}})$. Considering the case shown in figure 6(*a*) with $\hat{\rho} = 0.5$ and $U_l = 20$ m/s, Dimotakis speed gives $U_D \approx 53$ m/s. The interface velocity (at the base of the KH wave) in our simulation (figure 6*a*) indicates a value of $U_{int} \approx 52$ m/s, which agrees well with the expected Dimotakis wave speed.

### 2.5. *Data analysis*

Our goal is to study the vorticity dynamics as well as the liquid surface dynamics in order to understand breakup mechanisms at different flow conditions. To this end, $\lambda_2$ contours at different cross-sections of the domain are analyzed in time. Here, we briefly review the definition of the $\lambda_2$ method.

An objective definition of a vortex should permit the use of vortex dynamics concepts to identify coherent structures (CS), to explain formation and evolutionary dynamics of CS, and to explore the role of CS in turbulence phenomena. Jeong & Hussain (1995) define a vortex core as a connected region with two negative eigenvalues of $\boldsymbol{S}^2 + \boldsymbol{\Omega}^2$; where, $\boldsymbol{S}$ and $\boldsymbol{\Omega}$ are the symmetric and anti-symmetric components of $\boldsymbol{\nabla u}$; i.e. $S_{ij} = \frac{1}{2}(u_{i,j} + u_{j,i})$ and $\Omega_{ij} = \frac{1}{2}(u_{i,j} - u_{j,i})$. If $\lambda_1$, $\lambda_2$, and $\lambda_3$, are the eigenvalues such that $\lambda_1 \geqslant \lambda_2 \geqslant \lambda_3$, this definition is equivalent to the requirement that $\lambda_2 < 0$ within the vortex core, since $\lambda_3$ is always negative because the sum of the normal viscous stresses is zero. This definition is proven to meet the requirements for existence of a vortex core in different flow conditions (Jeong & Hussain 1995), while the vortex identification by the Q-definition (Kolář 2007) may be incorrect when vortices are subjected to a strong external strain (Jeong & Hussain 1995), as in our study.

## 3. Results and discussion

In this section, the vorticity dynamics associated with each of the breakup mechanisms are analyzed to explain the hole formation, the corrugation formation and the lobe/ligament stretching at different Reynolds and Weber numbers. For this purpose, one case is picked from each domain (see figure 1), and its vorticity dynamics are studied using the $\lambda_2$ criterion. We make use of some of our earlier findings (Zandian *et al.* 2016), where necessary; however, those results are shown from a different perspective, to convey our findings more clearly. The gas flows in the positive *x*-direction, from left to right, in all of the figures in this section.

### 3.1. *Vortex dynamics and surface dynamics*

To interpret the surface deformation via vorticity concepts and to more clearly delineate the complex 3D flow physics, the dual approaches of vortex dynamics and surface wave dynamics are employed. The language of wave dynamics focuses on crests, troughs, and lobes, while the vortex-dynamics terminology refers to vortex rollers (eddies), braids, and hairpins. The instability starts from the initially symmetric KH waves. The crest of the KH waves, when amplified and slightly stretched in the flow direction, is referred to as "KH crest". As the 3D character of instability develops, "KH crests" divide into distinct "lobes". The forward-most tip of the lobe is referred to as "spanwise crest", and the region between two adjacent crests is called "spanwise trough". In terms of the vorticity dynamics, the "KH crests" and "KH troughs" are equivalent to "vortex rollers" and "braids", respectively; see figure 7.

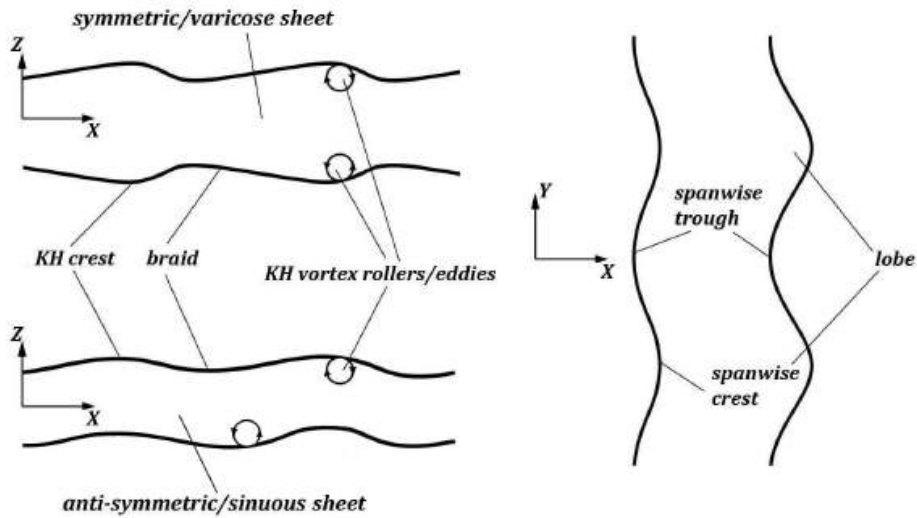

FIGURE 7. Terminology used for vorticity and liquid-gas interface deformations from side-view (left), and top view (right).

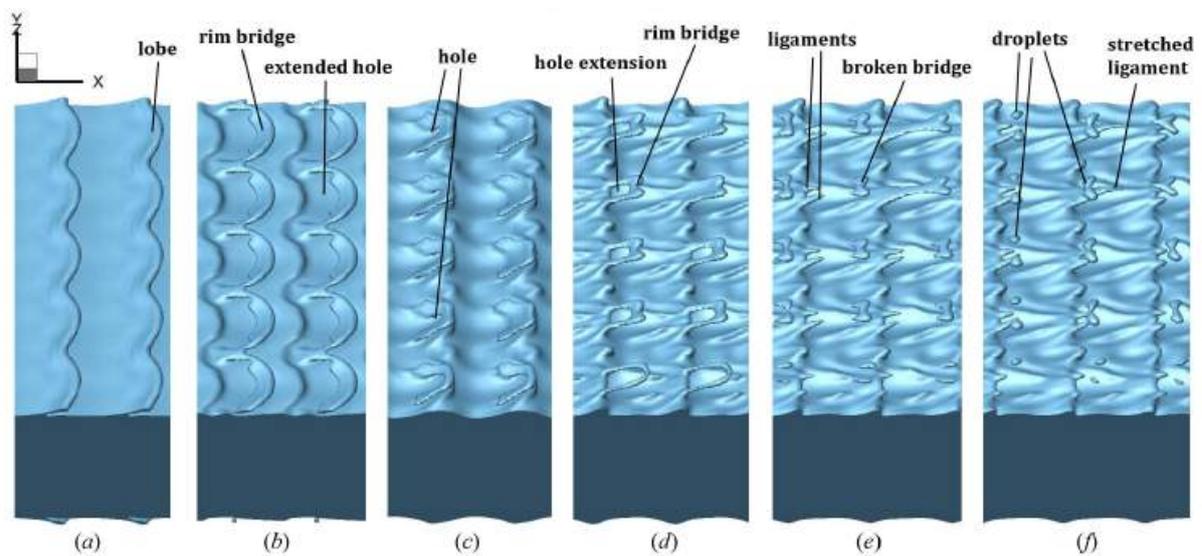

FIGURE 8. Liquid surface deformation in the *LoHBrLiD* mechanism; $Re_l = 320$, $We_g = 115\,000$ ($Oh_m = 1.06$), $\hat{\rho} = 0.5$, and $\hat{\mu} = 0.0022$, at $t = 18$ μs (*a*), 22 μs (*b*), 26 μs (*c*), 28 μs (*d*), 30 μs (*e*), and 32 μs (*f*).

The spanwise vorticity rolls up as KH rollers. The braid connects two adjacent rollers separated in $x$, which stretch the fluid in between. There are two modes of unstable waves, corresponding to the two surface waves oscillating exactly in or out of phase, commonly referred to as the sinuous (anti-symmetric) and varicose (symmetric) modes, respectively; see figure 7.

### 3.2. *Hole and bridge formations (LoHBrLiD mechanism)*

The *LoHBrLiD* mechanism occurs at medium $Re_l$ and high $We_g$; see figure 1. This process is shown in figure 8. The lobes form and thin on the primary KH wave crests. The middle section of the lobes (the braid), where the highest strain occurs, thins faster and thus perforates, creating a hole and a bridge on the lobe rim. Bridges become thinner as the holes expand. Finally, the bridges break and create one or two ligaments depending on the breakup location. The ligaments stretch and eventually break into droplets under capillary action.

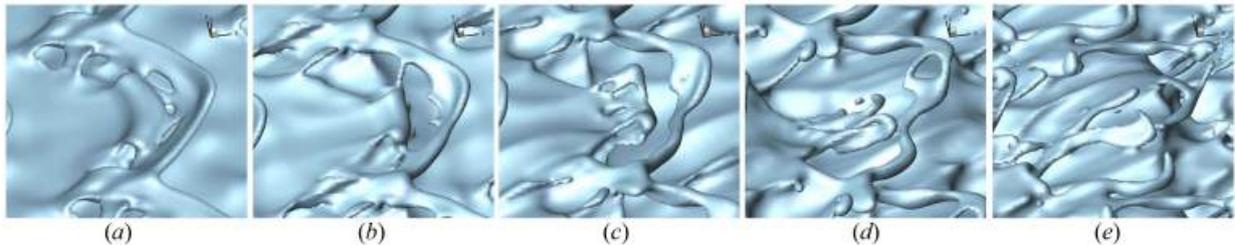

FIGURE 9. Close-up view of the hole formation and expansion on a single liquid lobe; $Re_l = 5000$, $We_g = 36\,000$, $\hat{\rho} = 0.1$, and $\hat{\mu} = 0.0066$. $t = 10$ µs (*a*), 11 µs (*b*), 12 µs (*c*), 13 µs (*d*), 14 µs (*e*).

At lower $We_g$, the surface tension force resists the formation of holes. In this range, usually one hole forms on the liquid lobe and stretches into a large one. At higher $We_g$, lobes thin much easier since the resistance due to the surface tension forces is not large enough to stabilize the growth of the instabilities. In this range usually several small scale holes appear at different locations on the lobe, as seen in the close-up view of figure 9(*a*). As these perforations grow, several of these holes merge to create larger holes (figure 9*b*), and a thick liquid bridge is created on the lobe rim (figure 9*c*). As the lobes continue to stretch and the holes continue to expand, the bridges become thinner (figure 9*d*) and finally break to create several ligaments (figure 9*e*). The mechanism of hole formation, to be discussed in this section, is believed to be the same at high and low $We_g$'s and density ratios; the only difference is in the time, location and size of the holes.

Case D2a (table 1), which falls in the $LoHBrLiD$ class, is chosen in this section to study and explain the vortex dynamics in the hole-formation mechanism. At the end of this section, Case D2b is studied to clarify the effects of lowering density ratio. Zandian *et al.* (2016) and Jarrahbashi *et al.* (2016) related the hole formation to the overlapping of the hairpin vortices that form on the braids. Their finding is confirmed here and more details in the vortex overlapping process are revealed. In this section, $\lambda_2$ contours are elucidated in two cross-sections in *x*–*z* planes - one passing through the spanwise crest and the other through the trough - along with the instantaneous liquid/gas interface (red lines) position at different times.

Figure 10(*a*) shows the formation of lobes and hairpin vortices that occur on the braid on both top and bottom sheet surfaces, at very early time $t = 6$ µs. The vortex structure is symmetric at this time and consists of a large vortex just downstream of the wave, which is hereafter called the *"KH vortex"* (shown by the white arrows), and hairpin vortices on the braid, between two adjacent KH rollers. The location of the vortices relative to the interface are the same on the spanwise crest and trough cross-sections at this time; i.e. the vortices on the spanwise trough cross-section (not shown here) are slightly upstream of those on the spanwise crest; see the top views of figure 10. The vortices have some undulations in *x* and manifest a hairpin structure similar to what was observed by Bernal & Roshko (1986). Lasheras & Choi (1988) attributed the appearance of three-dimensionality to the stretching along the principal direction of the positive strain. The maximum amplification of the vortex lines occurs near the braid region, where positive strain is the maximum. Therefore, they called these hairpin vortices "strain-oriented vortex tubes" and described a mechanism for evolution of the three-dimensional instabilities, where vortices enhanced by stretching are pulled more strongly into the streamwise direction until a series of hairpins extend from the underside of one roller to the top of its neighbor. A similar mechanism is seen here for the two-phase flow.

As shown in figure 10(*b*), the upstream hairpins are pulled downstream by the KH vortex on the outer side of the roller, while the downstream hairpins are pulled upstream on the inner side of the KH roller (easily understood from a frame fixed with the KH vortex). The direction of the streamwise hairpin stretch is determined by the global induction of the KH roller, as indicated by the black arrows in figure 10($b$). Martin & Meiburg (1991) explained that the sign of the streamwise vorticity component is determined by the competition between global and local inductions; i.e. between the overall effect of the vorticity field and the locally self-induced velocity of a vortex tube. While the global induction effect continues to determine the sign of the streamwise braid vorticity, the direction of the streamwise vorticity component in the crest region is determined by the local induction of the vortex tube (Martin & Meiburg 1991).

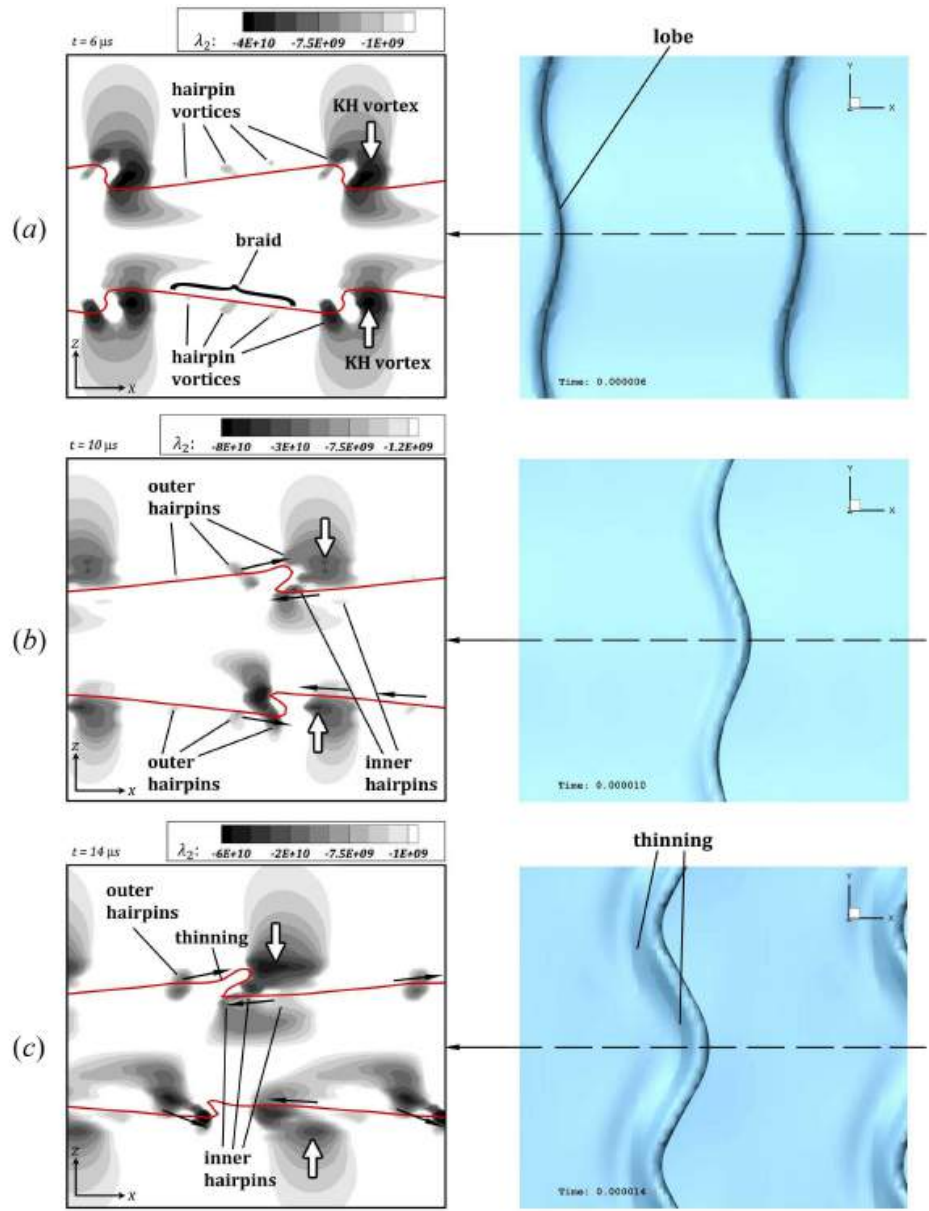

FIGURE 10. $\lambda_2$ contours on the spanwise crest (left), and the top view of the liquid surface (right), at $t = 6$ µs ($a$), 10 µs ($b$), and 14 µs ($c$) of Case D2a.

The motion of the upstream and downstream hairpins on the outer and inner sides of the KH roller - due to the roller's induced motion - causes these two hairpins to align spanwise and overlap - one layer locating on the outer surface of the lobe, i.e. on the streamwise wave crest, and the other layer on the inner side of the lobe. This hairpin overlapping has been observed and explained in both mixing layers (see figures 3 through 6 of Comte *et al.* 1992) and in two-phase round jets (see figure 9 of Jarrahbashi *et al.* 2016). Comte *et al.* (1992) discuss helical pairing of the oppositely-oriented hairpins and the formation of diamond-shaped vortex-lattice structures. As shown in the illustrative sketches of figure 11 and also described by Jarrahbashi *et al.* (2016), the liquid sheet between a pair of these overlapping, oppositely-pointed hairpins becomes thinner.

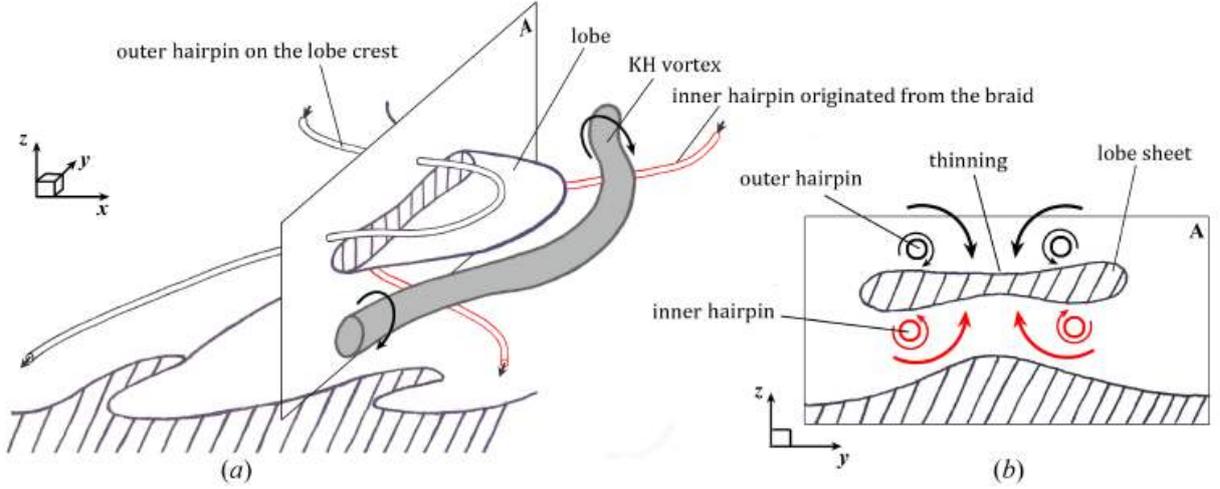

FIGURE 11. 3D Schematics showing the overlapping of the two hairpin vortices - one from the lobe crest (outer black tube, pointing downstream), and the other from the braid (inner red tube, pointing upstream) (*a*); *A* is the plane in which (*b*) is drawn; cross-sectional view of the *A*-plane, showing the thinning of the lobe sheet due to the combined induction of the two oppositely orientated overlapping hairpins (*b*). The vortex schematics are periodic in *x*- and *y*-directions.

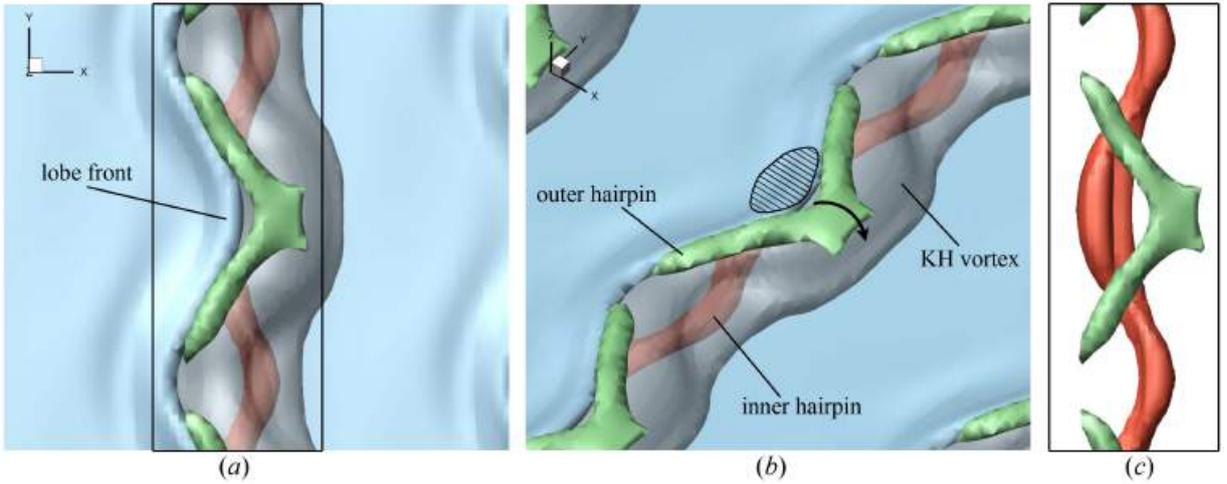

FIGURE 12. $\lambda_2$ isosurfaces in a top close-up view of a liquid lobe in Case D2a (*a*), and a 3D view of the same snapshot (*b*), at $t = 16$ µs. Top view of the box drawn in image (*a*), showing the overlapping of the red and green isosurfaces without the blockage of the gray KH isosurface and blue liquid surface (*c*). The isosurface values are: $\lambda_2 = -2 \times 10^{10}$ s$^{-2}$ (gray), $-3 \times 10^{10}$ s$^{-2}$ (green), and $-3 \times 10^9$ s$^{-2}$ (red). The gray isosurface is made transparent in subfigures (*a*) and (*b*) to display the inner hairpin underneath it. The hatched area denotes the approximate hairpin overlapping region, where the hole would appear on the lobe later.

Figure 11(*a*) schematically depicts two overlapping hairpin vortices in the liquid-gas interface region – one outer vortex originating from the lobe crest and stretching downstream (the slender black tube), and the other inner vortex originating from the braid and stretching upstream (the red tube). The KH vortex is shown by the thicker gray tube in this figure. Figure 11(*b*) shows a cross-sectional view of the vortex structure along with the lobe located between the two hairpin vortices on the *A*-plane of figure 11(*a*). The induced velocity of the two oppositely oriented overlapping hairpin vortices (see

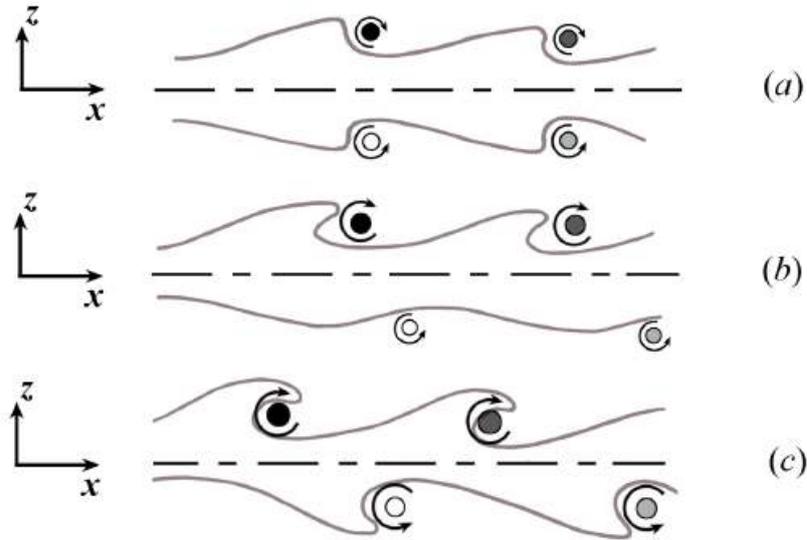

FIGURE 13. Schematic showing the transition of a symmetric thin sheet (*a*) to antisymmetric mode (*c*). The dashed lines denote the sheet center plane. The black and dark gray circles denote the two adjacent KH vortices on the top surface, and the light gray and white circles are two adjacent KH vortices on the bottom surface.

the qualitative streamlines shown by the black and red arrows in figure 11*b*) pushes the top surface of the lobe downward and the bottom surface upward, causing the lobe to become thinner in the middle and thus becoming vulnerable to puncture at that region. This occurs at $t = 16$ μs, on the top surface of the liquid sheet, as shown in figure 12. Three $\lambda_2$ isosurfaces are shown in this figure: the gray isosurface denotes the KH vortex (the gray tube in figure 11); the green isosurface denotes the outer hairpin (the black hairpin in figure 11); and the red isosurface denotes the inner hairpin (the red hairpin in figure 11). The $\lambda_2$ isosurface tracks the strength of the vorticity but circulation on its surface is not constant (as it is not a vortex surface); thus, it slightly differs from the surface of a vortex tube. The $\lambda_2$ magnitude of each isosurface is denoted in the figure caption. The KH vortex in figure 12 is the strongest (i.e., greatest circulation) and, at its core, has larger $\lambda_2$ values than the hairpins; however, the outer isosurface, with a lower $\lambda_2$ value, is depicted to give a better indication of its size. The hairpin overlapping region is denoted by a hatched area in figure 12(*b*). The overlapping of the outer and inner hairpins can be clearly observed in figure 12(*c*), where the gray KH vortex isosurface and the blue liquid surface have been removed from the box drawn in figure 12(*a*). The hatched area shows the zone where the lobe thinning occurs and the first hole forms a few microseconds later; see figures 14 and 15(*a*).

The sketches in figure 13 show the transition of a thin sheet from a symmetric mode to an antisymmetric mode with the qualitative location of the KH vortices at the two liquid surfaces. In the beginning, the top and bottom sheet surfaces as well as their KH vortices are symmetric with respect to the center-plane, as shown in figure 13(*a*). All plane jets in reality exit with symmetric perturbations because of the long-wavelength perturbations due to the upstream chamber's Helmholtz resonance modes or driving compressor blade wakes, but the antisymmetric mode has a much higher growth rate and thus eventually dominates. The true physical explanation of this transition is intriguing, but remains somewhat elusive, and deserves careful examination and explanation. Ashurst & Meiburg (1988) showed that the loss of symmetry in the two vortex layer calculation is related to the enhancement and reduction of relative streamwise displacements for filaments in the stronger layer by the addition of the second layer of vorticity.

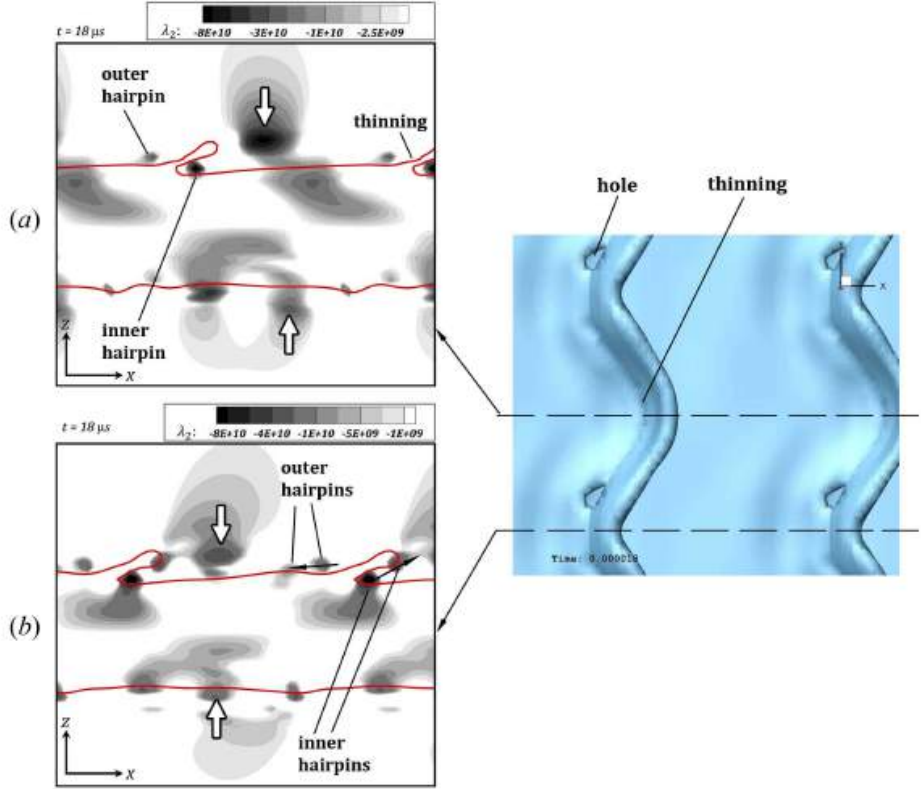

FIGURE 14. $\lambda_2$ contours on the spanwise crest (*a*), and the spanwise trough (*b*), and the top view of the liquid surface (right), at $t = 18$ µs of Case D2a.

Due to this transition towards antisymmetry, the KH rollers on one of the layers (the bottom layer in our case) are not able to stretch the bottom lobes and cause the hairpins overlapping at early times; thus, lobe formation and thinning are delayed on the bottom surface. This corresponds to the instant in figure 13(*b*) and the period 18–34 µs in our simulation, shown in figures 14 through 19. As the sheet gets thicker, i.e. the two vorticity layers get more independent, the transition towards antisymmetry is delayed, and both sides have time to roll up the surface waves and stretch them into lobes. For the rest of this section, our focus will be only on the top interface where lobe stretching and hole formation are evident at early times.

As figure 14 shows, the KH vortex starts to move away from the KH wave at 18 µs. However, it has had enough time to roll the lobe and make the overlapping occur on the top sheet surface. Notice that the KH vortex on the trough cross-section (figure 14*b*) is still closer to its original location compared to the same vortex on the crest cross-section (figure 14*a*). As described earlier, this results in further stretching of the hairpins near the sides of the lobe compared to its crest; hence, holes occur sooner near the lobe sides than at the lobe crest. This motivates the idea that the KH vortex location and strength are important in formation or inhibition of holes at early times. The KH vortex transition and structure is directly related to the liquid viscosity, thus the Reynolds number. At high $Re_l$, even though vorticity diffusion is slower, the fluids motion is less constrained by the viscous forces; thus, the KH vortices advect away from the interface much easier, and do not acquire enough time to roll up the lobes and stretch the hairpins. Hole formation also depends on the surface tension. The inertia should have enough strength to overcome the surface tension forces in order to perforate the lobe. In conclusion, the $LoHBrLiD$ mechanism becomes less probable as $We_g$ decreases or $Re_l$ becomes very high or very low; see figure 1.

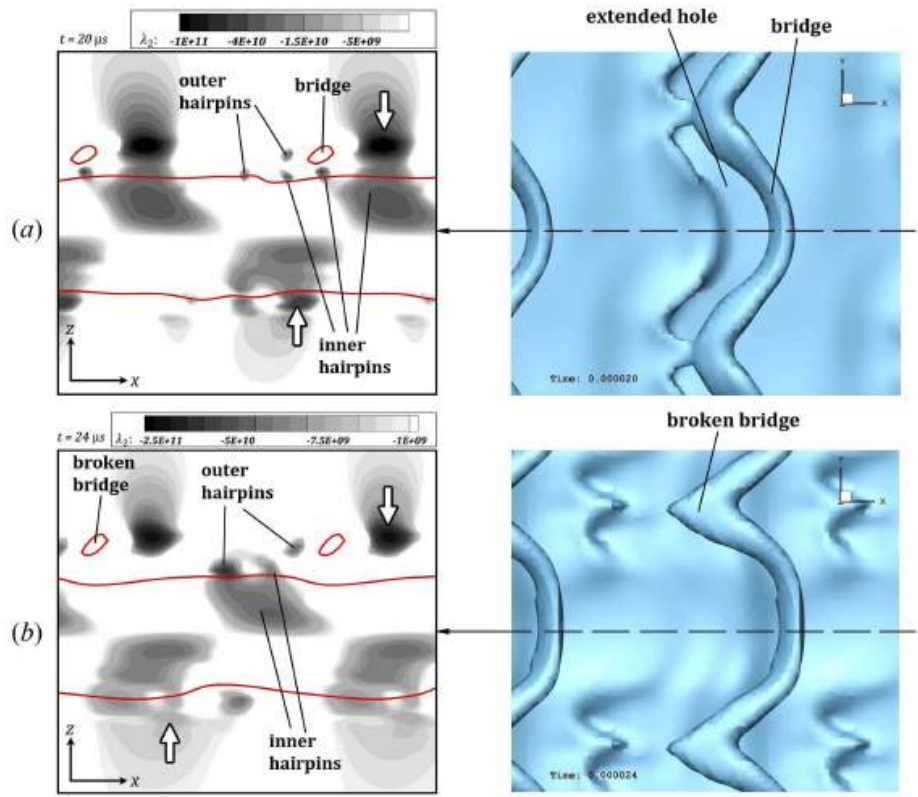

FIGURE 15. $\lambda_2$ contours on the spanwise crest (left), and the top view of the liquid surface (right), at $t = 20$ µs (*a*), and 24 µs (*b*) of Case D2a.

$\lambda_2$ contours of figure 15 clearly show the overlapping of the outer and inner hairpins at the location of the holes. The holes expand and create bridges at $t = 20$ µs (figure 15*a*). Meanwhile, the KH vortices advect both downstream and away from the interface. Later at 24 µs (figure 15*b*), the bridges break and create spanwise ligaments. If the bridge breakup would have happened at the spanwise crest instead of its trough, the resulting two ligaments would have been streamwise oriented, as in the sketch of figure 2.

The broken bridges undergo capillary instability and break further into droplets and smaller ligaments, as shown in figures 16 and 19. Meanwhile, the hairpin vortices that are overlapping near the spanwise trough, thin the bridges, as has been indicated by the red arrows in figure 16(*b*). This thinning along with the stretching of the bridges by the induced velocity of the KH roller deforms the bridge and makes it thinner and on the verge of further breakup. This completes the $LoHBrLiD$ breakup mechanism at primary atomization. The liquid interface at this time has become completely antisymmetric (figure 16) following the KH vortices that became antisymmetric earlier.

The role of vortices in the $LoHBrLiD$ breakup mechanism is summarized schematically in figure 17. This figure shows the liquid surface and also the qualitative location of the nearby vortices at four consecutive times. At an early time $t_1$, hairpin vortices form on the braids due to the strain caused by the neighboring KH vortices. The hairpins closer to the KH wave crest - shown by black lines - are stretched downstream, and the hairpins near the KH wave trough - shown by red lines - are stretched upstream by the induced motion of the KH rollers. The hairpin parts that are stretched downstream are rolled over the KH vortex tube, and are denoted by solid lines, while the hairpin parts that are stretched upstream are pulled under the KH vortex tube, and are denoted by dashed lines.

Later at $t_2$, the KH vortices deflect more under the induction of the hairpin vortices.

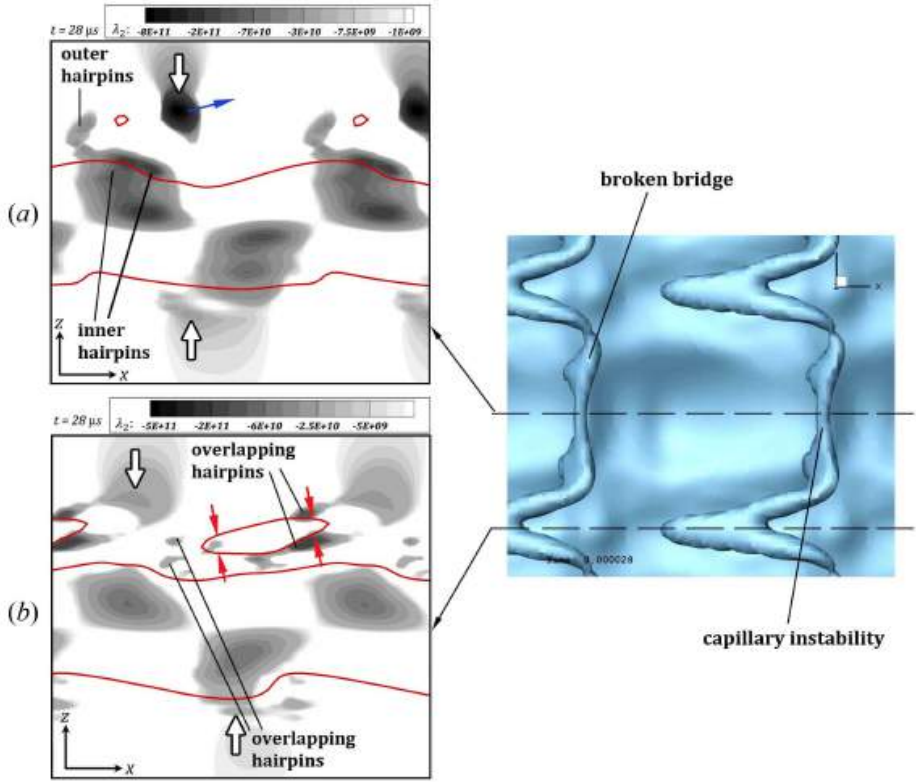

FIGURE 16. $\lambda_2$ contours on the spanwise crest (*a*), and the spanwise trough (*b*), and the top view of the liquid surface (right), at $t = 28$ μs of Case D2a.

The KH vortex stretches the lobes over itself and covers underneath the lobe surface at later times (see bottom images of figure 17). The black and red hairpins that roll over and under the KH vortex, respectively, overlap later at the center of the lobe as well as the two sides of the lobe; i.e. at the spanwise troughs. These overlapping regions have been denoted in figure 17 at $t_3$. The lobe sheet fills the vertical gap between these overlapping hairpins, causing the lobe to become thinner and thus vulnerable to puncture at the overlapping regions, as was described in figure 11.

Whether the liquid sheet subject to these conditions punctures or not depends on other flow conditions, particularly the surface tension. At high $We_g$ (high $Oh_m$), the hole formation prevails and the lobes perforate at the predicted locations, as shown at $t_4$ in figure 17. As the overlapping hairpin filaments continue to stretch, the holes also stretch and expand, creating even larger holes and thinner bridges. If $We_g$ is not large enough, the liquid lobe in the overlapping region can recover instead. In this case, hole formation is inhibited and the lobes stretch directly into ligaments via $LoLiD$ or $LoCLiD$ mechanisms, as discussed later.

The mechanism of hole formation at low density ratios is similar to the high $\hat{\rho}$ described in figure 17, with minor differences. Figure 18 shows the vortex structures and the lobe deformation at 90–94 μs for $\hat{\rho} = 0.05$ (Case D2b). Three $\lambda_2$ isosurfaces corresponding to the KH vortex (gray), the outer hairpin (green), and the inner hairpin (red) are shown in figure 18(*a*) at 90 μs. The liquid surface has been removed from this image to reveal all the vortex structures, but the black solid line indicates the location of liquid lobe front edge for comparison. The liquid surface at the same time is shown in figure 18(*b*). The outer and inner hairpins at this low $\hat{\rho}$ stretch and wrap around the KH vortex in opposite directions, and overlap on top and bottom of the lobe. The difference here is that the hairpins are more stretched and the overlapping region is more elongated in the streamwise direction, as shown by the hatched area in figures 18(*a,b*). The holes that are formed near the hatched area at later times (figures 18*c,d*) are also more elongated in the streamwise direction compared to the high $\hat{\rho}$ case shown in figure 15, where the holes were more spanwise oriented. This finding agrees with the predictions of hole location by Jarrahbashi *et al.* (2016) (see their figure 20). There when the gas density increased, the edge curvature decreased and the locations of the holes in neighboring lobes became closer to each other. This caused the neighboring holes to merge and expand the hole in the azimuthal (spanwise) direction.

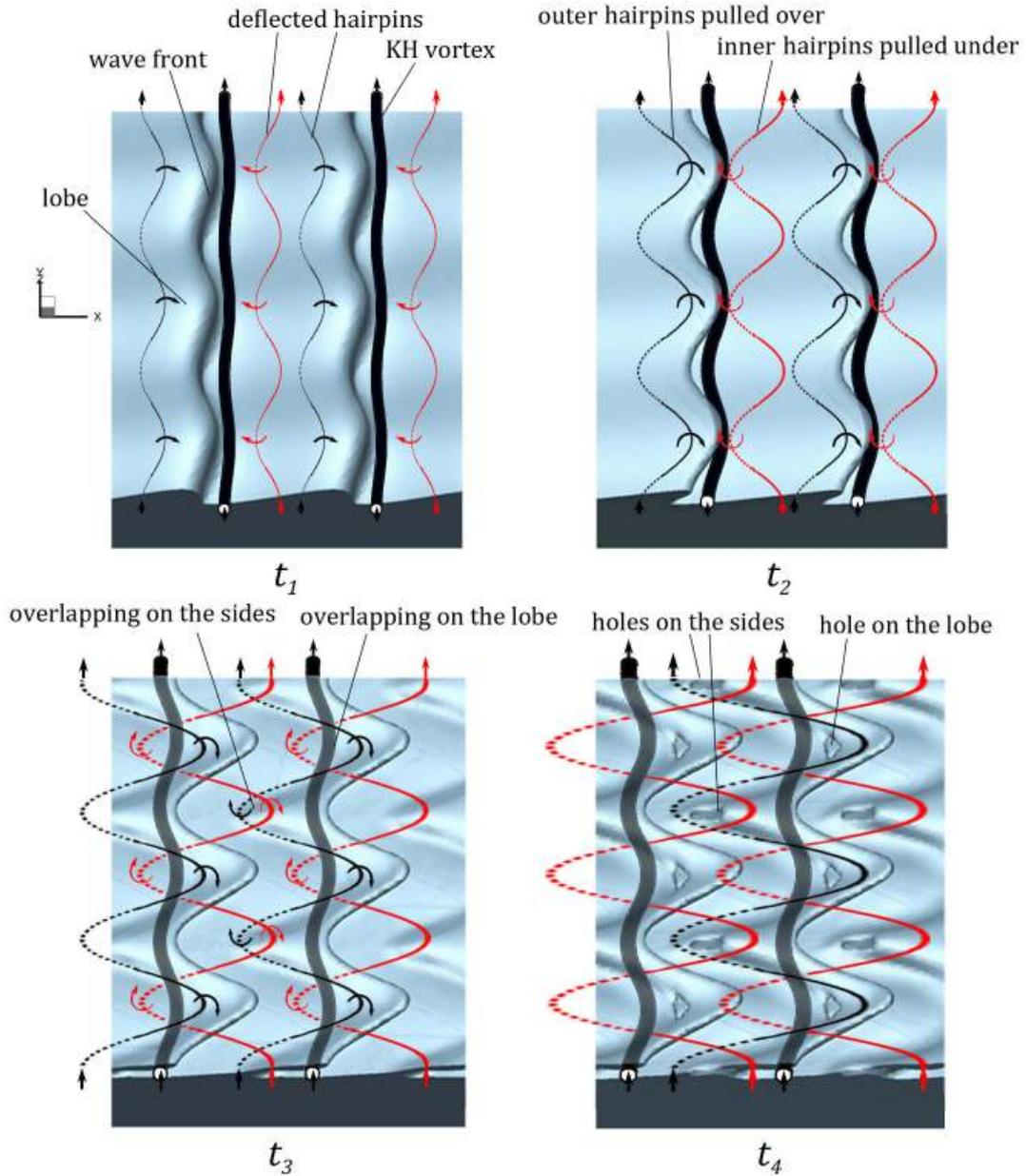

FIGURE 17. Schematics of the $LoHBrLiD$ process at four consecutive times. The liquid/gas interface is shown in blue, and the KH vortex by black tubes. The red and black lines denote the inner and outer hairpin vortices near the KH wave trough and crest, respectively. The solid or dashed lines denote where the hairpins are stretched upstream and inward, or downstream and outward, respectively.

The breakup mechanism involving hole and bridge formation and the role of vortical structures in this regard were discussed in this section via consecutive tracking of $\lambda_2$ contours near the interface. Vortex dynamics is able to explain the breakup process at high $We_g$ and medium $Re_l$. In the remainder of this section, $\lambda_2$ contours along with the interface location are shown at a few later time steps to track the deformation of the ligaments and also to see the evolution of the surface waves after the KH vortices have left the interface.

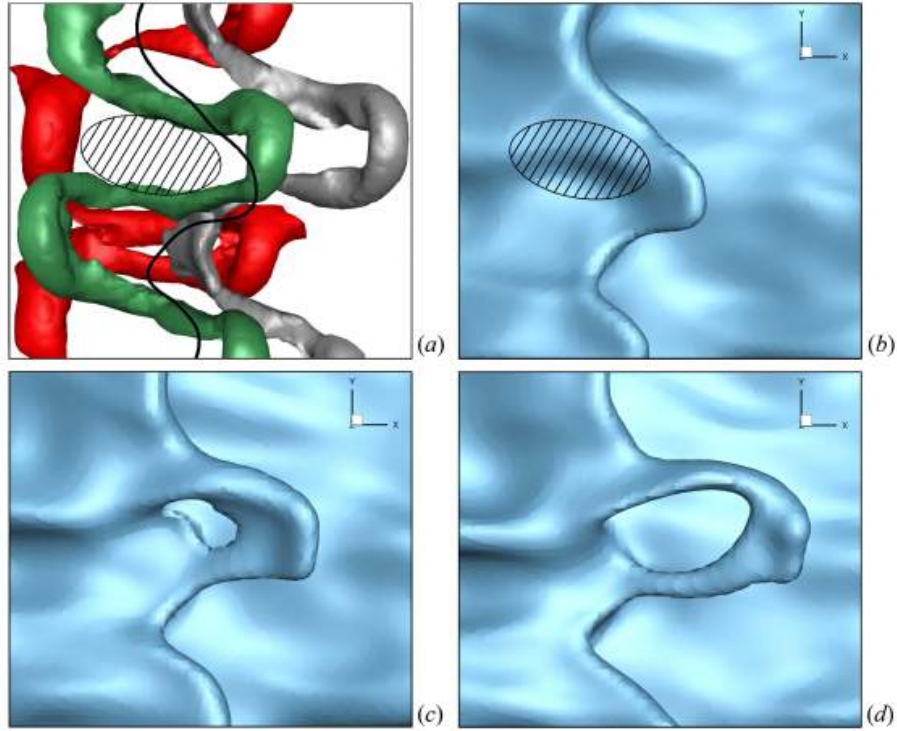

FIGURE 18. $\lambda_2$ isosurface in a top close-up view of a liquid lobe in Domain II at low density ratio (Case D2b) at 90 µs (*a*); the solid black line shows the lobe front edge location and the hatched area shows the location of hairpin overlapping. The isosurfaces represent: the KH vortex with $\lambda_2 = -4 \times 10^9$ s$^{-2}$ (gray), the outer crest hairpin with $\lambda_2 = -6 \times 10^9$ s$^{-2}$ (green), and the inner trough hairpin with $\lambda_2 = -3 \times 10^9$ s$^{-2}$ (red). Lobe surface showing the hole formation from a top view at 90 µs (*b*), 92 µs (*c*), and 94 µs (*d*) of Case D2b.

Figure 19 shows $\lambda_2$ contours of the liquid jet in the period 30–34 µs. Even though the original KH vortex (denoted by the white arrow) is now advected downstream and away from the waves, the vorticity contours on the braid stretch and collect at the crest of the new KH waves and create new KH vortices (figure 19*a*). These vortices roll a new KH wave and stretch new sets of lobes similar to the original ones; see figure 19(*b*). The direction of the induced motion by these new vortices are indicated by the black arrows in figure 19(*b*). The direction in which the fluids are swirling around the KH vortices are also denoted by the blue curly arrows. At $t = 34$ µs the sheet and the vortices are antisymmetric with half a wavelength phase difference between the top and bottom surfaces in the streamwise direction. The vortices on the bottom side are now nearly stationary with respect to the interface; hence, they have enough time to roll the waves and form stretched lobes on the bottom surface (figure 19*b*). As the new lobes get stretched, new pairs of hairpins form on their braids, and the whole *LoHBrLiD* process repeats, creating new holes, bridges, ligaments, and droplets. Meanwhile, the formerly broken bridges undergo capillary instability and thin at the necks and break into droplets, as shown in the right image of figure 19(*b*).

Overlapping of the oppositely oriented hairpins is the cause of lobe perforation, as described in this section; however, the locations of the perforations and the direction and rate of their growth depend on the overlapping location with respect to the lobe sheet. What was presented in this section was an ideally simple case to explain the process. In reality, the hairpin structures can be more irregular than what was presented here, which would shift the overlapping zones and hence change the location of the first perforations. More hole formation examples are presented by Jarrahbashi *et al.* (2016) and Zandian *et al.* (2016). Jarrahbashi *et al.* (2016) also showed the direction in which the holes merge and expand at high and low density ratios.

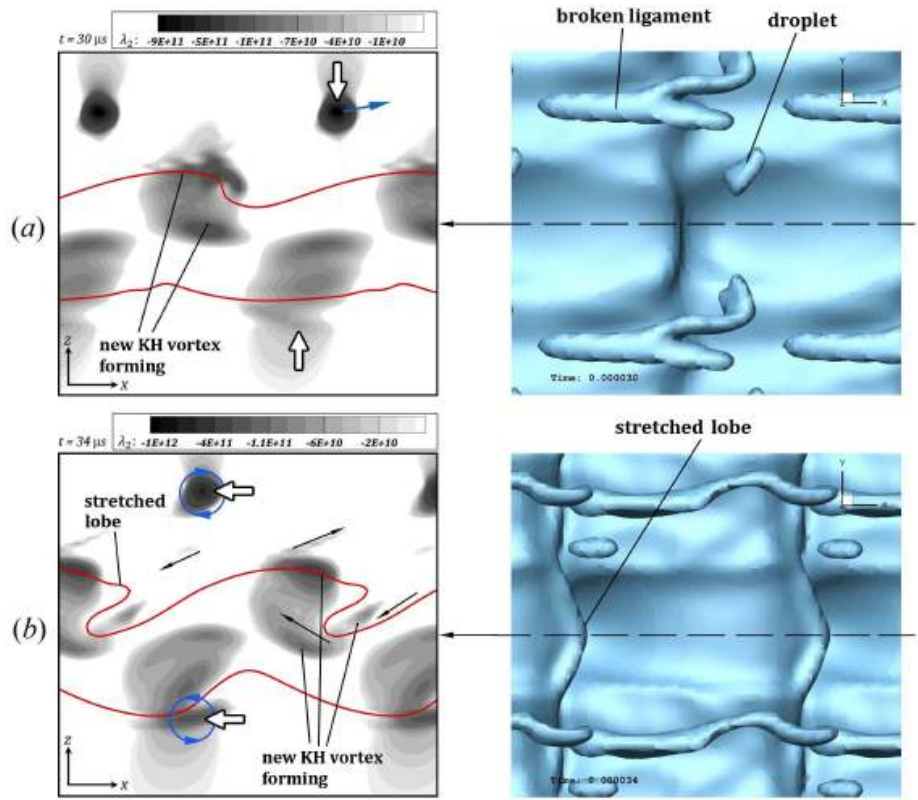

FIGURE 19. $\lambda_2$ contours on the spanwise crest (left), and the top view of the liquid surface (right), at $t = 30$ µs (*a*), 32 µs (*b*), and 34 µs (*c*) of Case D2a.

### 3.3. *Corrugation formation at high $Re_l$ (LoCLiD mechanism)*

The *LoCLiD* mechanism occurs at high $Re_l$ and low $We_g$ (low $Oh_m$), as indicated in figure 1. This process is shown in figure 20. The lobes form similar to the previous case, but do not stretch as much. Corrugations form on the lobes' front edge and stretch to create ligaments. Multiple ligaments are formed per lobe, typically shorter and thinner compared to the ligaments seen in the *LoHBrLiD* Domain. Eventually, the ligaments detach from the liquid jet and break up into droplets by capillary action. These droplets are consequently smaller than the ones formed in the *LoHBrLiD* mechanism. Sequential evolution of $\lambda_2$ contours on *y*-plane cross-sections are used in this section to delineate the physics of this process and more particularly describe why lobes do not perforate but get corrugated at higher $Re_l$ and lower $We_g$. Case D3a (see table 1) is analyzed in this section; at the end of this section, the effects of lower density ratio are described by analyzing Case D3b.

Figure 21 shows $\lambda_2$ contours on a *y*-plane passing through the spanwise crest at 6 µs and 10 µs. Since the vortex structure passing through the spanwise trough has similar features, albeit with some phase difference with respect to the vortices at the spanwise crest, it will not be shown here. In the beginning, the vorticity field contains a series of hairpin filaments on the braid and a stronger KH vortex just downstream of the KH wave crest (indicated by the white arrows). The hairpins are stretched by the neighboring rollers as discussed before. As also described by Martin & Meiburg (1991), since stretching is the most intense in the downstream half of the braid region, hairpin vortices acquire a larger streamwise component in the upstream neighborhood of a vortex roller, i.e. on the lobe crest, than its downstream side; i.e. on the trough. Hence, the hairpin filaments that are closer to the streamwise wave crest (denoted by the gray arrows) are stronger than the upstream braid hairpins. Ashurst & Meiburg (1988) also found that, while in one vortex layer the filaments at the center between two spanwise rollers experience the most stretching, if a second vortex layer is added, the filaments with the strongest spanwise modulation will be located closer to the downstream roller; i.e. three-dimensionality occurs first in the downstream half of the braid region. The vorticity field as well as the sheet itself are initially symmetric with respect to the center plane. The hairpin filaments reside near the interface, slightly inclined towards the gas zone.

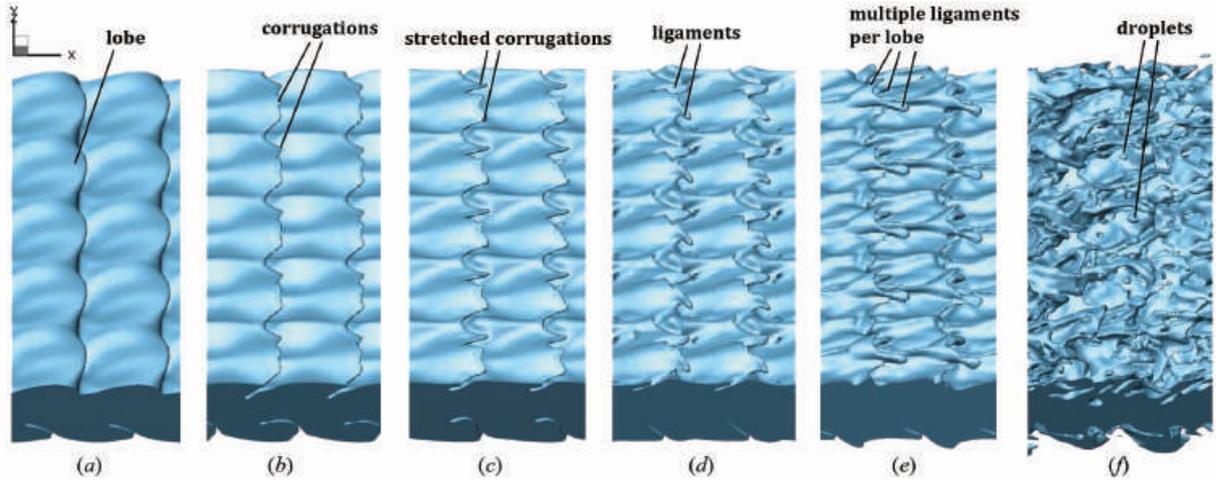

Figure 20. Liquid surface deformation following the *LoCLiD* mechanism in Case D3a, at $t = 44$ μs ($a$), 48 μs ($b$), 50 μs ($c$), 52 μs ($d$), 56 μs ($e$), and 60 μs ($f$).

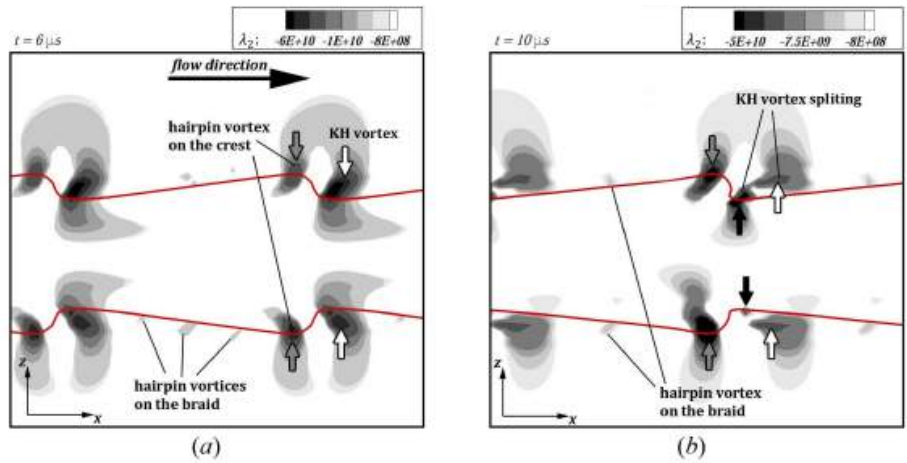

Figure 21. $\lambda_2$ contours on the spanwise crest cross-section, at $t = 6$ μs ($a$), and 10 μs ($b$) of Case D3a.

At high $Re_l$, inertia dominates the viscous forces and the fluid particles are more free to move around as the viscous forces do not produce enough resistance against their motion - especially in the gas zone. The higher velocity of the gas layer compared to the liquid layer causes the KH roller to split into two vortices at 10 μs (see figure 21$b$); this process is shown schematically in figure 22 and using the $\lambda_2$ isosurfaces in figure 23. The outer part of the KH vortex (indicated by the white arrow in figure 21$b$), which resides in the fast-moving gas layer, separates from the part that is inside the surface of the lower speed liquid (indicated by the black arrow in figure 21$b$). The splitting of the KH vortices starts at 10 μs (figure 21b) and continues until 12 μs (figure 23). The gray, green, and red $\lambda_2$ isosurfaces in figure 23 denote the KH vortex, the outer (crest) hairpin, and the inner (braid) hairpin, respectively (the magnitude of each $\lambda_2$ isosurface is indicated in the caption). The split slow- and fast-moving vortices are denoted by "s" and "f" in figure 22, respectively.

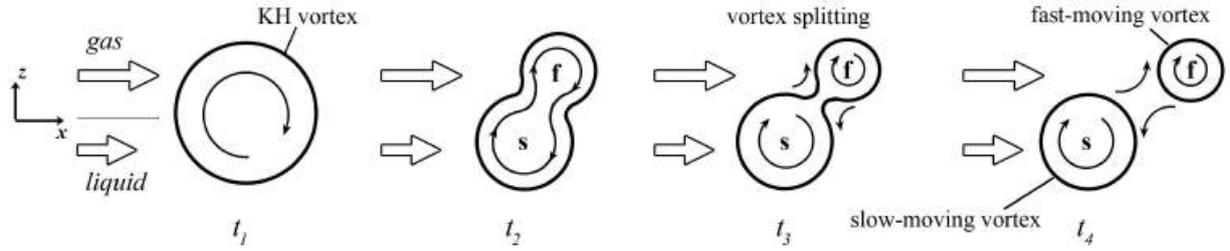

FIGURE 22. Schematic of how KH vortex splits into a slow-moving vortex (denoted by s) in the liquid and a fast-moving vortex (denoted by f) in the gas zone, at four consecutive times. The qualitative velocity magnitudes are denoted by the straight arrows in the gas and the liquid. The vortices are nearly uniform in the $y$ direction, normal to the paper.

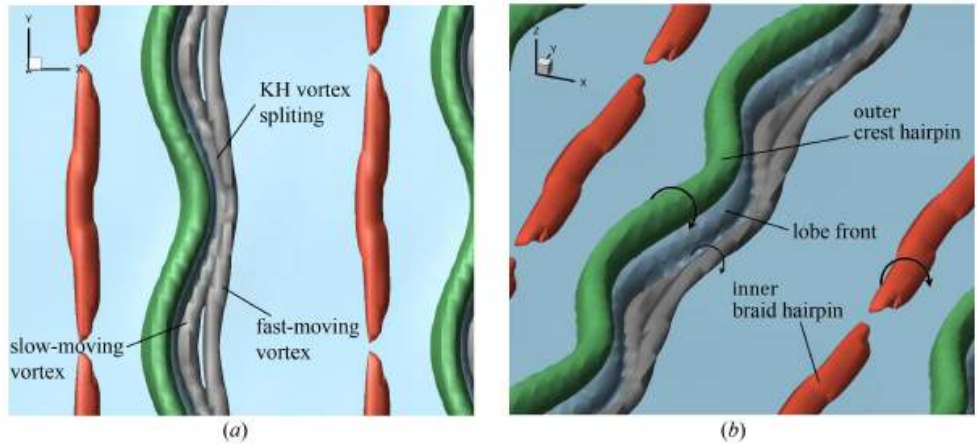

FIGURE 23. $\lambda_2$ isosurfaces in a close-up view of a liquid lobe in Case D3a, from a top view (a), and a 3D view (b), at $t = 12$ μs. The isosurface values are: $\lambda_2 = -8 \times 10^{10}$ s$^{-2}$ (gray), $-4 \times 10^{10}$ s$^{-2}$ (green), and $-10^{10}$ s$^{-2}$ (red). The liquid surface is shown in blue.

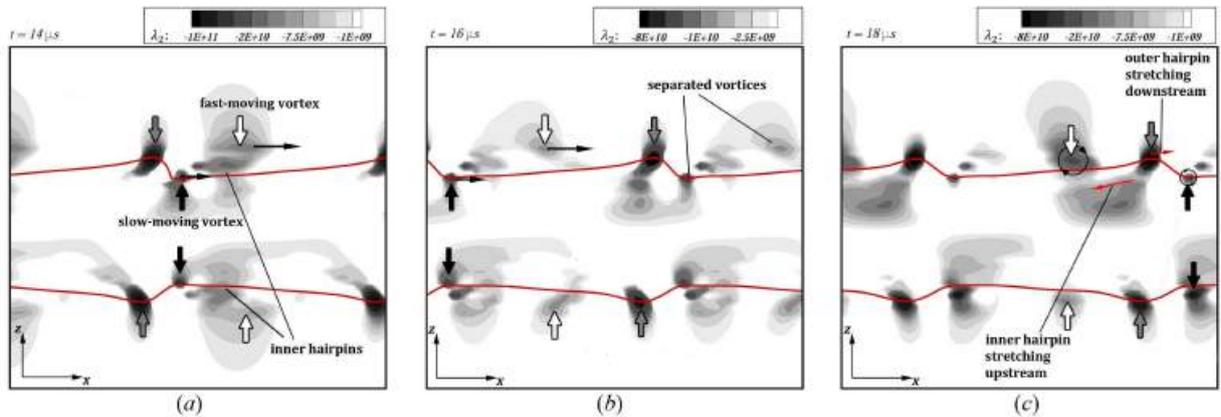

FIGURE 24. $\lambda_2$ contours on the spanwise crest cross-section, at $t = 14$ μs (a), 16 μs (b), and 18 μs (c) of Case D3a.

As demonstrated in figures 22 and 24(a), part of the KH vortex that resides in the faster moving gas, near the liquid interface, advects downstream with the gas, while the slow-moving vortex (indicated by the black arrow in figure 24a) advects more slowly with the interface velocity, remaining stationary relative to the liquid surface. The two vortices are completely separated at $t_4$ in figure 22, corresponding to $t = 16$ μs in our simulation (see figure 24b). This vortex separation has two significant consequences. (i) The slow-moving vortex (black vortex) downstream of the KH wave is not strong enough to curl the KH wave and stretch the lobe downstream over itself. Consequently, the outer hairpins do not overlap in $x$ with the inner trough hairpins, as in the $LoHBrLiD$ Domain; hence, the hole formation is inhibited at early times. (ii) The fast-moving vortex (the white vortex) gets closer to the downstream outer hairpin as it advects away from the upstream hairpins, residing in the vicinity of the trough; see figure 24(b). This successive variation in the distance between the hairpins and the fast-moving vortex induces a fluid motion that stretches the hairpins in the opposite directions, resulting in a less-orderly hairpin structure with more undulations, as shown schematically in figure 25.

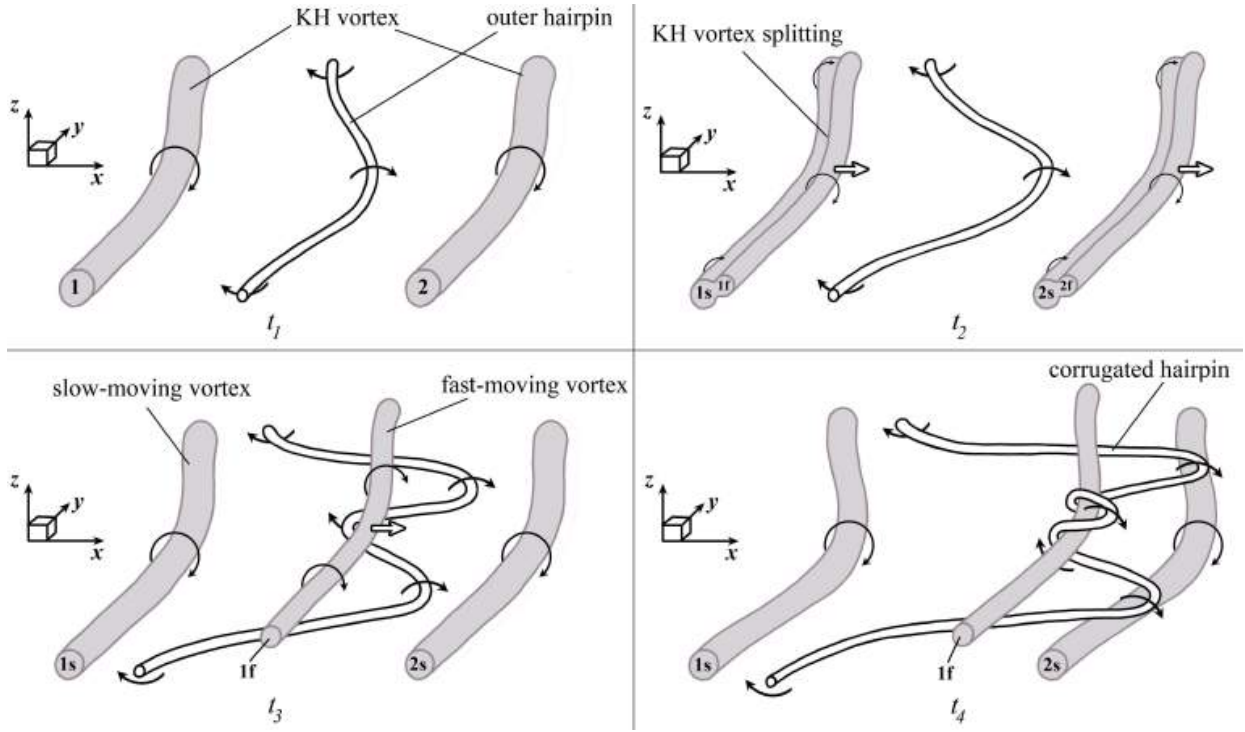

FIGURE 25. Schematic of the hairpin undulation formation under the influence of the fast-moving KH vortex (denoted by 1f) at four consecutive times $t_1$–$t_4$, from a frame of reference moving with the slow-moving KH vortices (denoted by 1s and 2s).

After splitting of the KH vortex into fast- and slow-moving vortices ($t_2$ in figure 25), the fast-moving vortex (1f) advects downstream over the outer hairpin ($t_3$ in figure 25). As the faster-moving vortex (1f) passes over the hairpin, its induced motion pulls the crest of the hairpin in the upstream direction under vortex 1f, causing the hairpin to undergo an undulation with smaller local wavelength ($\approx 45$ μm). Meanwhile, the tip of the hairpin (now having two crests) is stretched downstream over the neighboring slow-moving vortex (2s), and the trough of the hairpin is stretched upstream under the upstream slow-moving vortex (1s) ($t_3$ in figure 25). The streamwise stretching on different parts of the hairpins are indicated by the black curly arrows. As the faster-moving vortex (1f) moves further downstream, the newly formed undulation wraps around it and stretches downstream, as shown at $t_4$ in figure 25. Therefore, another turn is created on the hairpin vortex and the local wavelength of the undulations decreases (15–35 μm). As the fast-moving vortex keeps moving downstream, the hairpin corrugations stretch with it. It will be shown below that this hairpin structure results in less stretched (slower stretching) and more corrugated lobes. The direction of the streamwise stretch on the hairpin filaments – based on the velocity field induced by the two split rollers – is denoted by the red arrows in figure 24(*c*). The sheet is still symmetric at this moment, while the vortices have lost their symmetry with respect to the center-plane. The fast-moving vortex (1f) finally reaches the downstream slower-moving vortex (2s) and combines with it to create a stronger KH vortex, which stretches the now corrugated hairpin in the downstream direction.

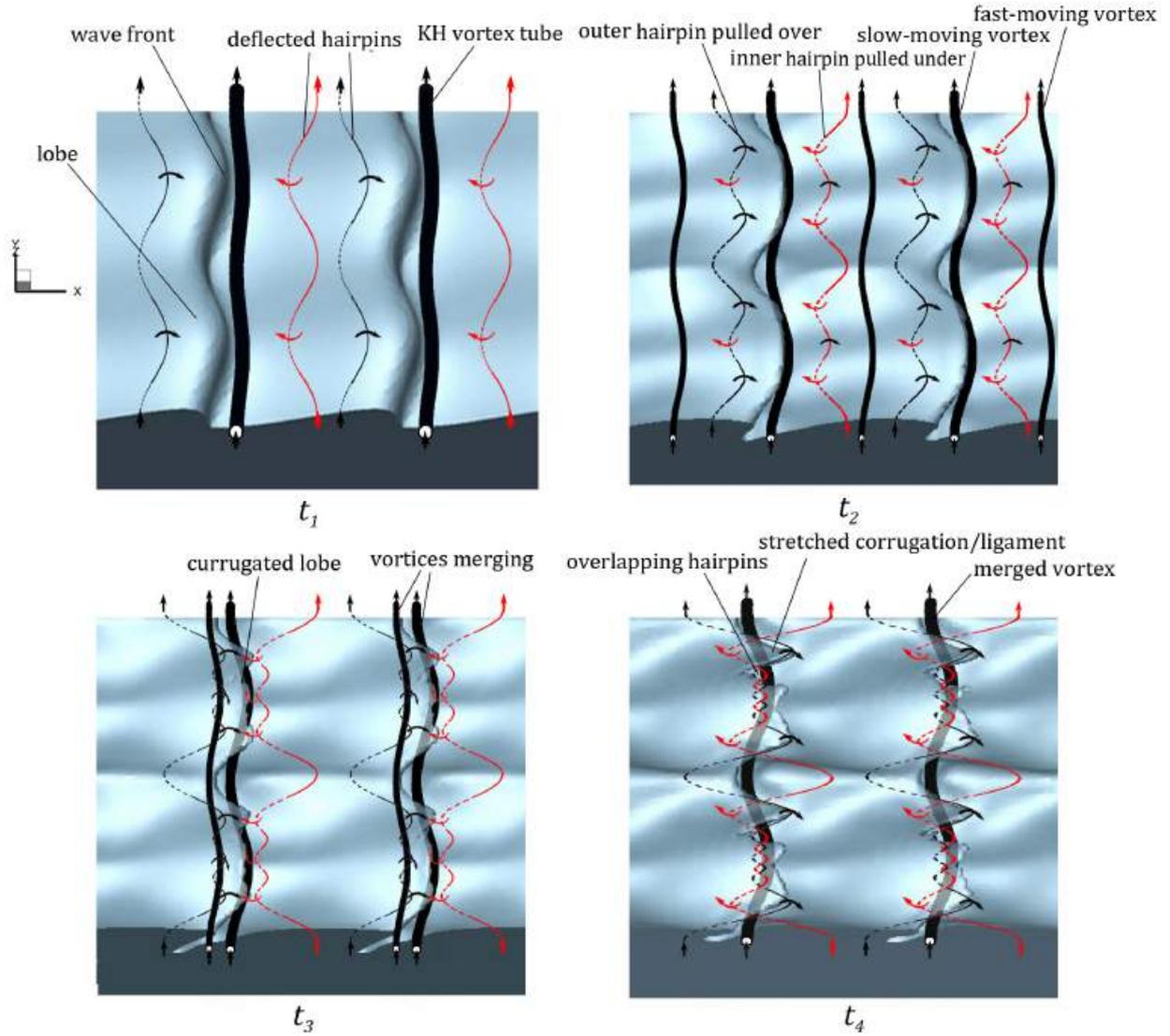

FIGURE 26. Schematic of the *LoCLiD* process at four consecutive times. The liquid surface is shown in blue, and the the KH vortex is denoted by the black tube. The red and black lines denote the inner and outer hairpin vortices near the trough and crest of the KH wave, respectively. The solid and dashed lines indicate the parts of the hairpins that are stretched upstream and inward, or downstream and outward, respectively.

To better understand the consequence of hairpin vortex structure on the liquid surface deformation at high $Re_l$ and low $We_g$, the evolution of the vortices and the interface in the *LoCLiD* process (Domain III) are schematically depicted in figure 26 at four consecutive instances. At an early time $t_1$, the braid regions connecting the emerging KH rollers become progressively more depleted of vorticity. The spanwise vortices on the braid deflect due to the induced motion of the neighboring KH rollers in both the upstream and downstream directions – creating the hairpin vortex structures with a spanwise size equal to the spanwise perturbation wavelength (100 µm). Two hairpins are formed on the braid – one located near the lobe crest, called the outer hairpin and denoted by the black line (corresponding to the green isosurface in figure 23), and the other slightly upstream near the trough, called the inner hairpin and denoted by the red line (corresponding to the red isosurface in figure 23). The deflected hairpin filaments form the lobes as they are stretched by the KH roller. So far, the process is similar to the $LoHBrLiD$ process.

As discussed earlier and demonstrated in figure 25, the hairpin vortices get corrugated under the influence of the fast-moving vortex that hovers over the interface after splitting of the KH vortex. This corresponds to $t_2$ and $t_3$ in figure 26. Only the deformation of the outer hairpin (black hairpin) was shown in figure 25; the inner hairpin (red hairpin) undergoes the same process, but in the opposite direction – stretching downstream and over the fast-moving vortex. The upstream turns and bends on the hairpins prevents further downstream stretching of the lobes. Consequently, the lobes are less stretched and more blunt compared to the lobes in Domain II (compare $t_2$ in figure 26 with $t_2$ in figure 17). The fast-moving vortex moves closer to the interface to merge with the downstream slow-moving vortex and later passes through the lobe and reaches the slower vortex below the lobe. In this process, the faster vortex carries along the corrugated outer hairpin that is wrapping around it. Consequently, the corrugated outer hairpins penetrate through the lobe and locate underneath the lobe in contrast to Domain II, where the outer hairpins were located on top of the lobe. Since this process occurs quickly, the corrugated hairpins do not have enough time to completely wrap around the faster vortex and create a combined vortex; therefore, after the merging of the two split KH vortex counterparts, only two pairs of undulated hairpins – which are separate from the KH vortex – remain on top and bottom of the reformed KH vortex, right below the lobe. The liquid surface approximately follows the hairpin structures with some delay at this high $Re_l$ range – as the vortex lines are nearly material lines. Because of these shorter hairpin wavelengths, corrugations with length scales comparable to the local hairpin wavelengths (15–25 µm) form on the front-most edge of the lobes, as shown in figure 26 at $t_3$. Both experimental observations (Liepmann & Gharib 1992; Lasheras & Choi 1988) and numerical simulations (Brancher *et al.* 1994; Danaila *et al.* 1997) for homogeneous jets show that the size of vortex pairs (lobes in two-phase flows) decreases with increasing Reynolds number.

Upon creation of a stronger KH vortex downstream of the KH waves at $t_4$ – after merging of the fast- and slow-moving vortices – the new KH roller, which is now located under the lobe (now a thicker tube) is strong enough to stretch the hairpins and the corrugations. The two hairpin layers overlap as illustrated in figure 26 at $t_4$. In this figure, the dashed lines represent the hairpins stretching upstream and on the streamwise trough, i.e. lower surface of the gas tongue, while the solid lines denote parts of the hairpins that are stretched downstream below the lobe; i.e. upper surface of the gas tongue.

The illustrative sketches of figure 27 show the corrugated hairpin structures and their position with respect to the lobe in the $LoCLiD$ process (corresponding to $t_4$ in figure 26). The black tube represents an outer hairpin on the lobe crest and the red tube represents an inner hairpin on the streamwise trough (originating from the braid). The two hairpin layers – after getting corrugated due to the induced motion of the two halves of the split KH vortex (see $t_2$ and $t_3$ in figure 25) – overlap under the lobe, as shown in figure 27(*a*). Following the fluid motion induced by the vortex pairs, the lobe front edge gets corrugated with comparable length scales to the hairpin undulation wavelength; see figure 27(*a*). The layer of the outer hairpin (black tube) is located underneath the lobe (on top of the gas layer), and the layer of the inner hairpin (red tube) is located on top of the interface at the trough; see figure 27(*b*). The induced flow creates undulations on both the bottom surface of the lobe and the trough surface. The combined induction of the two oppositely oriented overlapping hairpin layers thins the gas layer that fills the vertical gap underneath the lobe; i.e. the lobe collapses on the liquid sheet as the bottom surface of the lobe descends and the trough surface ascends; see the qualitative streamlines in figure 27(*b*).

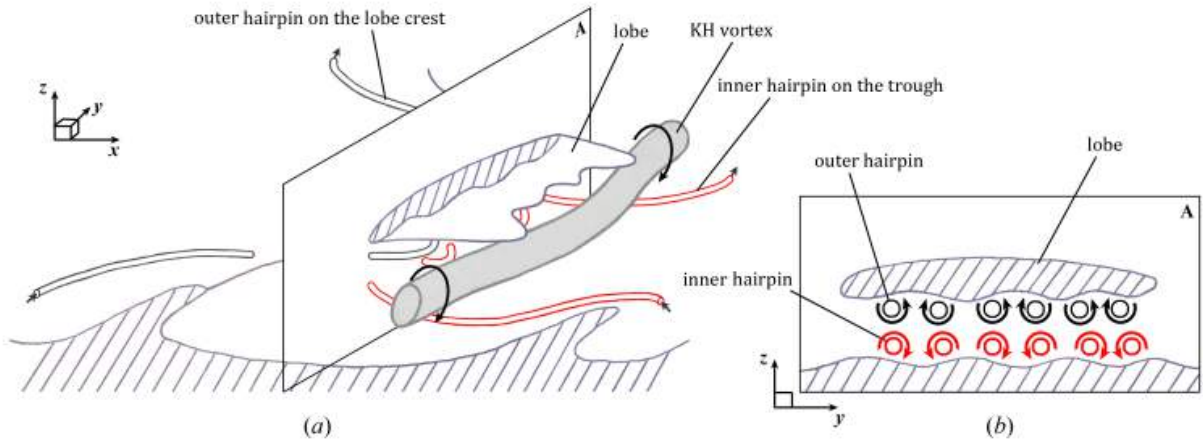

FIGURE 27. 3D Schematics showing the overlapping of the outer hairpin (black slender tube) and the inner hairpin (red tube) resulting in formation of lobe corrugations (*a*) – *A* is the plane in which (*b*) is drawn; cross-sectional view of the *A*-plane showing the corrugation formation and thinning of the gas tongue due to the combined induction of the overlapping hairpins (*b*). The vortex schematics are periodic in *x*- and *y*-directions.

As the counter-rotating pairs of hairpins stretch under the induction of the KH vortex, the corrugations on the lobes stretch with them and form thin ligaments (see $t_4$ in figure 26). The ligaments stretch downstream and break up into droplets as they undergo capillary instabilities (the ligament breakup mechanism is discussed in §3.4). In the meantime, the eddies cascade into smaller vortical structures as transition to turbulence occurs. The smaller vortices are moved towards the gas by the induction of the larger eddies. The outward movement of the vortices spreads the droplets in the normal direction, helps the expansion of the spray angle, and enhances the two-phase mixing.

The process of lobe-sheet collision after the merging of the two split KH vortices is shown in the side views of the liquid sheet along with $\lambda_2$ contours near the interface in figure 28. The antisymmetric vortices, which now are strong enough to stretch the lobes, create antisymmetric KH waves, at 44 µs (figure 28*a*). The hairpin vortices on the KH wave crest and trough stretch in the opposite directions following the induced fluid motion of the KH roller. Following the direction of the swirl, the hairpin projections on the trough are stretched upstream from the inner side of the roller, while the hairpin projections on the KH crest are stretched downstream from the outer side of the roller. The direction of streamwise stretch on the fluid elements is shown by the red arrows in figure 28.

Recall that at lower $Re_l$ the two overlapping hairpins lie above and below the lobe and cause its thinning. At higher $Re_l$, on the other hand, the outer and inner hairpins are inside the liquid (very close to the surface), and overlap on outer and inner sides of the gas layer that penetrates under the lobe. To understand this better, consider the opposite perspective; i.e. the layer of gas that fills the vertical gap underneath the KH waves can be considered as an upstream-pointed gas tongue, surrounded by the liquid; see figure 28(*b*). The combined induction of the overlapping hairpins thins the tongue similarly to the process described in figure 11. The thinning regions are indicated by the two thin black arrows in figure 28(*c*). Since it would be confusing to describe a hole formation in the tongue, we could interpret this phenomenon as the collapse of the lobe on the jet, encapsulating the gas in between.

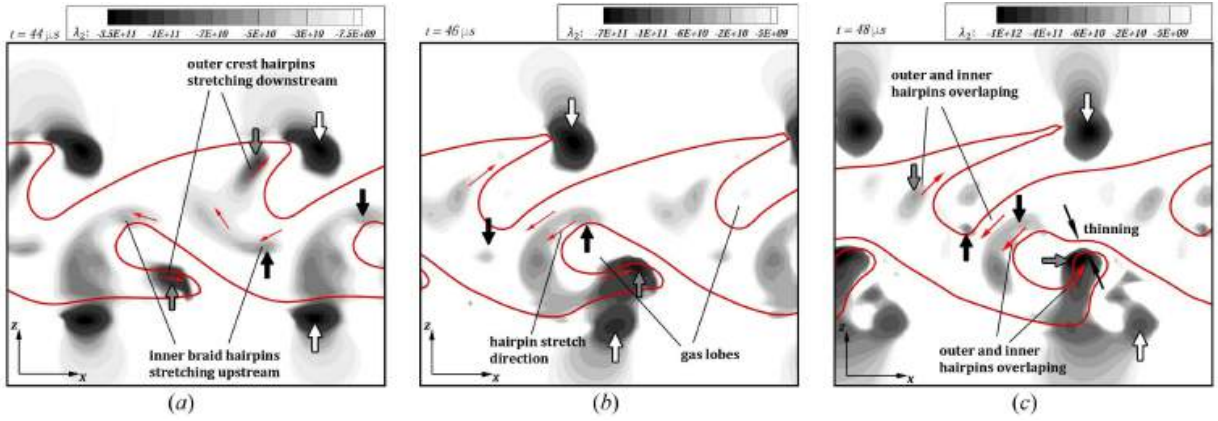

FIGURE 28. $\lambda_2$ contours on the spanwise crest cross-section, at $t = 44$ μs (*a*), 46 μs (*b*), and 48 μs (*c*) of Case D3a.

Figure 29 shows the top view of a corrugated lobe and the $\lambda_2$ contours at three spanwise (*y*–*z*) planes passing though the lobe body (plane *A*–*A*), through lobe corrugations (plane *B*–*B*), and through the braid (plane *C*–*C*), at 50 μs. The $\lambda_2$ contours clearly show the vortical structures suggested by the schematics of figure 27. Upstream of the lobe front, at plane *A*–*A*, one downstream stretching vortex pair is seen just underneath the lobe – indicated by the black curly arrows – and three upstream stretching counter-rotating vortex pairs are seen just above the trough surface – indicated by the red curly arrows. At this cross section, the red hairpin shows the three vortex pairs shown in figure 27(*b*), but the black hairpin still has the main two legs and does not manifest undulations in this plane. That is, the undulations have not stretched enough to reach the *A*–*A* plane yet. The effects of these vortex pairs on the liquid surface is evident in this figure. Moving slightly downstream, in plane *B*–*B*, both black and red hairpins manifest three vortex pairs, located under the lobe and above the trough, respectively. The $\lambda_2$ contours at this cross-section agree well with the scenario depicted in figure 27(*b*) on plane *A*. However, the vortex pairs are not nicely aligned in *y*, as shown in figure 27(*b*). The distance between the two legs of each vortex pair has been denoted on the figure. As indicated before, the vortex undulations with wavelengths in the range 15–35 μm are seen in this picture. The creation of bumps and craters on the liquid surface induced by these vortex pairs is also seen here. Further downstream, on plane *C*–*C*, only the two main legs of the red hairpin and parts of the KH vortex are seen. This is what would be expected based on the 3D schematic of figure 27(*a*). The distance between the two legs of this hairpin vortex is 70 μm at this cross section. The wavelength of the undulations and the scale of surface corrugations presumably depend on other parameters such as surface tension and liquid viscosity; however, analysis of the effects of these parameters is beyond the scope of this article and would be an interesting subject for a prospective study.

The mechanism of corrugation formation at lower density ratios is similar to what was shown above in this section. However, the hairpins are not as easily bent (corrugated more slowly) as in high density ratios and, consequently, the number of corrugations are less. Figure 30 shows the vortex structures along with the images of a lobe from a top view in the 64–68 μs period of Case D3b (see table 1). The vortex structures in figure 30(*a*) correspond to the liquid lobe in figure 30(*b*). The liquid surface has been removed from figure 30(*a*) to show all the vortices under the liquid surface; to facilitate the comparison between the vortex structures and surface structures, the lobe front edge is denoted in this neighboring liquid retards the rate of growth of the instability. That is, the reaction of the hairpins to the KH vortex is slowed.

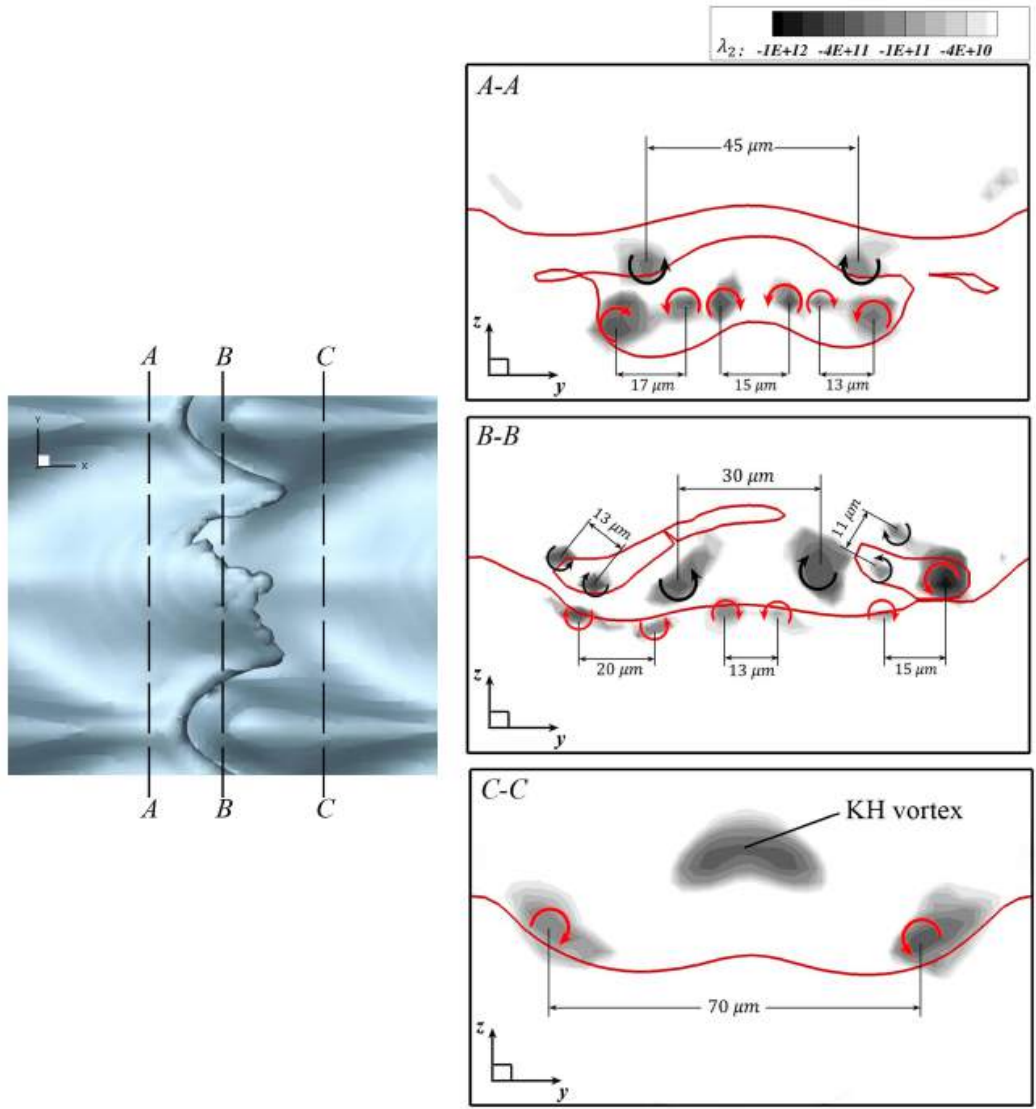

FIGURE 29. Top view of a liquid lobe in Case D3a at $t = 50$ µs (left), and the $\lambda_2$ contours on three $yz$-cross-sections (right); $A$–$A$ passing upstream of the lobe front (top), $B$–$B$ passing through the corrugations (center), and $C$–$C$ passing downstream of the lobe (bottom).

figure by the solid black line. The outer hairpin (green isosurface) and the fast-moving KH vortex still lie above the lobe surface at 64 µs, while the inner hairpin (red isosurface) and the slow-moving KH vortex lie underneath the lobe and closely downstream of the lobe front. Following the illustrations in figure 25 (at $t_3$ and $t_4$), the tip of the outer hairpin stretches upstream while the inner hairpin stretches downstream under the influence of the fast-moving vortex. This process creates the hairpin undulations and increases the number of counter-rotating vortex pairs along the lobe span. However, comparison of figures 30(*a*) and 29 reveals that the hairpin corrugation appears much more quickly at high gas density, as three pairs of counter-rotating hairpins are seen at an earlier time in figure 29. At low gas density, the hairpins are not easily deflected and bent; the reason is conjectured to be the lower gas inertia. While the hairpins are in the gas, the lower inertia of the gas slows down the process of hairpin deflection. By the time the outer hairpin and the fast-moving vortex move into the liquid, the hairpin deformation hastens and another turn forms at its trough (at the center of figure 30*a*), where the green hairpin stretches downstream, over the fast-moving vortex. Thereafter, the number of undulations increases and the corrugations stretch. The relatively higher inertia of the

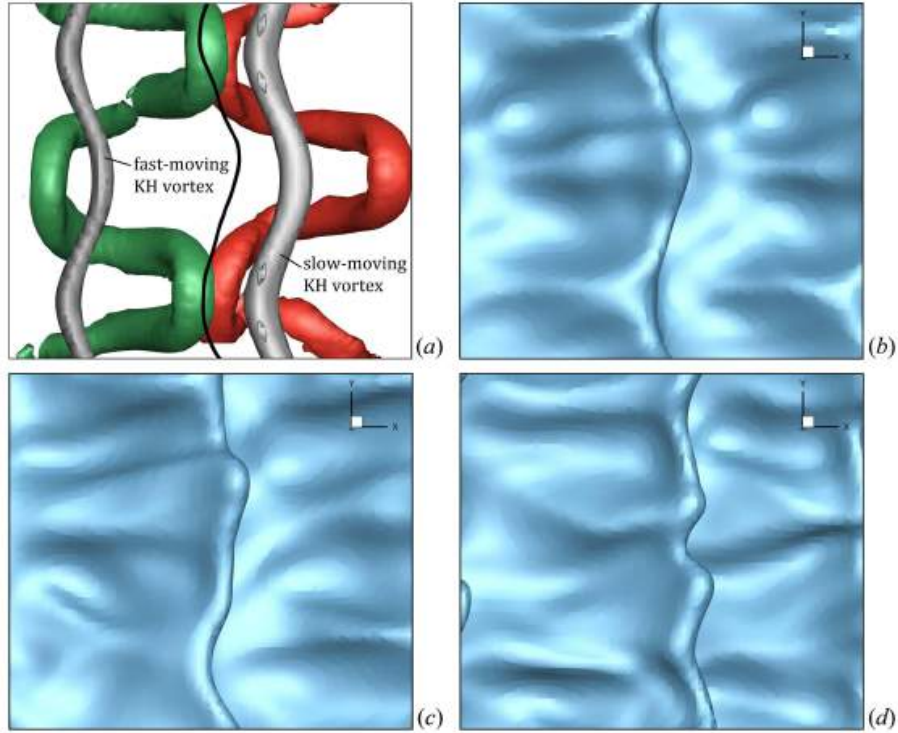

FIGURE 30. $\lambda_2$ isosurface in a top close-up view of a liquid lobe in Domain III at low density ratio (Case D3b) at 64 µs (*a*); the solid black line shows the lobe front edge. The isosurfaces represent: the fast- and slow-moving KH vortices with $\lambda_2 = -7 \times 10^{10}$ s$^{-2}$ (gray), the outer crest hairpin with $\lambda_2 = -4 \times 10^{10}$ s$^{-2}$ (green), and the inner trough hairpin with $\lambda_2 = -3 \times 10^{10}$ s$^{-2}$ (red). Lobe surface showing the corrugation formation from a top view at 64 µs (*b*), 66 µs (*c*), and 68 µs (*d*) of Case D3b.

Figure 30 confirms that the lobe deformation is in fact an outcome of the KH and hairpin vortex distortions, not the other way around. The vortex and liquid structures at 64 µs, shown in figures 30(*a,b*), prove that the vortex deformation precedes the surface deformation. At this time, the hairpins have been already distorted and manifest two pairs of counter-rotating streamwise vortices near the lobe front, while the lobe front itself does not show any corrugation. At later times (figures 30*c,d*), two corrugations start to form following the structure of the undulated hairpins. This transforms the lobe from a singular protrusion to two smaller ones growing out of the corrugated rim. While the lobe rim stretches under the influence of the KH vortex (after the two split counterparts merge), it also retracts towards the center by the capillary forces. This brings the hairpins and the corrugations closer to the center of the lobe, as seen in figure 30(*d*). This retraction is faster at lower $We_g$; compare figures 30 and 29.

### 3.4. *Lobe and ligament stretching at low $Re_l$ (LoLiD mechanism)*

At low $Re_l$ and low $We_g$, the surface tension force resists perforation. The liquid viscosity is also fairly high and damps the small scale corrugations on the lobe front edge. Consequently, the entire lobe stretches slowly into a thick and long ligament, which eventually breaks into large droplets. This terminates the *LoLiD* breakup mechanism, which prevails in Domain I, shown in figure 1. The *LoLiD* process is shown step by step in figure 31. The case shown in this figure (Case D1a) is taken for vortex analysis in this section. $\lambda_2$ contours on *y*-planes and $\lambda_2$ iso-surfaces in the interface vicinity are used in this section to explain the vortex dynamics of the *LoLiD* process.

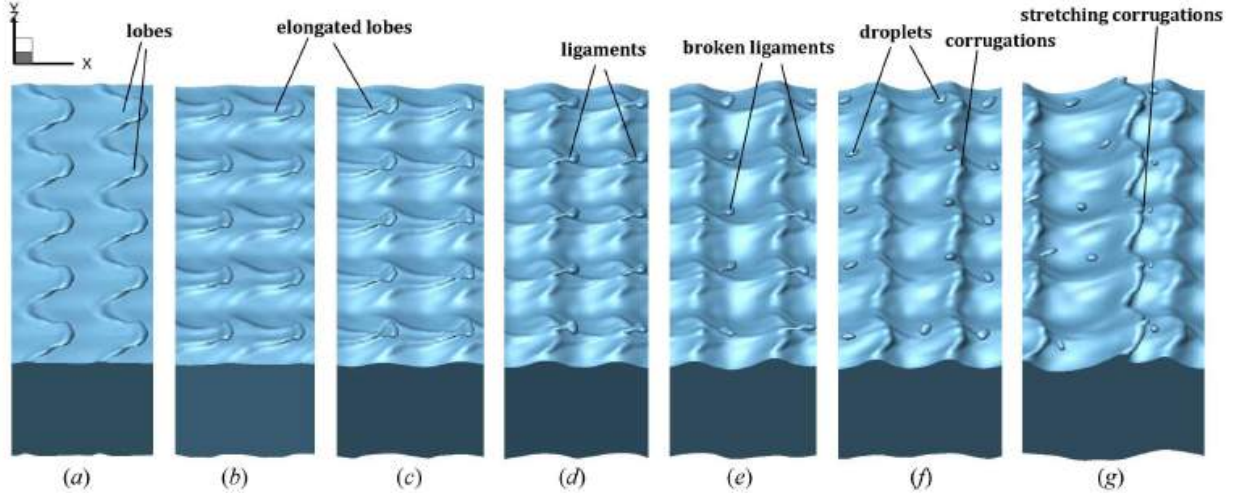

FIGURE 31. Liquid surface deformation following the *LoLiD* mechanism in Case D1a, at $t = 26$ µs (*a*), 36 µs (*b*), 40 µs (*c*), 44 µs (*d*), 46 µs (*e*), 48 µs (*f*), and 52 µs (*g*).

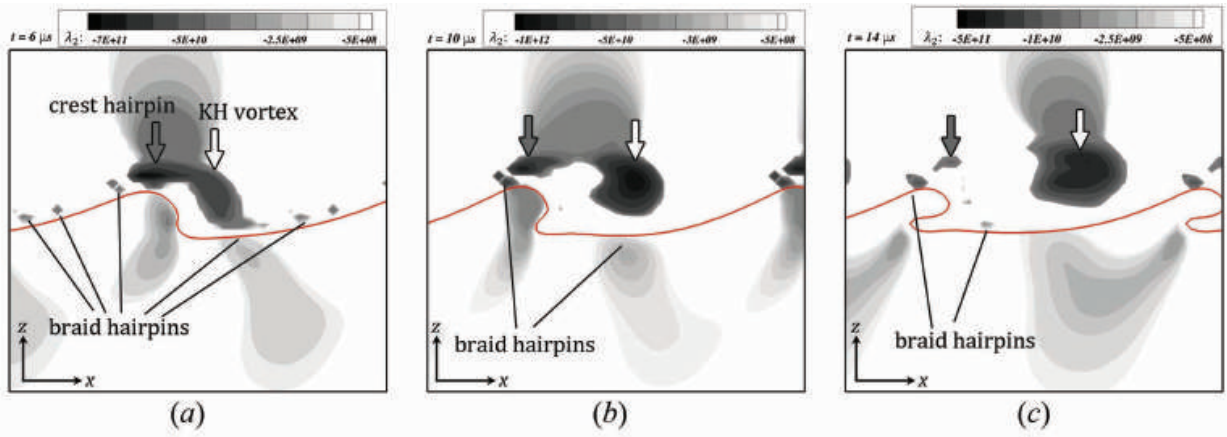

FIGURE 32. $\lambda_2$ contours on the spanwise crest cross-section, at $t = 6$ µs (*a*), 10 µs (*b*), and 14 µs (*c*) of Case D1a.

As shown in figure 32(*a*), the process starts with a large KH roller downstream of the KH wave (denoted by the white arrow), hairpin filaments on the braid, and a much stronger hairpin filament on the KH wave crest (denoted by the dark gray arrow). Because of the high gas viscosity, the KH vortex diffuses much faster into the gas and at 6 µs the vortex core is entirely in the gas zone. Since the gas phase has higher velocity compared to the liquid phase, the gas distant from the liquid surface advects faster with respect to the interface. Hence, a few microseconds later, say at 10 µs (figure 32*b*), the entire KH roller advects downstream with respect to the interface as its core moves farther from the interface via diffusion. The KH roller gains speed as it moves away from the interface. In contrast, the hairpin filaments that are closer to the liquid surface remain almost stationary with respect to the interface; see figure 32(*c*). The KH vortex gets larger with time as it diffuses; compare figures 32(*a-c*).

The KH roller reaches the neighboring downstream crest hairpin at 16 µs (figure 33*a*). Meanwhile, the braid hairpins overlap with the crest hairpins, constraining the lobe sheet in between. So far, the vortex dynamics manifest the conditions required for hole formation, i.e. overlapping of two oppositely-oriented hairpins on top and bottom of the lobe, as well as for corrugation formation; i.e. constant pull in opposite directions induced by a moving vortex hovering over the hairpins. However, none of these structures are seen on the lobe at this moment (see figure 31*a*). The inhibition of hole formation is due to the high surface tension. In such a low $We_g$, the inertia and viscous forces are not strong enough to overcome the surface tension force to stretch and thin the lobe; hence, the lobe perforation is inhibited. Also, because of the high liquid viscosity at such a low $Re_l$, the liquid surface deformation is much slower. Hence, the corrugation formation on the lobe edge does not occur as quickly as for the higher $Re_l$. Small scale corrugations are damped by both the high viscous and high surface tension forces. Thus, the lobe slowly stretches into a thick ligament. Moreover, the KH roller is also farther away from the interface in this case compared to the $LoCLiD$ process (compare figures 33*a* and 24*c*), which means that it has a much weaker influence on the hairpin filaments.

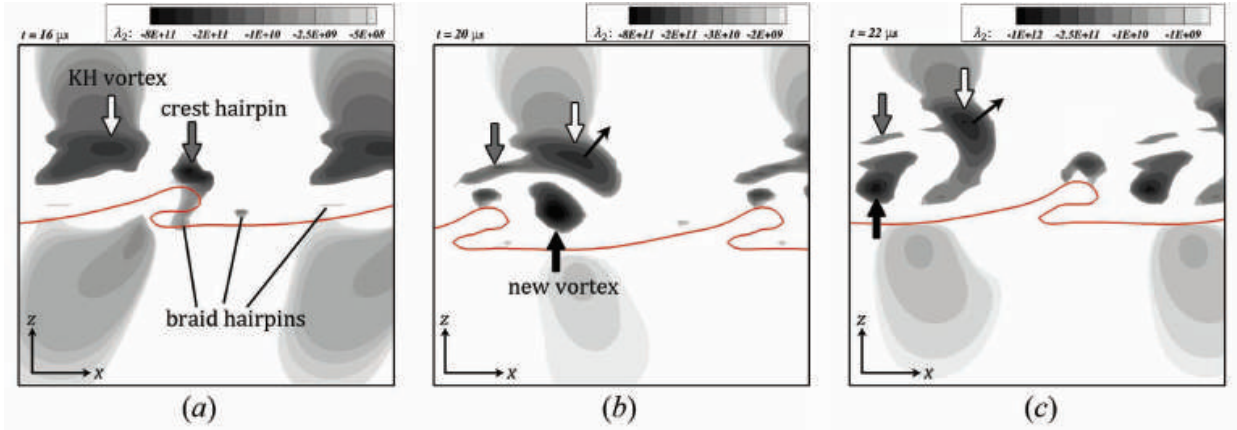

FIGURE 33. $\lambda_2$ contours on the spanwise crest cross-section, at $t = 16$ µs (*a*), 20 µs (*b*), and 22 µs (*c*) of Case D1a.

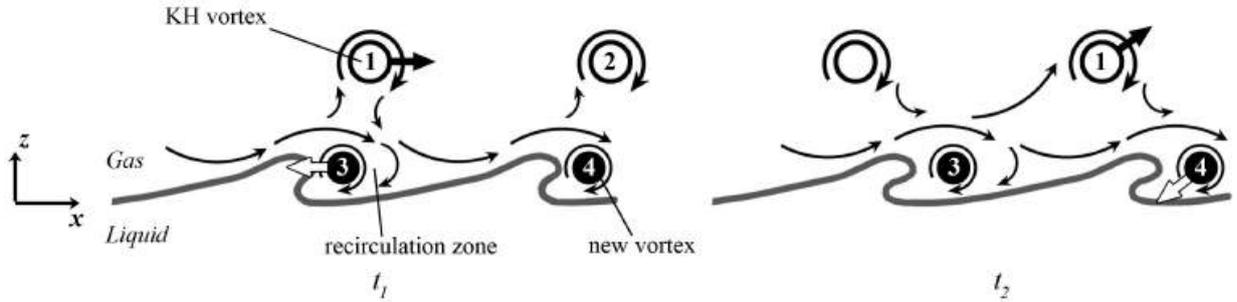

FIGURE 34. Schematic of the recirculation zone downstream of the KH wave. The hollow circles denote the KH vortex and the black circles denote the newly formed vortex in the recirculation zone. The induction of each vortex on the neighboring vortices is indicated by the straight arrows with the same shading as the source vortex. The curly arrows qualitatively indicate the streamlines near the interface.

The tip of the KH wave acts like a backward-facing step for the gas flow and a temporary recirculation zone forms immediately downstream of the wave, as shown in figure 34. This is very similar to the scenario seen at low density ratios by Hoepffner *et al.* (2011). They also showed that the head of the wave appears to the gas stream as a fixed obstacle, with the ensuing vortex shedding. The high strain rate at the braid depletes the vorticity upstream of the braid and collects the vorticity into a new vortex just downstream of the KH wave, in the recirculation zone at 20 µs. This new vortex is indicated by the thick black arrow in figure 33(*b*) and the black circles in figure 34. When the KH roller (vortex 1) passes over the KH wave front (see $t_1$ in figure 34), the mutual induction of this roller and the new vortex (vortex 3) causes the KH roller to advect farther downstream, further stretching the KH vortex. The local induction of each vortex on the neighboring vortices is indicated by the straight arrows of the same shading as the source vortex in figure 34. When the KH vortex (vortex 1) passes over the downstream wave at 22 µs (figure 33*c*), the mutual induction of the vortices pushes away the KH vortex, advecting it in the $z$-direction; see $t_2$ in figure 34. The stronger induction by vortex 4 pushes vortex 1 in the direction normal to the line connecting the centers of the two vortices (in the direction of the black arrow). Vortex 3 also induces a flow that would bring vortex 1 closer to the interface; however, this induction is much weaker as vortex 3 is more distant from vortex 1 compared to vortex 4. The new vortex wraps under and around the KH roller at its crest and later over the trough of the adjacent downstream KH roller, as shown in figure 35. Thus, this new vortex (the red tube in figure 35) transforms into a hairpin vortex that is stretched between two adjacent KH vortices and wraps around them at their crests and troughs.

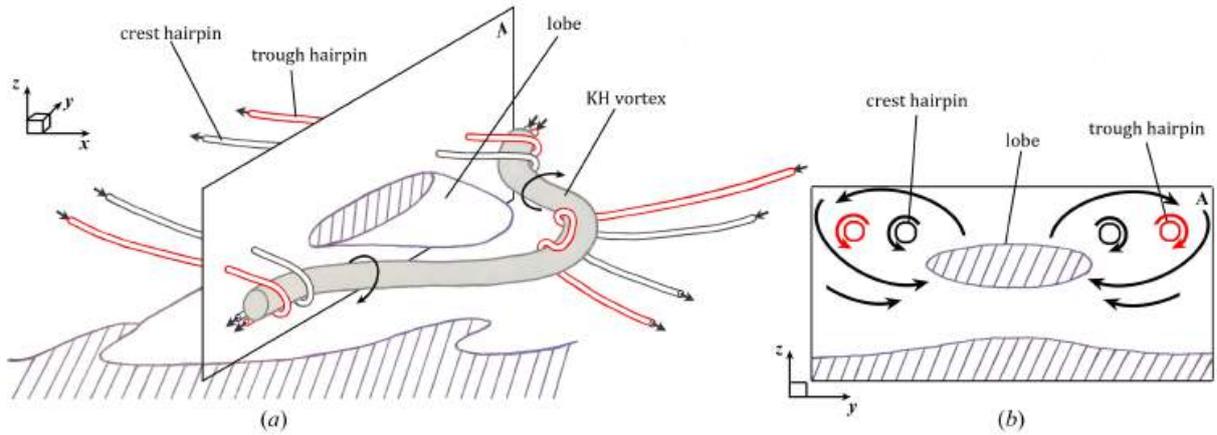

FIGURE 35. 3D Schematics showing the vortex structures in the *LoLiD* Domain (*a*) - *A* is the plane in which (*b*) is drawn; cross-sectional view of the *A*-plane, showing the spanwise squeezing of the lobe by induced flow of the hairpin vortices (*b*). The vortex schematics are periodic in $x$- and $y$-directions.

Schematics of the vortex structures corresponding to $t = 26$ µs (shown in figure 36*a*) are illustrated in figure 35. The KH vortex has a larger undulation in this domain compared to the other two domains and is also farther away from the interface in the gas zone (compare figure 35 with figures 11 and 27). Two pairs of counter-rotating hairpins – one on the lobe crest (the black tube), and the other on the trough (the red tube) – stretch and wrap around the KH vortex (figure 35*a*). Notice that the red hairpin in figure 35 is actually the new vortex that was formed at the wave front end and was indicated by the black arrows in figures 33 and 36. These hairpins are periodic in both spanwise ($y$) and streamwise ($x$) directions; i.e. the tubes that emerge at the bottom corner or the left side of figure 35(*a*), re-enter from the top corner or the right side of the sketch, respectively. As shown in the cross-sectional view of the *A*-plane passing through the lobe in figure 35(*b*), both the black and the red hairpins are located slightly above and on both sides of the lobe at this moment. While the flow induced by the KH vortex creates a streamwise flow on top and bottom of the lobe, the gas flow induced by these two counter-rotating hairpins (shown by the curly arrows is figure 35*b*) generates a spanwise flow towards the lobe midplane. Consequently, the lobe is both squeezed in the spanwise direction – via the induced flow of the hairpins – and stretched in the streamwise direction by the induced flow of the KH vortex. The gas flow induced by the hairpins also lifts the lobe in the $z$-direction; see figure 35(*b*).

While the KH vortex keeps diffusing due to the high gas viscosity, the dark gray and black vortices wrap around it and are stretched with it as it moves downstream (figure 36*b*). Meanwhile, another vortex forms downstream of the wave crest at 30 µs, indicated by the light gray arrow in figure 36(*b*). This vortex advects downstream closely hovering over the interface at 36 µs, while new vortices keep forming downstream of the KH wave crest at the recirculation zone to replace it; see figure 37. Hoepffner *et al.* (2011) were the first to identify this vortical mechanism at low density ratios. Similar to figure 37, they showed that two vortices exist in the gas – one is at the instant of leaving the shelter of the wave (the dark gray curly arrow in figure 37), and the second further downstream is the result of the previous shedding event (the light gray curly arrow in figure 37). The induced motion of these vortices entrains the gas and draws it under the protruding lobe (follow the velocity vectors in the gas zone near the interface in figure 37). As is well-known, the streamwise strain rate is highest at the saddle (Hussain 1986; Lasheras & Choi 1988). The location of the saddle point is marked by a star symbol in figure 37. The vortices that reach the saddle point stretch in the streamwise direction and transform into hairpin vortices with streamwise legs. The location of the black and dark gray hairpins that are wrapping around the KH vortex are denoted by straight arrows of the corresponding vortex color. Since these vortices are mostly streamwise at this cross-section, they do not have $\omega_y$ contours. The streamwise elongation of the vortices progressively aligns the two counter-rotating legs of the hairpins in the streamwise direction. The self-induction of the counter-rotating legs of the KH vortex and the hairpin vortices, moves the vortex tubes in the normal direction away from the interface, as shown schematically in figure 38.

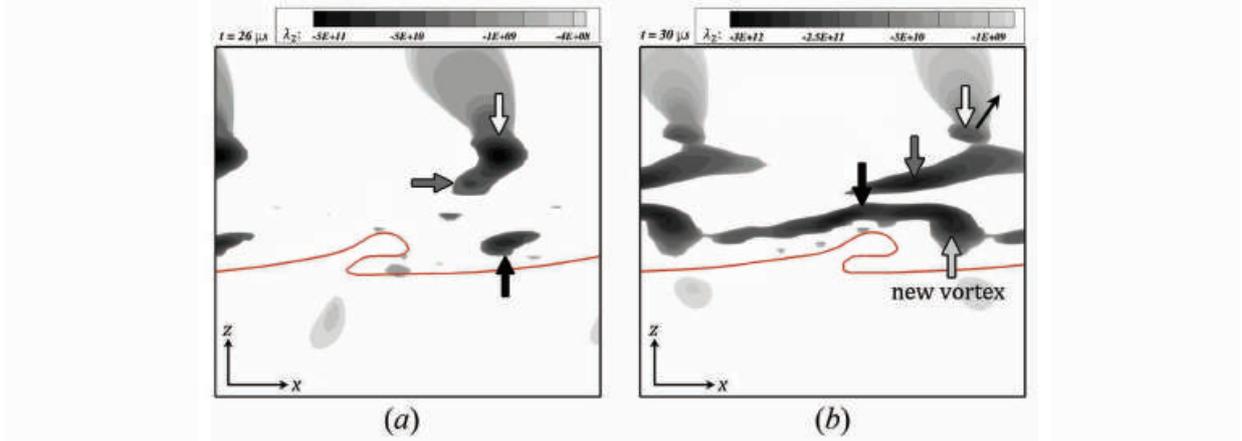

FIGURE 36. $\lambda_2$ contours on the spanwise crest cross-section at $t = 26$ µs ($a$), and 30 µs ($b$) of Case D1a.

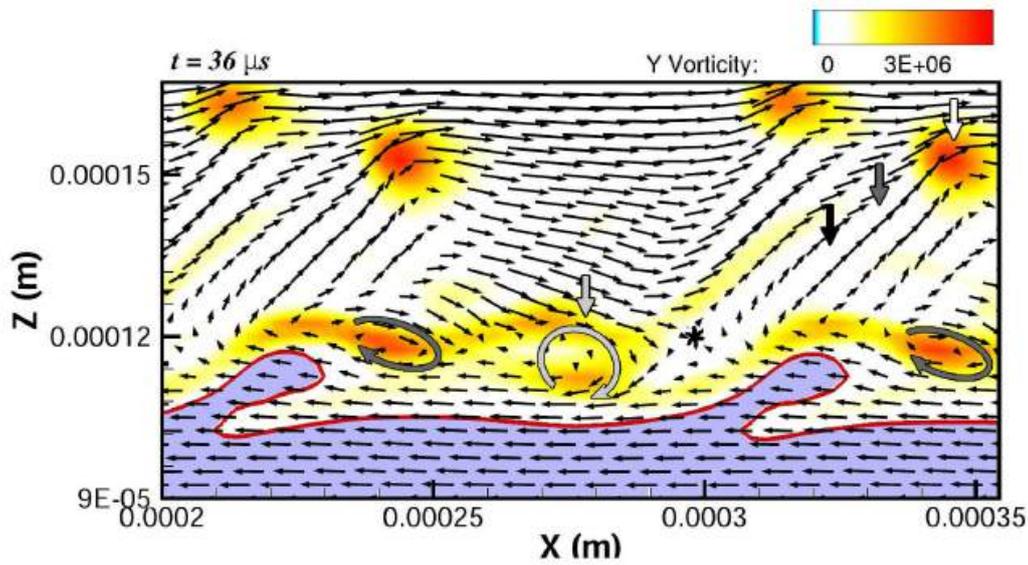

FIGURE 37. Velocity fluctuation vector field with respect to the average velocity of the vortex cores near the interface, and $\omega_y$ contours on the spanwise crest cross-section at $t = 36$ µs of Case D1a. The thick arrows refer to the vortices shown in figure 36($c$). The star symbol marks the saddle point, and the liquid is shown in light blue.

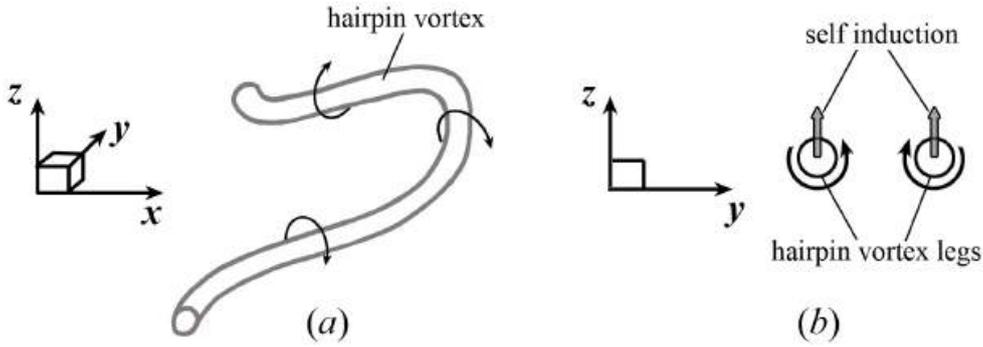

FIGURE 38. Schematic showing the self-induction of the streamwise counter-rotating hairpin legs (*a*), and in a *yz*-plane crossing the legs (*b*). The gray arrows in (*b*) show the direction of the self-induction.

The temporal evolution of the vortex structures in the *LoLiD* process is illustrated in figure 39 from a top view of the liquid surface. The liquid surface is illustrated in the right panel of this figure, and the $\lambda_2$ isosurfaces (vortex filaments) are depicted in the left panel. In the beginning of the process (figure 39*a*), the KH vortex (gray tube) and the crest hairpin filament (green tube) are completely spanwise oriented (in the *y*-direction). The braid hairpins (thinner green tubes) are also primarily spanwise with some undulations and are half a wavelength (180°) out of phase with respect to the crest hairpins in *y*. This is consistent with the findings of Jarrahbashi *et al.* (2016) for non-homogeneous round jets, and Danaila *et al.* (1997) for like-density jets. At 16 µs, the KH vortices stretch more; the hairpins on the braid also collect into the crest hairpin and manifest higher undulations (green isosurface in figure 39*b*). The trough vortex (red isosurface), which was first seen in figure 33(*b*) (indicated by black arrow) also manifests at this time. Both the green and the red vortex filaments slide underneath the KH roller at the spanwise wave crest (lobe crest), and wrap over it at the spanwise wave trough (lobe sides). The wave front has been marked with solid black line in figure 39 to help find the location of the lobe crest and trough with respect to the vortices. The KH roller in this process is more stretched than the other two mechanisms (compare figure 39 with figures 17 and 26).

The vortices become streamwise near the braid in both spanwise crest and trough at 26 µs (figure 39*c*). Except for the tip of the KH roller, which is still spanwise, the rest of the vortex structures are nearly streamwise-oriented. Both the green and the red hairpins wrap around the KH vortex on top of each other (see figures 35*a* and 39*c*). As the vortices get stretched in the streamwise direction, the lobes follow their shape and get thinner in the spanwise direction because of the induced gas flow by the hairpin legs; compare the shape of the lobes with that of the KH vortex just downstream of the lobe in figure 39(*c*) to observe their similarity. A neck starts to form on the lobe edge right between the two hairpins on the two sides of the lobe. This completely follows the mechanism introduced in figure 35.

Figure 40 shows the cascade and deformation of the vortex filaments in the period 6 µs – 52 µs. The $\lambda_2$ isosurfaces are colored by the streamwise velocity contours here. The vortices start from a spanwise orientation and gradually turn streamwise. The vortex structures cascade from thick and uniform vortices to thin and chaotic structures as they stretch. The gradual departure of the vortices away from the liquid surface is also shown in this figure, where the KH roller, indicated by the white arrow, starts on the liquid surface at 6 µs (figure 40$a$) and advects away from the interface as it gets streamwise due to the self-induction process described in figure 38.

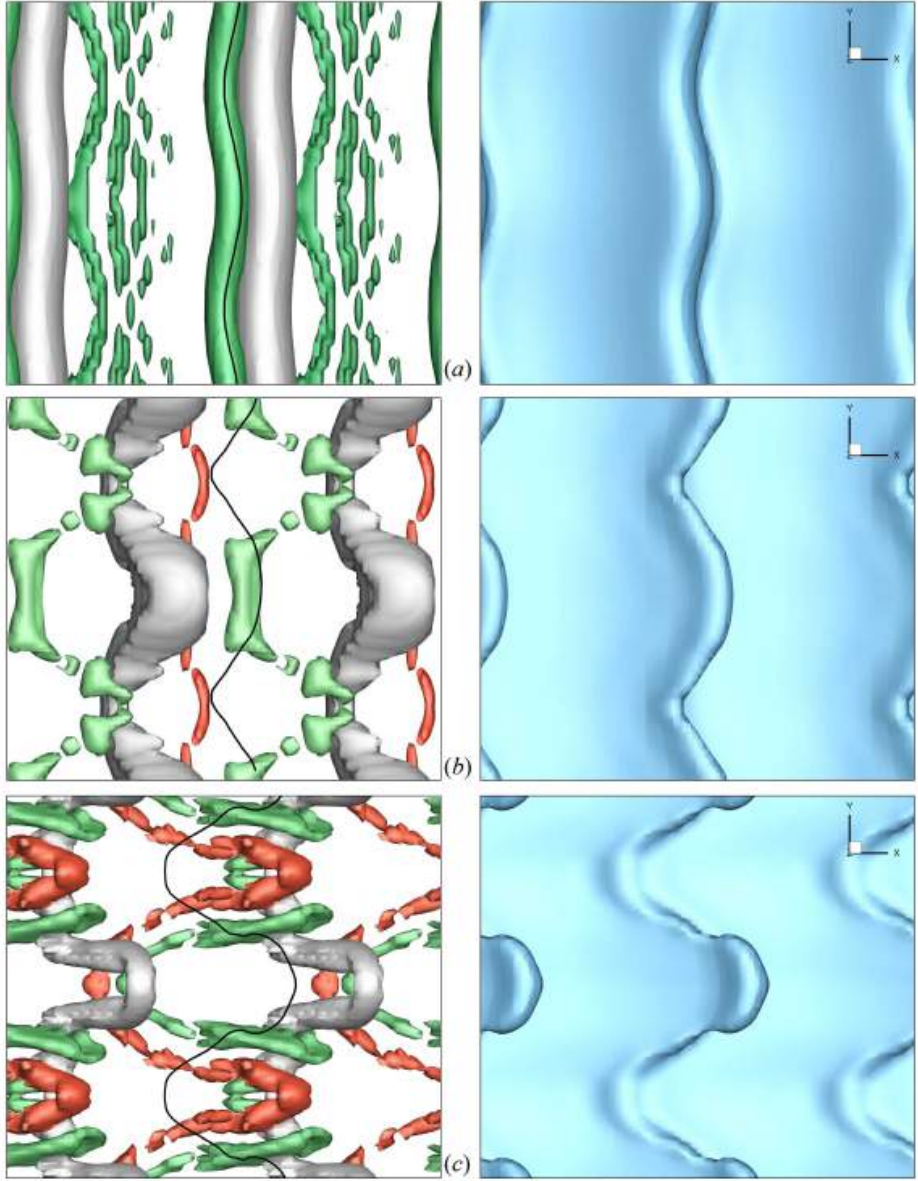

FIGURE 39. $\lambda_2$ isosurface in a top close-up view of a liquid lobe in Domain I (Case D1a) on the left and the liquid surface from a top view on the right, at $t = 6$ µs ($a$), 16 µs ($b$), and 26 µs ($c$). The solid black line on the left images shows the location of lobe front edge. The isosurfaces represent: the KH vortex with $\lambda_2 = O(-10^{10})$ (gray), the crest hairpin with $\lambda_2 = O(-10^{11})$ (green), and the trough hairpin with $\lambda_2 = O(-10^{11})$ (red).

At 40 µs (figure 40$d$), almost all of the vortex structures have become streamwise except for a short section at the tip of the KH roller. This confirms that the vortices become streamwise-oriented as they get close to the saddle point. Meanwhile, the vortices cascade into smaller structures due to turbulence; compare figures 40($c$) and 40($d$). As the legs of the hairpins get closer to each other, the mutual-induction of the legs lifts the hairpins in the normal direction (figure 38), and locates the hairpins above the lobes. The lobes get thinner as the streamwise counter-rotating legs induce a gas flow in the spanwise direction towards the lobe midplane, squeezing the lobe and transforming it into a ligament. Ligament creation is strongly correlated with its local velocity field, and is induced by the local shear, as indicated by Shinjo & Umemura (2010). Nearby vortices determine the ligament formation direction. Spanwise vortices form ligaments normal to the injection direction, i.e. spanwise, and streamwise vortices form ligaments parallel to the injection direction; i.e. streamwise. The influence of the vorticity field on the ligament orientation is consistent with the findings of Shinjo & Umemura (2010). Following the streamwise-oriented hairpin vortices in the *LoLiD* process, mostly streamwise ligaments form in Domain I.

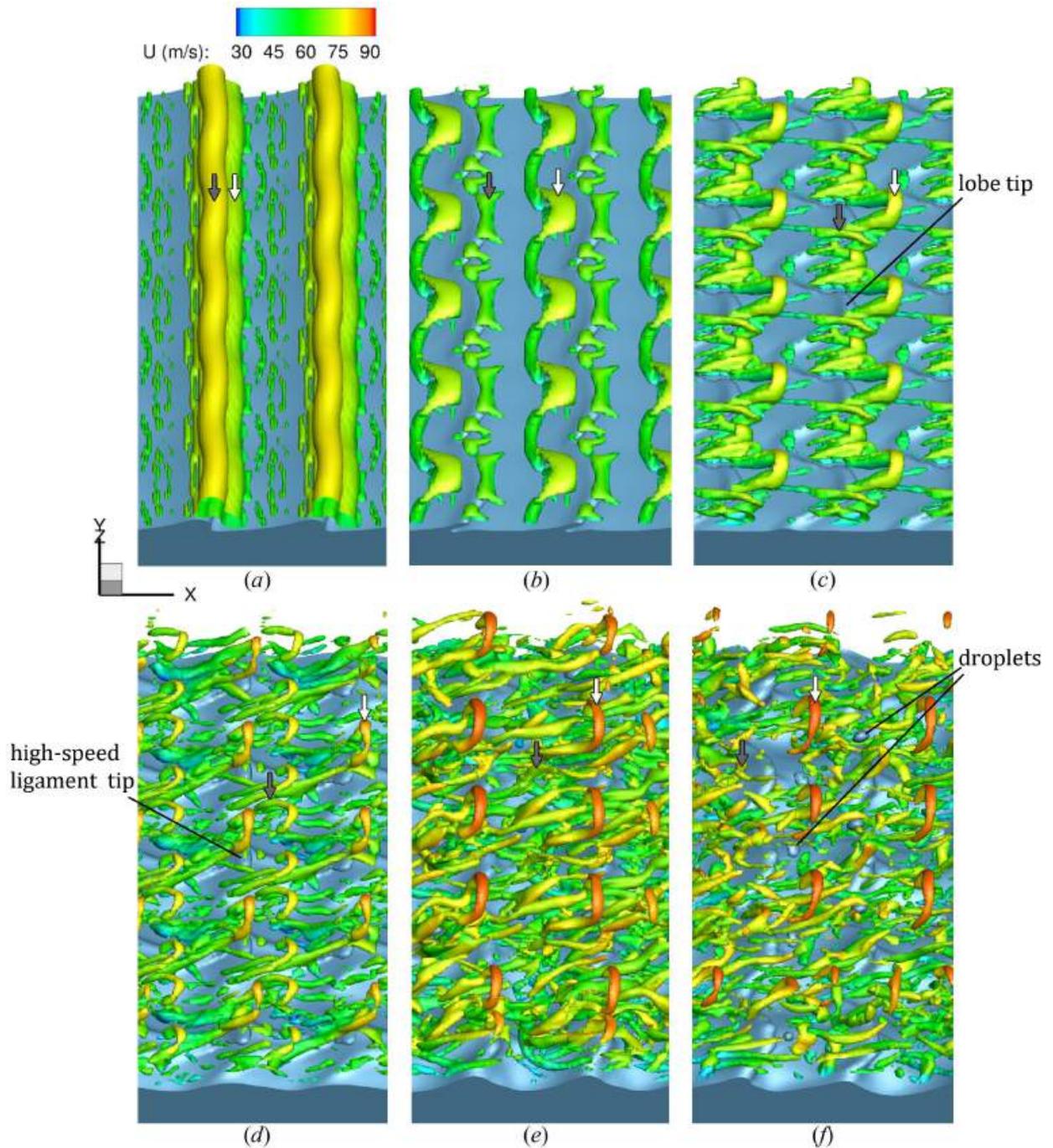

FIGURE 40. Temporal evolution of the vortices indicated by the $\lambda_2$ iso-surface colored by the streamwise velocity contours, at $t = 6$ μs (*a*), 16 μs (*b*), 26 μs (*c*), 40 μs (*d*), 48 μs (*e*), and 52 μs (*f*) of Case D1a. The liquid surface is shown in blue. The arrows refer to the vortices denoted in figures 32–36.

Velocity contours in figure 40(*e*) show that the streamwise velocity of the vortices increases at higher *z*-levels. This can be clearly seen from the gradient of colors from green (50 m/s) – near the liquid surface – to red (90 m/s) – at the tip of the KH vortex.

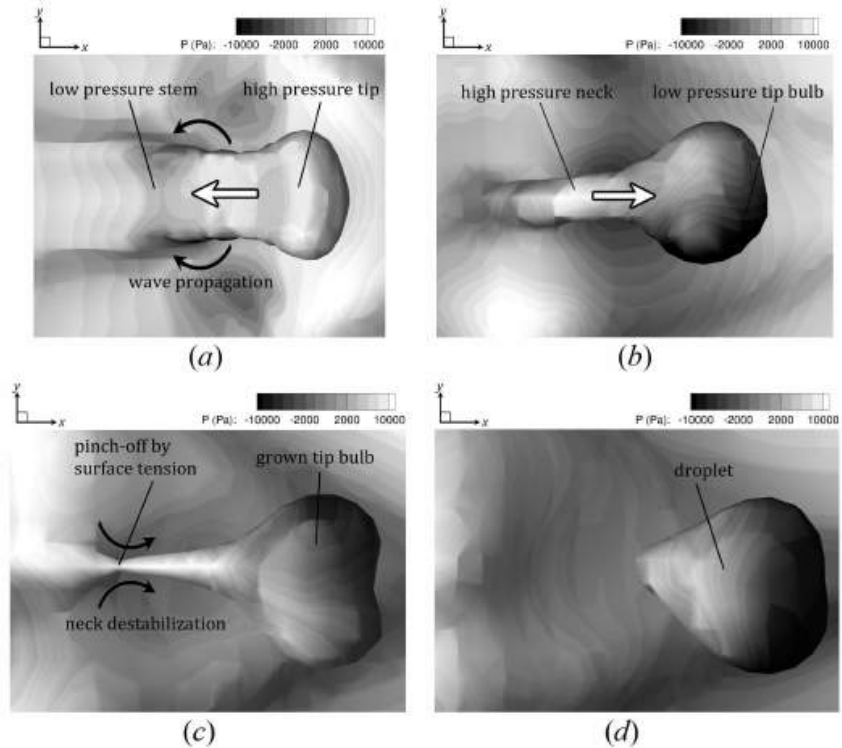

FIGURE 41. Pressure contours on the surface of a ligament at $t = 42$ μs (*a*), 44 μs (*b*), 45 μs (*c*), and 46 μs (*d*) of Case D1a, showing the ligament pinch-off following the short-wave breakup mode.

The liquid surface also experiences a similar velocity gradient, where the velocity at the tip of the lobe is higher compared to its root; i.e. where the lobe connects to the liquid sheet. This velocity gradient manifests how the lobe elongates under the streamwise strain and forms the ligament. The ligaments finally pinch-off and create droplets, as shown in figure 40(*f*). The ligament pinch-off follows the short-wave breakup mode introduced by Shinjo & Umemura (2010). The ligament acts as a very small round liquid jet emanating from the sheet surface. The ligament tip pressure is high in the beginning and contracts due to surface tension and pushes the inner part along the ligament axis (figure 41*a*). This motion emanates compression waves in the upstream direction and a neck forms. The tip bulb grows as it absorbs the liquid from the upstream liquid sheet by contraction (figure 41*b*). As the tip bulb size grows, its inner pressure drops. Consequently, the tip bulb sucks the liquid further from the neck, as shown in figure 41(*b*) – causing the neck to become narrower. When the neck becomes thin enough and its pressure high enough, the circumferential surface tension cuts the neck and a droplet pinches off (figure 41*c*). This process reiterates from the beginning, and the next droplets pinch off identically. The droplets fly away from the interface under the vortex induction and gain a higher velocity than the liquid sheet; therefore, the droplets advect downstream with respect to the jet (figure 40*f*).

The vortex structures and lobe deformation at low density ratio (Case D1b) are shown in figure 42. The lattice made by the KH vortex and the hairpin vortices is very similar to what was proposed in figure 35 for high density ratios. The only difference being that as gas density decreases, the KH vortices are less bound to the liquid and depart from the interface much easier and meanwhile stretch in both streamwise and normal $z$-directions (figure 42*a*). All the vortices in figure 42(*a*) exist in the gas zone and above the liquid interface. Following this vortex deformation, the lobe is squeezed and thinned in the spanwise direction, while being stretched and lifted by the KH vortex normally outward. This causes the ligaments to protrude more and quicker in the normal direction (figure 42*d*) compared to higher $\hat{\rho}$. This figure also proves that the vortex deformation leads the surface deformation, and not the other way around.

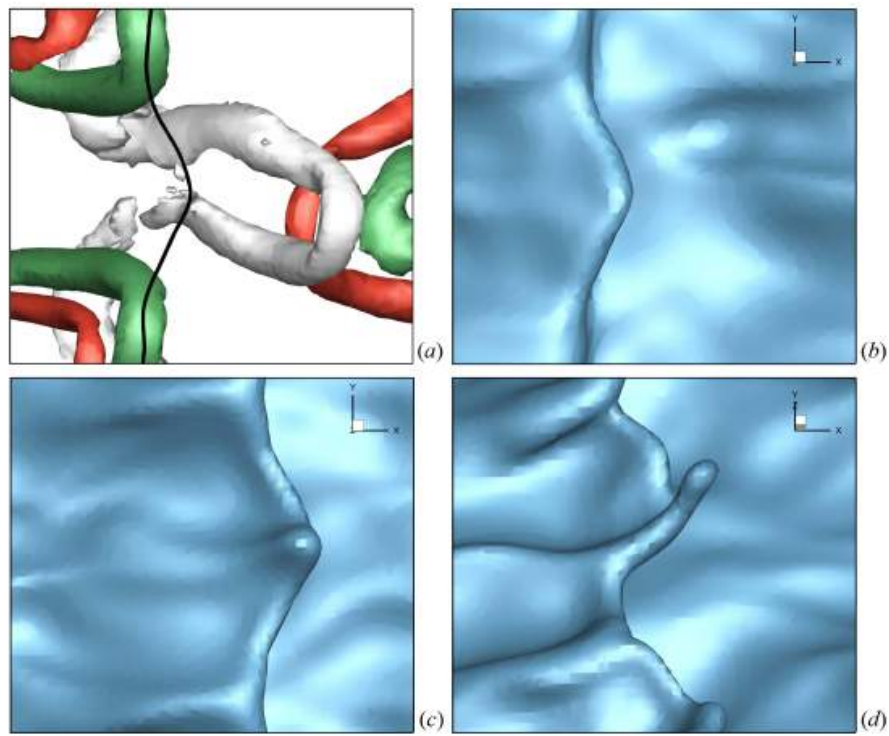

FIGURE 42. $\lambda_2$ isosurface in a top close-up view of a liquid lobe in Domain I at low density ratio (Case D1b) at 16 µs (*a*); the solid black line shows the lobe front edge. The isosurfaces represent: the KH vortex with $\lambda_2 = -10^{11}$ s$^{-2}$ (gray), the outer crest hairpin with $\lambda_2 = -2 \times 10^{11}$ s$^{-2}$ (green), and the inner trough hairpin with $\lambda_2 = -3 \times 10^{11}$ s$^{-2}$ (red). Lobe surface showing the corrugation formation from a top view at 16 µs (*b*), 18 µs (*c*), and a 3D view at 22 µs (*d*) of Case D1b.

As will be discussed in §3.5, when density ratio gets higher, i.e. higher $We_g$, the vortices are closer to the interface and are less stretched, and more spanwise rather than streamwise oriented. Hence, the gas flow that is induced by the KH vortex shears the lobe from its top and bottom sides. In this scenario, the lobe thins in the normal direction instead of the spanwise direction, as shown schematically in figure 43(*a*). The blue curly arrows illustrate the qualitative streamlines in this situation, and the red straight arrows denote the direction of lobe squeeze. The shear due to the gas flow induced by the spanwise vortices stretches the lobe in streamwise direction and thins the lobe and makes it more vulnerable to puncture. This is consistent with the map in figure 1; as $We_g$ increases, the breakup mechanism shifts from $LoLiD$ to $LoHBrLiD$.

At lower density ratio and lower $We_g$ (Domain I), the vortices are more stretched in the streawise direction. In this domain ($LoLiD$ mechanism), the streamwise legs of the hairpin vortex on the two sides of the lobe induce a gas flow in the spanwise direction towards the lobe midplane, which squeezes and thins the lobe in the spanwise direction and transforms it into a thick ligament, as shown schematically from the top view of a lobe in figure 43(*b*). A neck forms at the location of the red arrows. The induced motion of the hairpin legs also lifts the lobe in the *z*-direction. Eventually, the ligament pinches off at the neck and a droplet forms following the short-wave mode discussed above.

In summary, whether the lobe thins in the *z*-direction and perforates or thins in the *y*-direction and forms a ligament depends on the orientation of the vortices in the vicinity of the lobe. The spanwise vortices result in hole formation and spanwise bridges,

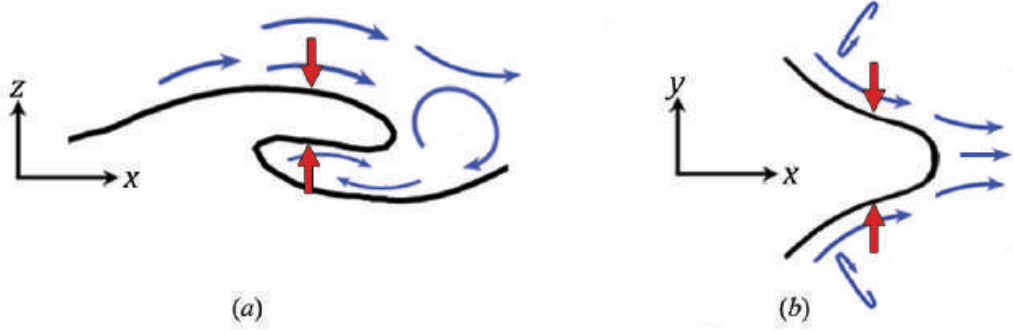

FIGURE 43. Schematic showing the lobe thinning in the normal direction in the *LoHBrLiD* process from a $xz$ plane (*a*), and lobe thinning in the spanwise direction in the *LoLiD* process from a $xy$ plane (*b*).

while streamwise vortices result in spanwise lobe compression and streamwise ligament stretching.

### 3.5. *Streamwise vorticity generation*

The streamwise vorticity ($\omega_x$) is crucial in initiation of the three-dimensional instability on liquid jets. $\omega_x$ generation via vortex stretching and vortex tilting – i.e. strain-vorticity interactions – and baroclinic effects are studied in this section for a low density ratio of 0.05 and a high density ratio of 0.5. The contributions of the different terms in the vorticity equation to $\omega_x$ generation are compared at two distinctly different density ratios, to understand the role of density ratio in the liquid-jet breakup. Specifically, the generation of the three-dimensionality on the liquid sheet interface is addressed in this section. Jarrahbashi & Sirignano (2014) performed a similar analysis for the round jet at several different density ratios. They showed for a density ratio of 0.01 that the baroclinic effect, i.e., the Rayleigh-Taylor (RT) instability, is the dominant cause of the initiation of three dimensional structures. This is consistent with the suggestion of Marmottant & Villermaux (2004) who performed their experiment at lower pressures and gas density. However, for a density ratio of 0.1 or greater, Jarrahbashi & Sirignano (2014) showed that the azimuthal tilting and radial tilting of the ring vortices are the dominant effects in streamwise vorticity generation.

The complete vorticity equation is

$$\frac{D\boldsymbol{\omega}}{Dt} = (\boldsymbol{\omega} \cdot \boldsymbol{\nabla})\boldsymbol{u} - \boldsymbol{\omega}(\boldsymbol{\nabla} \cdot \boldsymbol{u}) + \boldsymbol{\nabla} \times \left(\frac{\boldsymbol{\nabla} \cdot \boldsymbol{\tau}}{\rho}\right) + \frac{1}{\rho^2}\boldsymbol{\nabla}\rho \times \boldsymbol{\nabla}p + \boldsymbol{\nabla} \times \boldsymbol{F_\sigma}, \quad (3.1)$$

where $\boldsymbol{u}$ and $\boldsymbol{\omega}$ are the velocity and vorticity vectors, respectively. $\boldsymbol{\tau}$ is the viscous stress tensor, and $\boldsymbol{F_\sigma}$ is the surface tension force. Since the fluids are incompressible in this study, the second term on the right hand side is zero. A simple dimensional analysis shows that the viscous diffusion term (the third term on the right hand side) scales as $\mu U/\rho \Delta^3$, which for a typical case considered in our study, e.g. $\mu = O(10^{-3})$ kg/(m.s), $U = O(10)$ m/s, $\rho = O(10^3)$ kg/m$^3$, and a mesh size of $\Delta = 2.5$ μm, gives a magnitude of $\approx O(10^{10})$ s$^{-2}$. The surface tension term (the last term) scales as $\sigma \kappa/\rho \Delta^2$; which for a typical case with $\sigma = O(10^{-2})$ N/m, and a radius of curvature of 100 μm ($\kappa = 10^4$ m$^{-1}$), also gives a result in the order of $10^{10}$ s$^{-2}$. As will be shown in this section, these two terms are 2 to 3 orders of magnitude smaller than the other terms in equation 3.1, and therefore have negligible contributions to vorticity generation; thus, the rate of change of $\omega_x$ is approximately

$$\frac{D\omega_x}{Dt} = \omega_x \frac{\partial u}{\partial x} + \omega_y \frac{\partial u}{\partial y} + \omega_z \frac{\partial u}{\partial z} + \frac{1}{\rho^2}\left[\frac{\partial \rho}{\partial y}\frac{\partial p}{\partial z} - \frac{\partial \rho}{\partial z}\frac{\partial p}{\partial y}\right], \quad (3.2)$$

where $\omega_x$, $\omega_y$, $\omega_z$, and $u$ denote the streamwise, spanwise, and cross-stream (normal) vorticities, and streamwise velocity, respectively. The terms on the right-hand-side denote streamwise stretching, spanwise tilting, normal tilting, baroclinic effect due to normal pressure gradient, and baroclinic effect due to spanwise pressure gradient, respectively. Density gradient is normal to the liquid interface; i.e. approximately in the $z$ direction. The spanwise density gradient, i.e. $\partial \rho/\partial y$, is negligible compared to $\partial \rho/\partial z$, since the sheet cross-section remains fairly rectangular during early instability development. Yet, this term accounts for the baroclinic effect, which can deform the interface in the spanwise direction.

In the cases studied in this section, no initial perturbation is imposed on the liquid surface. Except for numerical errors which for our purpose correspond to small random physical disturbances, all terms in the $\omega_x$ generation equation (equation 3.2) are initially zero. Namely, the first and the third terms are zero because there is no vorticity components in the $x$ and $z$-directions initially, and the second term is zero since the streamwise velocity is uniform in the spanwise direction. The baroclinic terms are identically zero since density and pressure gradients in the $y$-direction are initially zero.

Since the streamwise ($\omega_x$) and normal vorticities ($\omega_z$) cannot be generated, but can be enhanced, the main source of $\omega_x$ generation at early times is either the spanwise vorticity tilting or the baroclinic torque, which become non-zero as a result of small perturbations of $u$ and $p$ in the spanwise direction. This intuition is consistent with the results of Jarrahbashi & Sirignano (2014) in round jets at early times, where for a wide range of density ratios, the baroclinic torque and the azimuthal vortex tilting terms are dominant for the first 5 µs of their computations; however, they might be overtaken later by other terms. Baroclinicity becomes more pronounced at lower density ratios, since the density gradient across the interface is higher.

In the data analysis, the gradients have been calculated and averaged over the computational interface thickness that equals three mesh points in the $z$ and tangential directions. The terms in equation (3.2) are:

- Streamwise vortex stretching:  $\omega_x \dfrac{\partial u}{\partial x}$
- Spanwise vortex tilting:  $\omega_y \dfrac{\partial u}{\partial y}$
- Normal vortex tilting:  $\omega_z \dfrac{\partial u}{\partial z}$
- Baroclinic vorticity generation:  $\dfrac{1}{\rho^2}\left[\dfrac{\partial \rho}{\partial y}\dfrac{\partial p}{\partial z} - \dfrac{\partial \rho}{\partial z}\dfrac{\partial p}{\partial y}\right]$

Since the peak of the streamwise vorticity occurs at the wave braids (as shown by figure 49*a*), the absolute value of the four above-mentioned terms are averaged only at the braid region. Both top and bottom surfaces have been considered in this measurement.

Two different density ratios have been analyzed in this section. The non-dimensional characteristics of these two cases are: $Re_l = 2500$, $We_l = 14\,400$, $\hat{\mu} = 0.0066$, and $\hat{\rho} = 0.05$ and $0.5$. The sheet thickness is $h = 50$ µm, and no initial perturbation is imposed on the liquid/gas interface.

Figure 44 shows the contribution of each term in the generation of $\omega_x$ for the high and low density ratios, in the first 20 µs. Baroclinicity (circles) is the most important factor at low density ratio, since the density gradient normal to the interface is higher, and also the local density in the gas zone near the interface is lower (see the baroclinic vorticity generation expression given above). However, at high density ratio, baroclinicity is the least significant. Baroclinicity is only slightly larger than the streamwise stretching (squares) in the beginning of the computations for high density ratio, but it is outrun by this term at about 13 µs and remains the lowest of all terms thence. The baroclicinc vorticity generation term is an order of magnitude smaller than the vortex stretching and tilting terms at the end of the computations.

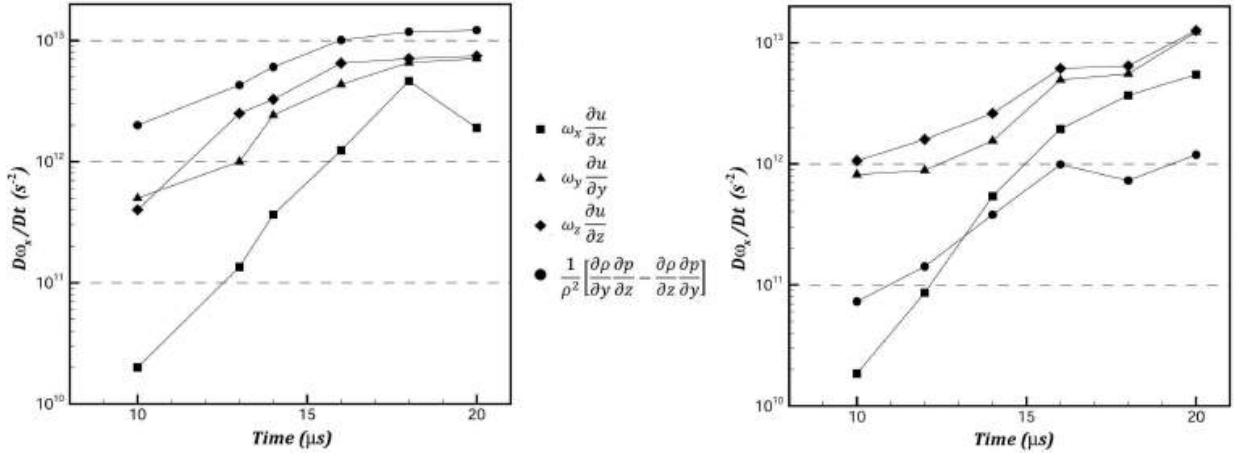

FIGURE 44. Contributions of streamwise vortex stretching (squares), spanwise vortex tilting (triangles), normal vortex tilting (diamonds), and baroclinicity (circles) to the generation of $\omega_x$ at the liquid surface for two density ratios; $Re_l = 2500$, $We_l = 14\,400$, $\hat{\mu} = 0.0066$; $\hat{\rho} = 0.05$ (left), and $\hat{\rho} = 0.5$ (right).

In both low and high density ratios, the spanwise and normal tilting seem to be more important than the streamwise stretching for the first 20 microseconds. The spanwise tilting is high because the initially spanwise vortex lines are gradually tilted in the streamwise direction. The normal tilting is also high since the velocity gradient is much higher in the normal direction ($\partial u/\partial z$), in the beginning of the computations. The streamwise stretching however has the highest growth rate and almost reaches the magnitude of vortex tilting at about 20 µs. Later on, as $\omega_x$ grows, the vortex stretching becomes more significant. These results are consistent with the findings of Jarrahbashi & Sirignano (2014) for round liquid jets. Jarrahbashi & Sirignano (2014) also concluded that, generally, the importance of baroclinic vorticity generation (RT instability) has been overemphasized in the literature, especially at very high pressures, and other important aspects of vorticity dynamics and similarities with injection into an alike fluid have been neglected. As described by them and also evident in our figures, the vortex tilting terms are the largest at early times; however, new findings show some cancellations, discussed below.

Figures 45 and 46 show the contours of the four $\omega_x$ generating terms at 13 µs, on a $y$-plane and $x$-plane, respectively. The spanwise and normal vortex tilting terms (figures 45b,c and 46b,c) are stronger than the streamwise stretching, but with opposite signs. As demonstrated in figure 44, the two vortex tilting terms are also nearly equal. A closer look at their equations explains the reason:

$$\omega_y \frac{\partial u}{\partial y} = \boldsymbol{\frac{\partial u}{\partial z} \frac{\partial u}{\partial y}} - \frac{\partial w}{\partial x} \frac{\partial u}{\partial y}, \tag{3.3}$$

$$\omega_z \frac{\partial u}{\partial z} = \frac{\partial v}{\partial z} \frac{\partial u}{\partial z} - \boldsymbol{\frac{\partial u}{\partial y} \frac{\partial u}{\partial z}}. \tag{3.4}$$

The two terms in bold font are exactly the same but have opposite signs. Thus, the only difference in the magnitudes of the spanwise and normal tilting comes from the other terms (the second term in equation 3.3 and the first term in equation 3.4). However, since these terms deal with gradients of $v$ and $w$ of $O(1)$, which are two orders of magnitude smaller than $u$ of $O(10^2)$, they are much smaller than the boldface terms, at the beginning of the computations. Thus, the deviation of the vortex tilting terms from the boldface terms is very small, and the two terms nearly cancel each other early on. Based on this, our earlier conclusion (and also that of Jarrahbashi & Sirignano 2014) should be modified: the spanwise and normal vortex tilting terms, even though the largest among the $\omega_x$ generating terms, are not the most important in $\omega_x$ generation, since they nearly cancel each other. The streamwise vortex stretching and baroclinic effects (RT instability) are the most important in generation of $\omega_x$, at high and low density ratios, respectively.

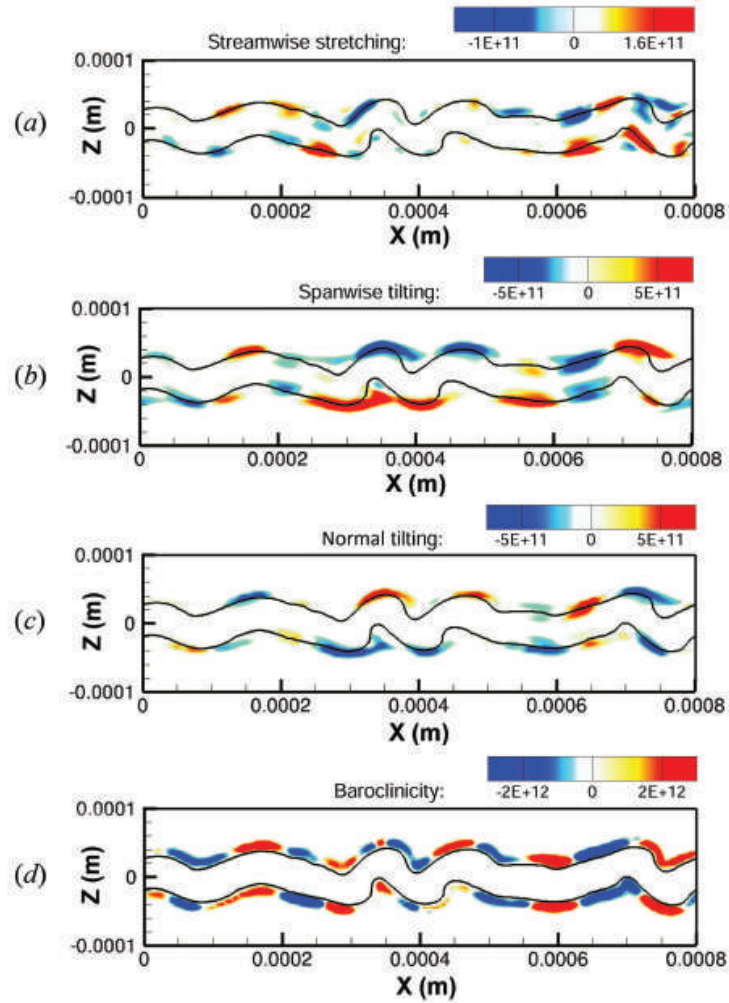

FIGURE 45. Contours of streamwise vortex stretching (*a*), spanwise vortex tilting (*b*), normal vortex tilting (*c*), and baroclinic generation (*d*) on a *y*-plane at $t = 13$ µs for $\hat{\rho} = 0.05$; $Re_l = 2500$, $We_l = 14\,400$, $\hat{\mu} = 0.0066$.

Figure 45(*a*) also confirms that the vortex stretching originates from the braids first, as the strain due to the adjacent primary vortical structures is highest at the saddle (braid) and the ribs are aligned along the diverging separatrix (Hussain 1986). Note that most of the colored spots in figure 45(*a*) are on the braids and not the wave crests. This is consistent with the experimental observations of earlier researchers (Bernal & Roshko 1986; Lasheras & Choi 1988; Liepmann & Gharib 1992) for uniform-density flows. The location and direction of the stretch can also be seen in figure 47, which shows the fluctuation velocity vectors relative to the average KH vortex velocity on a blown-up section of the liquid jet at 13 µs. It is evident that the saddles with the highest strain rate are on the braids between two adjacent vortices, where the fluctuation velocity vectors depart in the opposite directions. The stretch direction at the saddle point is shown by the green arrows in figure 47. The saddle points are in the gas phase, close to the

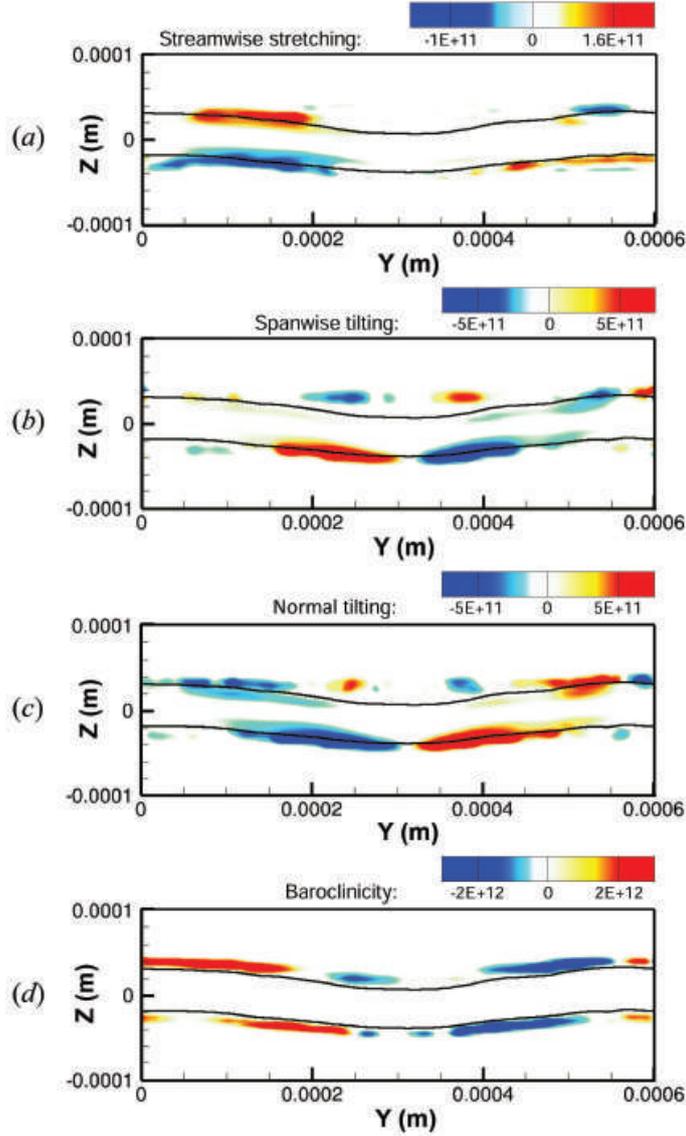

FIGURE 46. Contours of streamwise vortex stretching (*a*), spanwise vortex tilting (*b*), normal vortex tilting (*c*), and baroclinic generation (*d*) on a *x*-plane at $t = 13$ µs for $\hat{\rho} = 0.05$; $Re_l = 2500$, $We_l = 14\,400$, $\hat{\mu} = 0.0066$.

interface. The streamwise vortex stretching is highest at the saddle points (see figure 45*a*), where the flow is primarily discrete ribs (Hussain 1986). The fluid elements are stretched along the interface, i.e. along the diverging separatrix shown by the green arrows, and compressed normal to the interface at the saddle points. The center of the spanwise vortices (rolls) are at the crest of the interface waves, as denoted by the velocity vectors and the vorticity contours. Interestingly, the vorticity peak coincides with the interface at the crests, but not at the troughs, where the vorticity has migrated into the gas phase further away from the interface. Note that the vorticity contours look like crescents and not circular as the vorticity is not uniformly distributed as in a vortex column (like the typical Oseen vortex; see Pradeep & Hussain 2006); the non-centric vorticity distribution is due to the curved KH rollers.

The stretching and tilting terms are centered at the interface (see figures 45*a*–*c* and 46*a*–*c*), but the baroclinic torque term $\frac{1}{\rho^2}\nabla\rho \times \nabla p$ is always larger in the gas phase (see figures 45*d* and 46*d*). Baroclinic generation, being proportional to $1/\rho^2$, is two orders of magnitude larger in the gas phase compared to the liquid phase (for $\hat{\rho} = 0.05$). As density ratio increases, the difference between the gas and liquid densities decreases; thus, the contours of baroclinicity get closer to the liquid interface. This is seen in the baroclinicity contours of $\hat{\rho} = 0.5$ in figure 48. Since the density ratio is an order of magnitude higher than that of figure 45(*d*), the local density in the gas zone is much higher, hence the baroclinicity in the gas is lower and closer to its value in the liquid. Thus, the contours are closer to the interface (compare figures 48 and 45*d*). This contributes to the $\omega_x$ peak being closer to the interface and growing larger compared to the low density ratios, hence creating and stretching more lobes at higher density ratios. Also, the peak of baroclinic torque is an order of magnitude smaller in figure 48 compared to figure 45(*d*), since the density gradient normal to the interface is lower at higher density ratios.

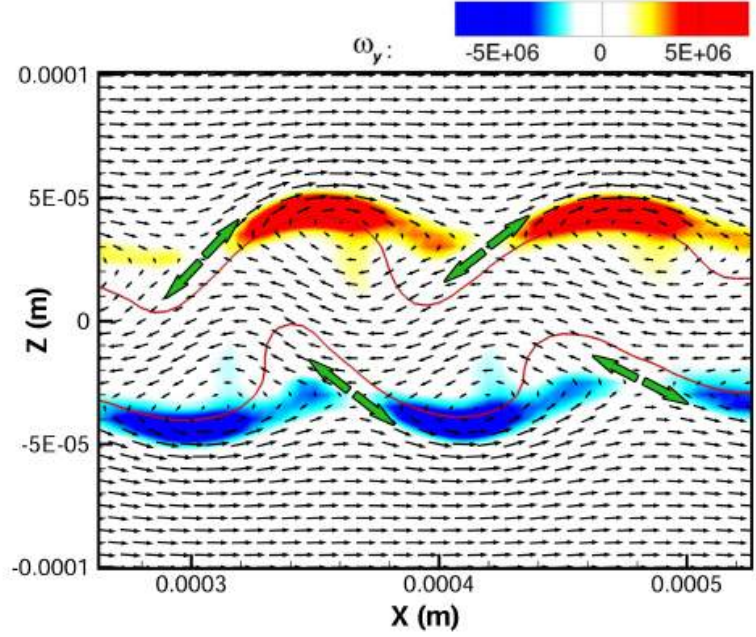

FIGURE 47. Perturbation velocity vectors superimposed on $\omega_y$ contours at $t = 13$ µs on a $y$-plane; $Re_l = 2500$, $We_l = 14\,400$, $\hat{\rho} = 0.05$, and $\hat{\mu} = 0.0066$. Green arrows denote the maximum stretch along the diverging separatrix. The liquid/gas interface is indicated by the red line.

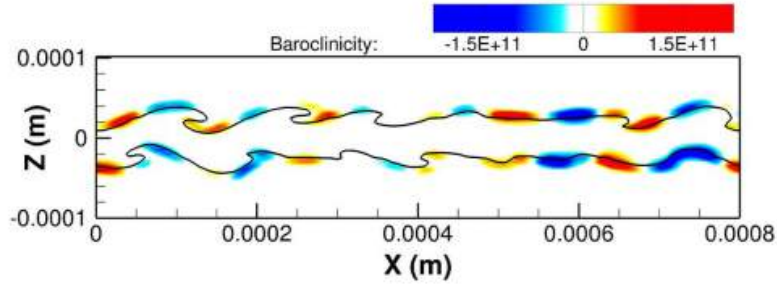

FIGURE 48. Contours of baroclinicity on a $y$-plane at $t = 13$ µs for $\hat{\rho} = 0.5$; $Re_l = 2500$, $We_l = 14\,400$, $\hat{\mu} = 0.0066$.

As shown in figure 45(*d*), the baroclinicity contours change sign (i.e. color) continuously in $x$. This pattern is very similar to the hairpin pattern seen in the $\omega_x$ contours presented by Jarrahbashi & Sirignano (2014). The $\omega_x$ and $\omega_y$ contours for the current case are illustrated in figure 49. Comparison of the contours of $\omega_x$ and baroclinicity (compare figures 49*a* and 45*d*) shows that they follow a very similar pattern. Hence, we can conclude that baroclinicity is the most important factor in creation of the hairpin vortex structure at low density ratios. The role of the baroclinic torque in deformation of the surface waves in a stratified shear layer is studied in detail by Schowalter *et al.* (1994), and is consistent with our results. The vector field of figure 47 shows that $\omega_x$ and $\omega_y$ are highest at the braids and wave crests, respectively. This is also evident in the vorticity contours of figure 49.

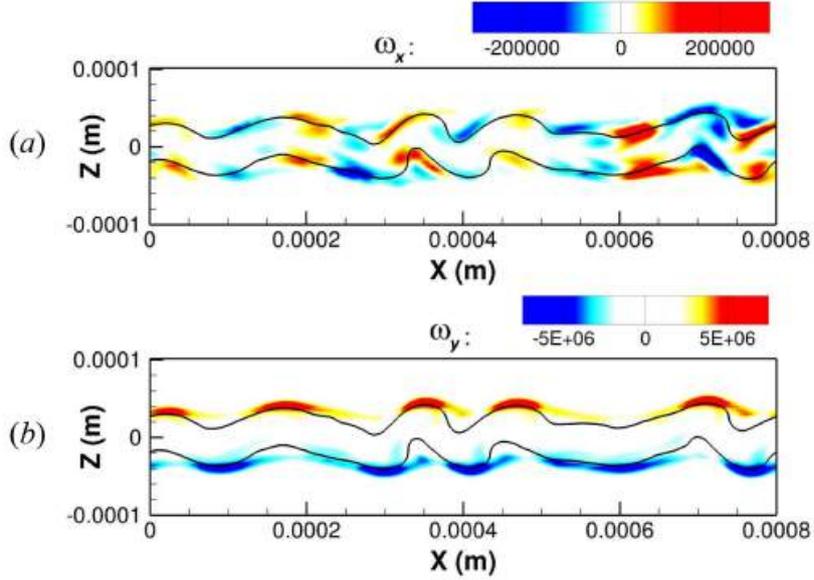

FIGURE 49. Contours of streamwise vorticity $\omega_x$ (a), and spanwise vorticity $\omega_y$ (b) at $t = 13$ μs for $\hat{\rho} = 0.05$; $Re_l = 2500$, $We_l = 14\,400$, $\hat{\mu} = 0.0066$.

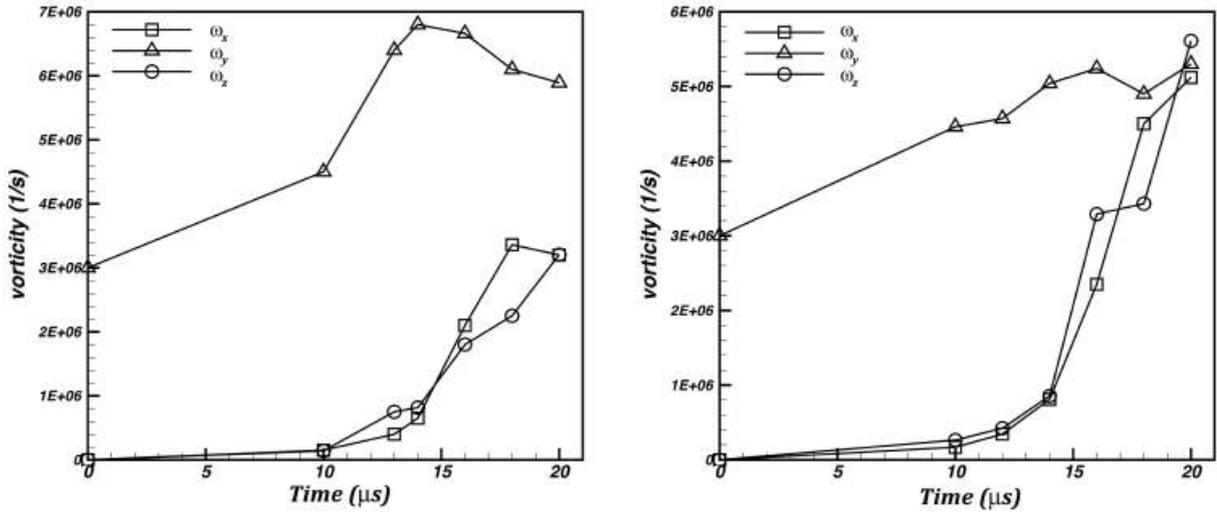

FIGURE 50. Temporal variation of the vorticity components for $Re = 2500$, $We = 14\,400$, $\hat{\mu} = 0.0066$; $\hat{\rho} = 0.05$ (left), $\hat{\rho} = 0.5$ (right).

In order to understand how fast the liquid sheet deforms and manifests 3D instabilities, the magnitudes of the vorticity components are examined through time. The 3D instabilities are directly related to the magnitudes of $\omega_x$ and $\omega_z$ against $\omega_y$, which exists from the beginning when the flow is still 2D. As mentioned earlier by Jarrahbashi & Sirignano (2014), Jarrahbashi *et al.* (2016), and Zandian *et al.* (2016), the streamwise vorticity is the main cause of the three-dimensional instabilities and interface distortion.

The absolute value of each vorticity component, averaged over the entire liquid/gas interface, is plotted in figure 50 for low and high density ratios. The spanwise vorticity $\omega_y$ (KH vortex) is the only component that exists initially. In both cases, $\omega_y$ grows for a certain time, until about 15 μs, and then onward keeps more or less the same order of magnitude. The spanwise vorticity is larger for lower density ratio.

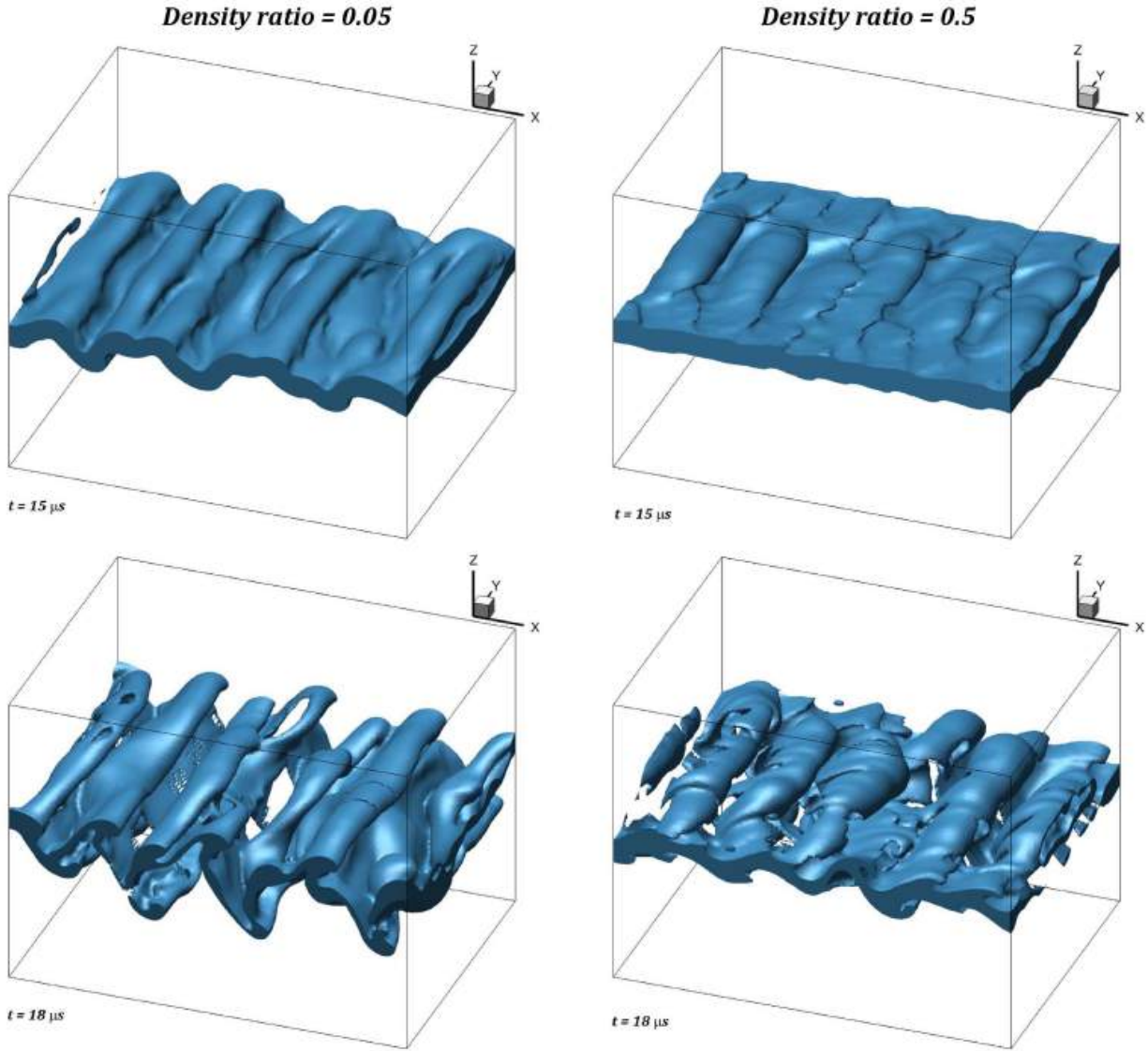

FIGURE 51. Liquid/gas interface at $t = 15$ μs (top) and 18 μs (bottom) for $Re_l = 2500$, $We_l = 14\,400$, $\hat{\mu} = 0.0066$; $\hat{\rho} = 0.05$ (left), $\hat{\rho} = 0.5$ (right).

$\omega_x$ and $\omega_z$ grow very slowly up to 10 μs, after which, there is a sudden increase in their growth rate. In both cases, $\omega_x$ and $\omega_z$ are of the same order of magnitude, which indicates that initially spanwise vortex filaments are lifted in the normal direction and tilted in the streamwise direction at almost equal rates. The growth rate of $\omega_x$ is higher at higher density ratios. $\omega_x$ reaches the same order of magnitude as $\omega_y$ at $t = 20$ μs, for the high density ratio. For the lower density ratio, however, the growth is slower. Hence, three-dimensionality manifests sooner at higher gas densities. For $\hat{\rho} = 0.05$, the magnitude of $\omega_x$ and $\omega_z$ are still half of $\omega_y$ at 20 μs; notice that $\omega_y$ always remains much higher. Thus, $\omega_y$ is still dominant and 2D deformations build up while the streamwise vorticity grows. $\omega_x$ growth has consequent impacts on the surface dynamics. In order to understand this, the liquid surface has been compared at two instances for both low and high density ratios in figure 51. The boxes in this figure show the computational domain edges. The boundary remains far away from the liquid surface; hence the results are not influenced by the domain size.

The higher $\omega_x$ growth rate at high density ratio causes the liquid surface to undergo

3D instabilities faster, and there are more streamwise lobes seen at high $\hat{\rho}$ than at low $\hat{\rho}$. At 15 µs, the surface of the sheet with $\hat{\rho} = 0.05$ is still roughly two-dimensional, while the high density-ratio case manifests more 3D deformations, and streamwise lobes are apparent on top of the primary KH waves. On the other hand, the higher $\omega_y$ compared to $\omega_x$ at low $\hat{\rho}$ causes the liquid sheet to become antisymmetric much faster (compare the top images of figure 51). Recall that as explained in §3.2, transition towards antisymmetry expedites when the two vorticity layers - on top and bottom surfaces - become stronger, as $\omega_y$ grows.

The difference in the vorticity dynamics between the two cases has significant effects on the characteristics of the jet instabilities. As can be seen at $t = 18$ µs (figure 51), the low $\hat{\rho}$ case can be characterized by roll-up of the KH waves, which creates more spanwise-aligned liquid structures and fewer stretched lobes; the entire sheet thins faster, and the liquid sheet breaks sooner. On the other hand, at high $\hat{\rho}$, the liquid structures orient streamwise more and manifest more lobes. The lobes are more stretched and thinned due to the larger $\omega_x$ of the KH vortex legs and are more prone to perforation. Hence, the hole-formation mechanism is expected to prevail over a larger area in the parameter space of $We_l$ versus $Re_l$, at higher density ratios. This is consistent with the zones in figure 1, and is in accordance with the density ratio effects discussed by Zandian *et al.* (2017).

The top and bottom liquid surfaces tend towards an antisymmetric mode, whether we start with a flat surface or symmetric perturbations. The antisymmetric behavior is eventually favored since a planar jet is more unstable to the antisymmetric mode than the symmetric mode in the parameter range of interest. The transformation towards antisymmetry occurs sooner as density ratio is lowered.

Marmottant & Villermaux (2004) "suggest" that, for a coaxial round jet, the transverse (azimuthal) deformation of the wave crests depends on surface tension ($We_g$), density ratio ($\hat{\rho}$), and thickness of the vorticity layer. They propose the possibility that the transverse instability of the wave crest is the result of lobe and ligament formation due to opposing shear and surface tension. Unsteady motions at the sheet rim confer transient accelerations to the liquid perpendicular to the liquid-gas interface, which trigger a RT type of instability, producing indentations of the rim, which later result in ligaments.

In a physically different configuration for leading rim of a transient liquid sheet, Agbaglah *et al.* (2013) also propose that the dynamics of a receding liquid sheet initiates a RT instability due to surface tension. They show that the growth rate of this transverse RT instability increases as the liquid sheet decelerates. Hence, they suggest that this instability is dominant at low $We_g$, where the deceleration is strong enough so that the retraction forces overcome the liquid/gas inertia. Our mechanisms for the lobe and ligament formation – via vortex interactions – would not be in effect for a configuration where the oppositely-oriented hairpin pairs do not exist – such as the leading rim of a liquid sheet studied by Agbaglah *et al.* (2013). Our results here do not indicate that capillary action plays a major role in the deformation of the crest rim to create three-dimensional structures, e.g., lobes, corrugations and ligaments. Once ligaments are formed, capillary action becomes important.

### 3.6. *Vortex dynamics of round jets*

In this section, a qualitative comparison is made between the vortex dynamics of the planar jets (studied here) and round jets (from previous studies) to show that the causes of different liquid structures, e.g. lobes, holes, corrugations, and ligaments, are the same from the vortex dynamics perspective. For this purpose, the round liquid jet computational analysis of Jarrahbashi & Sirignano (2014) and Jarrahbashi *et al.* (2016)

are mainly incorporated, which are most pertinent to our vortex dynamics analysis. In a few cases, numerical results of Shinjo & Umemura (2010) are also addressed.

Jarrahbashi & Sirignano (2014) used the vorticity dynamics to explain the formation of lobes and ligaments in their computations. They mainly used the vorticity contours and the vortex lines projection on the liquid surface in their analysis. Even though the explanation for the hole and corrugation formation was less detailed in their analysis, the vortex dynamics related to lobe formation and ligament elongation in their round jets is very similar to our findings. Similar to the planar jets, Jarrahbashi & Sirignano (2014) and many researchers before them (Liepmann & Gharib 1992; Martin & Meiburg 1991; Brancher *et al.* 1994; Shinjo & Umemura 2010) observed the origination of the streamwise vorticity in the braid region of the round jet first. Later, as the vortices stretch in the streamwise direction, counter-rotating streamwise vortex pairs are formed. These are shown to be three-dimensional hairpin vortices that wrap around the KH rollers. The streamwise vorticity in the ring originates from the upstream braid adjacent to the ring. The hairpin vortex in the ring region is 180° (half wavelength) out of phase with respect to that of the braid region, similar to what we observed in our planar sheet. The lobe locations on the cone crests is correlated with locations of the hairpin vortices in the ring (crest) region.

A very similar mechanism involving hairpin vortices is proposed by Jarrahbashi & Sirignano (2014) for the formation of ligaments. They show that the streamwise vorticity projection on the ligament surface changes sign inside the ligament. Therefore, counter-rotating vortex structures exist both near the lobes in the gas phase and inside the elongated ligament. The side-jet phenomenon in homogeneous jets is similar to the formation of the ligaments in two-phase jet flow based on these hairpin vortical structures. The effects of the pressure gradient on the ligament elongation and breakup is also addressed by Jarrahbashi & Sirignano (2014). The absolute value of the pressure decreases toward the center of the ligament at the smallest cross-sectional area; however, the pressure decreases radially outward at the tip of the ligament. Therefore, there is an oscillation of the pressure gradient inside a ligament that produces a waviness on its surface. According to Shinjo & Umemura (2010), capillary waves in the short-wave mode propagate inside the ligament and decrease the diameter of the ligament locally. Surface tension pinches the ligament at minimum cross-sectional area. However, the large streamwise vorticity observed inside the ligament might produce a large angular momentum and compete with the capillary force. Jarrahbashi & Sirignano (2014) define the competition between the vorticity effects and the capillary forces through a local Weber number, which is the ratio of the radial pressure gradient due to the transverse velocity difference caused by the streamwise vorticity inside the ligament and the radial pressure gradient due to surface tension, i.e. $We = \rho_l \omega_x^2 R^3 / \sigma$; where $R$ is the radius of the ligament. These two pressure gradients have different signs. If this $We$ is much larger than one, the vorticity effects inside the ligament can be more important than the capillary force. In computations of Jarrahbashi & Sirignano (2014) $We$ is very close to unity, hence inertia can be as important as capillary forces in the ligament breakup.

In a later study, Jarrahbashi *et al.* (2016) described the mechanism of hole and bridge formation in a round jet using the vorticity dynamics. They arrived at the same conclusion of hairpin overlapping for hole formation as described herein (figure 52). However, the spanwise measure is replaced by the azimuthal angle for round liquid jets. At high $Re_l$ of interest, inertial effects dominate and vortex lines and material lines are almost identical. Thus, the hairpins are stretched around the braid and curled at the front of the crest following the fluid motion. In the vicinity of the lobe, the downstream stretching hairpin from the ring passes over the upstream stretching hairpin from the braid as the curling action continues, causing the phase variation sketched in figure 52(*a*). Jarrahbashi *et al.* (2016) also observed that the hairpin vortices approach each other and form a diamond-shaped region of the type shown by Comte *et al.* (1992), very similar to the vortex lattice observed experimentally for single-phase planar mixing layers (see figures 3 and 4 of Comte *et al.* 1992). The hole formation process relates to the increase in circulation originated from the hairpin vortices that envelope the lobe and make it thinner with time, as was explained earlier. Hole formation begins on a lobe sheet where the $We$, based on lobe thickness and relative gas-liquid velocity, is too large for capillary action to be the initiating mechanism. As discussed earlier (figure 11), where two hairpin vortices with same-sign circulation overlap, the liquid sheet between them becomes thinner because their mutual induction moves the material lines closer to the same radial position and a hole can form (figure 52*b*).

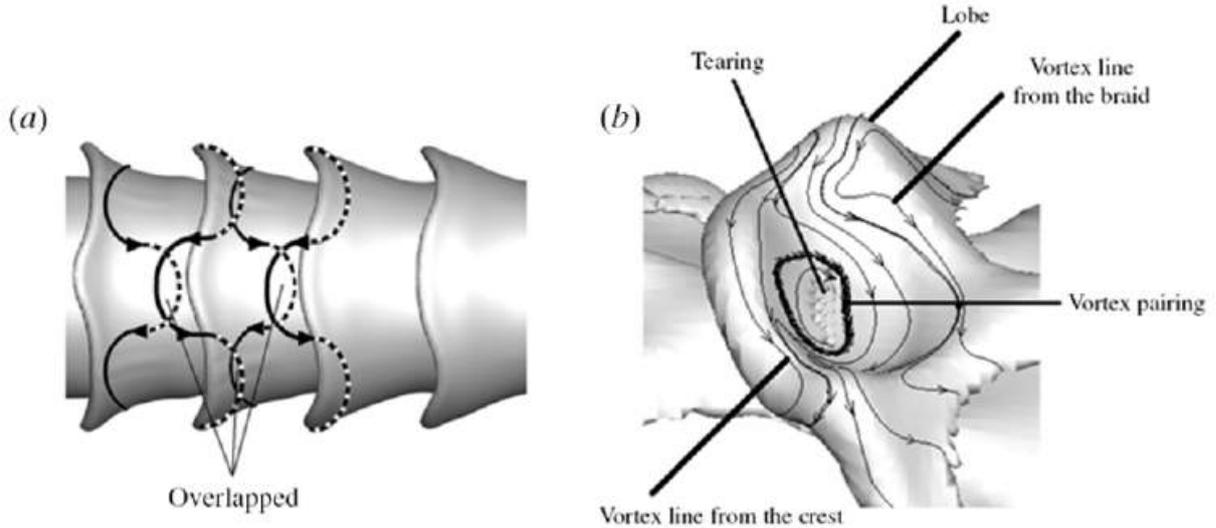

FIGURE 52. Isosurface of the liquid/gas interface accompanied by schematic of hairpin vortices on the ring and braid region (*a*); magnified lobe with projected vortex lines, showing two families of hairpin vortices from the braid and crest and their pairing at the center of the lobe leading to the lobe tearing and hole formation (*b*). $\hat{\rho} = 0.1$, $Re_l = 1600$, $We_g = 23\,000$. The solid and dashed lines show hairpin portions stretching downstream and upstream, respectively. Gas flows from right to left. Recast from Jarrahbashi *et al.* (2016).

Jarrahbashi *et al.* (2016) express that, although the hole location can be predicted by hairpin overlap, other flow parameters, e.g. density ratio, viscosity ratio and surface tension, play a role in changing the flow details and the hole formation process. For example, when the surface area of the lobe increases and its edge curvature decreases, the locations of the holes in neighboring lobes will be closer to each other, and the holes merge to create larger holes (Jarrahbashi *et al.* 2016).

The hole formation and ligament creation from the extension of the holes and tearing of the rim were observed in the computations of Shinjo & Umemura (2010) for the same range of $Re$ and $We$ numbers as Jarrahbashi *et al.* (2016). However, the lobe puncture was stated to be due to the impact of the droplets that formed earlier from the breakup of the mushroom-shaped cap on the jet core. Hence, their hole formation was inertially driven not vortically. Our temporal study produces no cap but better represents the behavior of spatial development after the cap has moved far downstream.

Jarrahbashi *et al.* (2016) relate the formation of small-scale corrugations to the smaller and less orderly alternating streamwise vorticity regions near the lobes for lower $Re_l$. However, they do not present a detailed explanation of the corrugation formation and stretching, nor do they explain the reason for such less orderly vortical structures near the ring region. Hence, even though our corrugation formation for planar jets is also related to smaller scales of the hairpin filaments (at high $Re_l$), this mechanism cannot be directly compared with their observations.

## 4. Conclusions

The present study has focused mainly on the vortex dynamics of planar liquid jets. A vortex has been defined using the $\lambda_2$ criterion. The relation between the surface dynamics and the vortex dynamics is sought to explain the physics of different breakup mechanisms that occur in primary atomization, by conducting DNS with level-set and volume-of-fluid surface tracking methods.

Vortex dynamics is able to explain the hairpins formation. The interaction between the hairpin vortices and the KH vortex explains the perforation of the lobes at moderate $Re_l$ and high $We_g$, which is attributed to the overlapping of a pair of oppositely-oriented hairpin vortices on top and bottom of the lobe. The formation of corrugations on the lobe front edge at high $Re_l$ and low $We_g$ is also explained by the structure that hairpins gain due to the induction of the split Kelvin-Helmholtz (KH) vortices. At low $Re_l$ and low $We_g$, on the other hand, the lobe perforation and corrugation formation are inhibited due to the high surface tension and viscous forces, which damp the small scale corrugations and resist hole formation. The hairpin vortices stretch in the normal direction while wrapping around the KH vortex. The induced gas flow squeezes the lobe from the sides and forms a thick and long ligament. In summary, the vortex dynamics analysis helps explain the three major atomization cascades at different flow conditions. The atomization mechanisms for the planar jet are qualitatively identical to the round-jet mechanisms.

Baroclinicity is the most important factor in generation of the streamwise vortices and manifestation of three-dimensional instabilities at low density ratios. At higher density ratios, the streamwise vortices are mostly rendered by streamwise vortex stretching. The streamwise vorticity growth is higher at higher density ratios, resulting in a faster appearance of three-dimensional instabilities. As density ratio is reduced, fewer lobes with less undulation form; hence, hole formation prevails more at higher density ratios. The relation between vortex dynamics and surface dynamics aids prediction of liquid-structure formations at different flow conditions and different stages of the primary atomization. This is very important in prediction and control of the droplet size distribution in liquid jet primary atomization.